\documentclass[12pt]{iopart}
\usepackage{graphicx}
\usepackage{xcolor}
\usepackage{cite}

\begin{document}

\title[Heterogeneous Anomalous Transport]{Heterogeneous anomalous transport in cellular and molecular biology}

\author{Thomas Andrew Waigh$^1$ \& Nickolay Korabel$^2$ }

\address{$^1$ Biological Physics, School of Physics and Astronomy,
University of Manchester, M13 9PL, UK}
\address{$^2$ Department of Mathematics, The University of Manchester, M13 9PL, UK}
\ead{t.a.waigh@manchester.ac.uk}
\ead{nickolay.korabel@manchester.ac.uk}
\vspace{10pt}
\begin{indented}
\item[]December 2021
\end{indented}

\begin{abstract}
It is well established that a wide variety of phenomena in cellular and molecular biology involve anomalous transport e.g. the statistics for the motility of cells and molecules are fractional and do not conform to the archetypes of simple diffusion or ballistic transport. Recent research demonstrates that anomalous transport is in many cases heterogeneous in both time and space. Thus single anomalous exponents and single generalized diffusion coefficients are unable to satisfactorily describe many crucial phenomena in cellular and molecular biology. We consider advances in the field of \emph{ heterogeneous anomalous transport} (\emph{HAT}) highlighting: experimental techniques (single molecule methods, microscopy, image analysis, fluorescence correlation spectroscopy, inelastic neutron scattering, and NMR), theoretical tools for data analysis (robust statistical methods such as first passage probabilities, survival analysis, different varieties of mean square displacements, etc), analytic theory and generative theoretical models based on simulations. Special emphasis is made on high throughput analysis techniques based on machine learning and neural networks. Furthermore, we consider anomalous transport in the context of microrheology and the heterogeneous viscoelasticity of complex fluids. HAT in the wavefronts of reaction-diffusion systems is also considered since it plays an important role in morphogenesis and signalling. In addition, we present specific examples from cellular biology including embryonic cells, leukocytes, cancer cells, bacterial cells, bacterial biofilms, and eukaryotic microorganisms. Case studies from molecular biology include DNA, membranes, endosomal transport, endoplasmic reticula, mucins, globular proteins, and amyloids.
\end{abstract}

%
%
%
%
%

\section{Introduction}
At the micron and submicron scale, particles suspended in simple fluids (e.g. liquids and gases) experience substantial random fluctuations in their displacements that are driven by collisions with the fluid molecules which surround them that are in turn driven by thermal forces. The displacement fluctuations are often independent of each other and the central limit theorem implies that the statistical distribution of their displacements tends to a Gaussian distribution provided a sufficient number of displacements are considered. This is \emph{classical diffusion} at the molecular scale and its experimental verification with small latex particles in solution led to a Nobel prize for J. B. Perrin in 1926. The experiments were based on the groundbreaking theories of A. Einstein (1905) and others (Langevin, Smoluchowski etc.). Mathematicians find that the central limit theorem applies to a wide range of basic probability distributions (and thus physical phenomena), but it can breakdown under some circumstances, particularly if the underlying distributions have fat tails \cite{bouchaud1990anomalous}. Such phenomena are now known to have important implications in a broad range of fields including finance,  
epidemiology, soft matter, quantum physics, condensed matter physics and anomalous transport in cellular and molecular biology \cite{metzler2000random,metzler2004restaurant,klages2008anomalous,barkai2012single,hofling2013anomalous,bressloff2014stochastic,metzler2014anomalous,banks2005anomalous}. 
The later subject will be the focus of the current review. 

Heterogeneous anomalous transport (HAT) sounds like an exotic phenomenon, but it is surprisingly common. It is the rule for the majority of molecules inside cells, the motion of individual cells and their collective motility. 
Anomalous transport of motile particles is defined broadly as the violation of any of the conditions of the central limit theorem. It is most prominently characterised by either non-Gaussian probability density functions for the particles' displacements and/or the non-linear power-law growth of the mean squared displacement of the particles on the time \emph{t},
\begin{equation}
\label{MSD}
\mbox{MSD(t)} 
\sim D_{\alpha}(r,t) t^{\alpha(r,t)}.
\end{equation}
Constant anomalous exponents $\alpha$  ($0<\alpha<2$) and the generalized diffusion coefficients $D_{\alpha}$ are the basic parameters used to describe homogeneous anomalous transport. For $\alpha = 1$ the motion is diffusive and for $\alpha = 2$ the motion is ballistic. However, the dependence of $\alpha(r,t)$ and $D_{\alpha}(r,t)$ on space ($r$) and time ($t$) in HAT gives rise to distinct phenomena. Equation (\ref{MSD}) is insufficient to fully characterise HAT since different phenomena can produce the same behaviour of the MSD. Therefore, we will discuss other quantities which can be used to more fully characterise HAT in experiments. Furthermore, machine learning techniques are rapidly developing that allow HAT to be probed avoiding (\ref{MSD}) which is prone to numerical errors, especially for short trajectories. 

Scaling laws, such as (\ref{MSD}), have a long history in physics, including developments in soft matter physics e.g. how the size of a polymer chain depends on the number of monomers \cite{de1979scaling,rubinstein2003polymer}. Fundamental reasons that physical laws often follow power laws are unknown and are probably unknowable \cite{feynman2011six}. 
Fractional powers are often a signature of fractal behaviour (with intermediate dimensionalities that are between integer values). Trajectories of diffusing particles are fractals, so it is perhaps no surprise that methods to quantify their behaviour (e.g. MSDs) also have self-similar scaling. HAT is thus closely connected with the study of multifractals, since heterogeneity causes the fractal exponents to vary. Multifractal analysis showed lots of initial promise in a broad range of research fields \cite{Mandelbrot1983}, 
but has not been extensively used in the interpretation of experiments due to challenges in making robust meaningful measurements. However, machine learning techniques combined with large high-resolution data sets are changing this situation \cite{munoz2021objective}. 

Our main goal is to give an overview of the physical mechanisms which lead to homogeneous anomalous diffusion, provide a manifesto for their uses to explain transport phenomena in molecular and cellular biology (Section 2), and discuss
the anomalous diffusion framework with an emphasis on its extensions to heterogeneous anomalous transport (Section 3). In Section 4 we explore the ever-growing number of applications of these methods to describe diffusion in heterogeneous media, along with experimental methods, microrheology, and measurable observables (Section 5). In Sections 6 and 7 we will focus on examples from cellular and molecular biology to constrain the review to a manageable size, although there is a much wider range of applications.
Before we embark on this journey we discuss the situations where HAT should be expected in place of homogeneous anomalous diffusion and its implications in various fields.

\section{Where and when to expect HAT}
 HAT occurs when the properties of anomalous transport change in space or time. The motility of particles can no longer be described by single constant anomalous exponents $\alpha$ and generalized diffusion coefficients $D_{\alpha}$ in  (\ref{MSD}). Instead, every trajectory accrues a space- and time-dependent anomalous exponent $\alpha(r,t)$ and generalized diffusion coefficient $D_{\alpha}(r,t)$. Thus, HAT is characterized by distributions of anomalous exponents and generalized diffusion coefficients. 
These distributions can vary with the time and length scale considered.  

The physical mechanisms which lead to homogeneous anomalous diffusion and the mathematical apparatus of fractional integration, which proved to be very useful for its description, are reviewed in many papers and books \cite{bouchaud1990anomalous,metzler2004restaurant,klages2008anomalous,barkai2012single,hofling2013anomalous,meroz2015toolbox,sokolov2005diffusion,sokolov2012models,zaburdaev2015levy}.
A variety of physical effects can give rise to heterogeneous anomalous diffusion in molecular and cellular biology. 
Effects are primarily attributed to either \emph{space} (molecular crowding, particle shape, particle size, caging and quenched 
disorder e.g., in glasses, gels and 
jammed systems) or \emph{time} (localized effects of a physiological process, activity of molecular machines e.g., enzymes or motors, transient binding reactions, escape out of confinement, aging, and fluctuations due to particle flexibility). In practice, the effects of space and time can rarely be separated and both need to be considered simultaneously i.e. the separation is slightly artificial and is directed by computational simplicity. 

Advances in single particle tracking (SPT) methods have provided the field of HAT with some robust experimental foundations in cellular and molecular biology. Measurements have created single-particle trajectories over long time scales (long tracks) that have anomalous statistics. Often the tracks have frequent switches between different anomalous transport regimes which indicate heterogeneity in time e.g. due to the switching behaviour of motor proteins \cite{saxton1997single,kusumi2014tracking,manzo2015weak}. The observed heterogeneity is not explained away by ensemble averaging over multiple particles, time averaging, or experimental noise. 
Other experimental techniques, such as fluorescence correlation spectroscopy (FCS) \cite{stolle2019anomalous,weiss2003anomalous}, dynamic light scattering \cite{berne2000dynamic} (or the more modern microscope-based equivalent, differential dynamic microscopy \cite{giavazzi2009scattering}), quasi-elastic neutron scattering \cite{kneller2005quasielastic} and nuclear magnetic resonance (NMR) \cite{calandrini2010fractional} have given further useful, but more indirect, evidence for HAT.

In general, it is challenging and currently impossible to find a universal description
for the dynamics of molecules in living cells due to the high concentrations of extremely complex molecules that cells contain that are constantly interacting with one another. Each biological molecule thus experiences a huge number of interactions with other molecules and the majority of these interactions are badly characterised due to experimental limitations. Rather than attempting to create universal soft matter models, which are likely to fail, studies of the statistics of anomalous transport provide agnostic model-free characterisation of the stochastic processes involved (an important first step for the description of the biological molecule). Some early applications of soft matter models in cellular/molecular biology have neglected accurate characterisation of the statistics of motion and are poor descriptions as a result.

\subsection{Polydispersity and diffusing diffusivity}

In dilute solutions every flexible molecule, and thus an ensemble of flexible molecules, will experience heterogeneous transport since changes in each molecule's size will occur in response to thermal forces, which will modulate the instantaneous diffusion coefficients of their centres of mass; this is the phenomenon of \emph{diffusing diffusivity}. Thus, a single isolated flexible molecular chain will experience heterogeneous diffusion. This is in addition to heterogeneity displayed on the ensemble level due to variations in the size and chemistry of a population of molecules (called polydispersity and chemical heterogeneity respectively), e.g. due to the imperfect synthesis of the molecules. Heterogeneous diffusion due to variations of size is commonly characterised in physical chemistry laboratories, e.g. via light scattering experiments where polydispersity indices are defined \cite{berne2000dynamic}. Heterogeneous diffusion due to flexibility has been much less explored than that due to chemical heterogeneity \cite{yamamoto2021universal,miyaguchi2017elucidating}, but will contribute to a wide range of phenomena in soft matter physics causing a broadening of dynamic processes, e.g. during reptation of flexible polymeric chains in concentrated solutions.

\subsection{Viscoelasticity}

From another perspective, the majority of complex fluids demonstrate heterogeneous viscoelasticity on micron and nanometer length scales and this is directly associated with HAT. Such behaviour is explored in the field of \emph{microrheology} \cite{waigh2005microrheology,waigh2016advances}. The manner in which materials respond to stress and strain is heterogeneous in \emph{space} due to their complex structuration on small length scales. This in turn impacts the efficiency of transport of cells and molecules as they move through the heterogeneous microenvironments created by the complex fluids from which they are constructed. Molecular motors acting at the molecular scale, gelation, glassy phenomena, jamming or driven flows at larger length scales can lead to a \emph{time} dependence of the mechanical responses (e.g. thixotropy or aging) and thus heterogeneous viscoelasticity can occur in both \emph{time} and \emph{space}. For equilibrium systems, the generalized fluctuation-dissipation theorem (GFDT) provides a rigorous link between HAT and heterogeneous viscoelasticity 
\cite{mason1995optical}. Specifically, HAT that follows (\ref{MSD}), a power law dependence of the mean square displacement on time, corresponds to a heterogeneous power-law fluid. The creep compliance ($J(t)$) of a microsphere of radius $a$ can be calculated from the GFDT,
\begin{equation}
\label{Compliance}
J(t)=J_0 t^{\alpha}= (\pi a/k_B T), 
\end{equation}
where \(k_B T\) is the thermal energy and \(J_0\) is a constant \cite{xu1998compliance,bonfanti2020fractional}.
Analogous expressions are expected to hold for non-equilibrium systems connecting HAT with viscoelasticity. 

A similar equation to (\ref{MSD}) can be written for angular diffusion \cite{hunter2011tracking}, which holds in the small time limit,
\begin{equation}
\label{MSDgamma}
\langle \Delta \varphi ^2 \rangle \sim D_\gamma t^\gamma, 
\end{equation}
where $\langle \Delta \varphi ^2 \rangle$ is the mean angular displacement of a microsphere, $D_\gamma$ is the generalized angular diffusion coefficient and $\gamma$ is an anomalous exponent. Heterogeneity is also observed for anomalous angular transport and it can be quantified in angular microrheology experiments \cite{cheng2003rotational}. Extensions have been made to consider diffusing diffusivity of angular motion in colloidal suspensions \cite{jain2017diffusing}. 

\subsection{Turbulence}

The turbulence of fluids provides a range of challenging statistical problems \cite{frisch1995turbulence}. Many cells are subjected to turbulent flows e.g. microorganisms in marine environments or blood cells in the aorta. Biological molecules can also experience classical turbulence e.g. biopolymer chains can give rise to the fireman's hose effect in turbulent flows. Other varieties of intermittent non-linear flow phenomena have also been defined, such as \emph{elastic turbulence} (e.g. shear flows of semi-dilute DNA solutions \cite{malm2017elastic}) and \emph{active turbulence} (e.g. in bacterial suspensions \cite{dunkel2013fluid}). Curiously, both these extended classes of turbulence occur in low Reynolds number flows (opposite to classical turbulence) and they have distinct statistical properties for particle motions e.g. the spectra of their velocity fluctuations is distinctive. 

In the context of HAT, turbulence is novel in that it corresponds to an accelerated type of motion and the mean square displacement can have a super-ballistic scaling e.g. in classical turbulence the anomalous exponent, $\alpha$, is 3 in equation (\ref{MSD}) \cite{shlesinger1993strange} i.e. \( \langle R^2(t) \rangle \sim t^3 \), which is called \emph{Richardson's law}. This stochastic accelerated scaling behaviour has been explained using a L\'evy walk model \cite{shlesinger1987levy}. Beyond turbulent flows, experimental observation of accelerated particle motion in cellular and molecular transport is very rare, presumably due to the systems' low Reynold's numbers and thus overdamped dynamics.

Multifractal scaling is a modern paradigm to explain the universal features of turbulence \cite{meneveau1991multifractal}  i.e. the flows are heterogeneous in time and space. How to quantify HAT in turbulence predominantly remains an unsolved problem, although neural networks are providing some new tools \cite{brunton2020machine}.

\subsection{Wavefronts in reaction-diffusion systems}

Reaction-diffusion (RD) equations are often used to describe pattern formation in cellular and molecular biology \cite{mendez2010reaction,keener2009mathematical}. A prototypical example is
$\partial c/\partial t=D \partial^2 c/\partial x^2 +f(c)$,
which is Fick's second law for diffusion of a molecule with concentration, \emph{c}, and diffusion coefficient, \emph{D}, plus a reaction term, $f(c)$. 

Mathematically, the additive contribution of the reaction term in RD equations is also applicable to describe anomalous reaction-diffusion processes with no memory \cite{volpert2013fronts}. However, for systems with subdiffusion of the continuous time random walk variety, the addition of a reaction term is physically inconsistent \cite{henry2006anomalous}, and no universal description of anomalous reaction-diffusion processes in terms of partial differential equations is currently available. More hope is currently offered by numerical solutions of anomalous RD models e.g. using agent-based models. Furthermore, in cellular and molecular biology often small numbers of molecules determine pattern formation via RD equations and extensive work has been performed to understand the impact of fluctuations in the number of molecules involved on Brownian RD processes \cite{erban2020stochastic}.

Heterogeneity due to the curvature of substrates, varying diffusion coefficients, varying local geometry (porous materials and gels) and varying density will all affect the propagation of the RD wavefronts. A specific example is electrical signalling in bacterial biofilms \cite{blee2019spatial} e.g. the speed of propagation of wavefronts of potassium ions in a mushroom-shaped biofilm will be affected by the local curvature of the mushrooms. We anticipate that agent-based models combined with HAT will be particularly useful for studying wavefronts in reaction-anomalous diffusion.

\section{Anomalous diffusion framework and its extension to HAT}

The abundance of Brownian (normal diffusive) processes in nature is a consequence of the central limit theorem (CLT). Under conditions of finite mean $\mu$ and variance $\sigma^2$, with independent and identically distributed random displacements $X_1, X_2, ..., X_n$, the CLT implies the convergence of their partial sums $S_n=X_1 + X_2 + ... + X_n$
to a standard normal (Gaussian) distribution, $\mathcal{N}(0,1)$, with zero mean and unit variance,
\begin{equation}
\label{CLT}
\frac{S_n-n\mu}{\sigma \sqrt{n}} \rightarrow \mathcal{N}(0,1), \; \;  n \rightarrow \infty.
\end{equation}
Such normal distributions are thus ubiquitous in physical phenomena, as their name implies e.g. normal distributions describe the probabilities of random errors in the majority of physical measurements. 

Anomalous diffusion corresponds to physical phenomena in which the CLT no longer holds in its standard form \cite{bouchaud1990anomalous}. The CLT has been generalized to a larger class of stable distributions. If there is a sequence of constants $a_n$ and $b_n$, the generalized CLT (GCLT) states that the sum of a large number of independent identically distributed random variables, including those with infinite variance, converge to L\'evy stable distributions \cite{mainardi2007levy} 
\begin{equation}
\label{GCLT}
\frac{S_n-b_n}{a_n} \rightarrow \mathcal{S}_{\alpha}(\beta,\gamma,\delta), \; \;  n \rightarrow \infty,
\end{equation}
where $\alpha$ is a characteristic exponent, $\beta$ is a skewness parameter, $\gamma$ is a scale parameter and $\delta$ is a location parameter. The only stable distributions that have closed forms for their densities are the normal distribution $\alpha=2$, the Cauchy distribution $\alpha=1$, and the L{\'e}vy distribution $\alpha=1/2$. The characteristic exponent $0 < \alpha \le 2$ controls the thickness of the distribution tails. For $\alpha < 2$, the PDF of the displacements ($ x$) is asymptotically proportional to $|x|^{-\alpha-1}$ as $x \rightarrow \pm \infty$. Therefore,  stable distributions, except the normal distribution, have long tails. 

A large variety of different mathematical models for anomalous transport have recently been developed and their number is ever increasing. Therefore a full account is beyond the scope of this review. Here we mention several approaches in molecular and cellular biology and their promising generalizations to HAT.

$\bullet$ \emph{Scaled Brownian motion (SBM).} Perhaps one of the simplest interpretations of the non-linear growth of the mean squared displacement (\ref{MSD}) can be given in terms of decelerating Brownian diffusion with a diffusivity $D$ that changes in time as a power law, MSD$(t)\simeq D(t)\ t$ with $D(t) \sim t^{\alpha-1}$. This defines a non-stationary process with the probability density function that coincides with that of the fractional Brownian motion (FBM) equation (\ref{FBM_PDF}) below \cite{lim2002self,jeon2014scaled,chechkin2017brownian,metzler2014anomalous}. In contrast to FBM, SBM is non-ergodic (see the discussion of ergodicity in section 5.13) \cite{fulinski2011anomalous}. Although it might be difficult to demonstrate such a dependence of diffusivity on time in experiments, clearly this interpretation belongs to the domain of HAT. A further generalization of SBM towards HAT can be realized by making the diffusion constant a function of both time and space, $D(r,t)$. A deterministic dependence of $D(r,t)=\vert r\vert^{\beta}t^{\alpha-1}$ was studied in \cite{cherstvy2015ergodicity}.
 
$\bullet$ \emph{Fractional Brownian motion (FBM).} FBM is a self-similar Gaussian
process with stationary increments which is characterized by the self-similarity Hurst exponent $H$. FBM can be modelled via the Langevin equation driven by the fractional Gaussian noise (fGn), $\xi_{fGn}(t)$ \cite{metzler2014anomalous}:
\begin{equation}
\label{FBM}
\frac{dx}{dt} = \xi_{fGn}(t).
\end{equation}
FGn is a stationary Gaussian process defined by the correlation function:
\begin{equation}
\label{FGN}
\left< \xi_{fGn}(t)\xi_{fGn}(t+\tau) \right> = 2H(2H-1) D_H \vert\tau\vert^{2H-2}.
\end{equation}
where $D_H$ is a diffusion coefficient and $\tau$ is the time interval. This correlation function is negative for $0<H<0.5$, zero for $H=0.5$, and positive for $0.5<H<1$ making the $\xi_{fGn}$ antipersistent for $H<0.5$ and persistent $H>0.5$. Since $\xi_{fGn}$ is a Gaussian process, the probability density function (PDF) for FBM is given by the Gaussian distribution:
\begin{equation}
\label{FBM_PDF}
P(x,t) = \frac{1}{\sqrt{4 \pi D_H t^{2H}}} \exp\left(-\frac{x^2}{4 D_H t^{2H}}\right).
\end{equation}
The MSD can be obtained as $\left< x^2(t) \right>=\int x^2 P(x,t) dx$ and reads:  
\begin{equation}
\label{FBM_PDF}
\left< x^2(t) \right>= 2 D_H t^{2H}.
\end{equation}
The anomalous diffusion
exponent $\alpha$ is related to the Hurst exponent by $\alpha = 2H$.

Recently, a variety of FBM-like stochastic models driven by fGn with random diffusivities were considered in \cite{wang2020fractional,wang2020unexpected,mackala2019statistical}. Also, FBM can be generalized to HAT via \emph{multi-fractal fractional Brownian motion} (mFBM) with a time-dependent Hurst exponent $H(t)$ ($H(t)$ is a deterministic function of time) \cite{peltier1995multifractional,benassi1997elliptic,adler2010geometry}. It has been used in modelling turbulence \cite{benzi1984multifractal} and time series data for ECGs \cite{falconer2004fractal, harte2001multifractals, mandelbrot2013multifractals} or generalized gray Brownian motion defined as FBM with random generalized diffusion coefficients drawn from the Mittag-Leffler distribution \cite{mura2008characterizations,mura2008non}. Generalizations of FBM characterized by time random Hurst exponents $H(t)$ \cite{ayache2005multifractional,molina2016fractional} and random $H(t)$ and $D_H(t)$ were successfully applied for the description of the heterogeneous motion of endosomes \cite{han2020deciphering,korabel2021local,korabel2023ensemble} and the motion of cells in Drosophila embryos \cite{korabel2022hemocytes}. 

The fractional Langevin equation (FLE) shares many common features with FBM. Similar to generalizations to FBM discussed above, FLE with time- \cite{jeon2014scaled,cherstvy2016anomalous} and space-dependent diffusion coefficients \cite{cherstvy2013population} are good candidates to model HAT. 

$\bullet$ \emph{Continuous time random walk (CTRW)}. The CTRW is a process that consists of instantaneous jumps, $\delta r$, drawn from the probability distribution, $\phi(\delta r)$, interspersed with periods of waiting drawn from the waiting time probability distribution, $\psi(\tau)$ \cite{montroll1965random,klafter2011first}. Depending on the existence of the mean and variance of these distributions, a wide variety of behaviours from sub- to superdiffusion can be modelled. For example, the power-law distributed waiting times $\psi(\tau) \simeq \tau_0^{\alpha}/\tau^{1+\alpha}$ with $0< \alpha < 1$ which leads to subdiffusion $\left< r^2(t) \right>= 6 D_{\alpha} t^{\alpha}$ was used to interpret diffusion of tracer microbeads in actin networks \cite{wong2004anomalous}, the motion of colloidal particles \cite{xu2011subdiffusion}, and diffusion in the plasma membrane of living human cells \cite{weigel2011ergodic}.

\emph{Heterogeneous CTRW} with a spatially dependent exponent $\alpha(x)$ was considered in \cite{chechkin2005fractional,korabel2010paradoxes,korabel2011anomalous,roth2020inhomogeneous}. Furthermore, models with
variable distributions for the steps \cite{dentz2012diffusion,grebenkov2018heterogeneous,barkai2020packets} and  characterized by space- and density-dependent  anomalous exponents \cite{korabel2010paradoxes,korabel2011boundary,korabel2011anomalous,kosztolowicz2012subdiffusion} are promising to characterize HAT in heterogeneous glassy systems.
 
$\bullet$ \emph{L\'evy walks (LW)}. The simplest realization of a LW is a continuous time random walk model in velocity space  \cite{metzler2014anomalous,zaburdaev2015levy}. It describes a particle that travels in a certain direction with constant velocity $v$ for a random period of time $\tau$ drawn from the running time probability distribution $\psi(\tau)$. After that period, the particle instantaneously and randomly changes its direction and the process is repeated. The position of the particle is given by $r(t) = \int_0^t v(t') dt'$. If the running time distribution is a power law $\psi(\tau) \simeq \tau_0^{\alpha}/\tau^{1+\alpha}$ with $1< \alpha < 2$, the LW has subdiffusive behaviour $\left< r^2(t) \right>= 6 D_{\alpha} t^{3-\alpha}$. More complex LWs can involve different velocities (either constant for each running segment drawn from the distribution of velocities or randomly fluctuating) \cite{zaburdaev2008random} or periods of rests between runs \cite{klafter2011first}.

\emph{Heterogeneous L\'evy walks} were proposed by introducing the space dependence of the directional switching rate. To model the interactions of walkers and the emergent superdiffusive regime, the switching rate was controlled by the local density of walkers \cite{fedotov2017emergence}.  Another model contained some statistical features of LW but describes HAT using an underdamped 
Langevin equation where each particle was characterized by its own relaxation time and velocity diffusivity  \cite{sliusarenko2019finite}. 

$\bullet$ Various \emph{subordinated mixtures} of models for anomalous transport have also been developed with multiple sources of anomalous transport \cite{weigel2011ergodic,tabei2013intracellular,kozubowski2006fractional,fox2021aging} e.g. FBM with CTRW. 

$\bullet$ \emph{Distributed and variable order time and space fractional diffusion equations.} 
Chechkin and colleagues studied the diffusion-like equation with a time-fractional derivative of the distributed order whose diffusion exponent varies with time \cite{chechkin2003distributed,chechkin2012natural,sandev2015distributed}. This describes a random process in which the exponent of the MSD decreases  (retarding sub-diffusion) or increases (accelerating sub-diffusion) in time. It was also used to describe ultraslow and anomalous aggregation \cite{fedotov2019asymptotic}. Distributed order space fractional diffusion equation can describe accelerating superdiffusion \cite{chechkin2012natural}.

$\bullet$ \emph{Soft-matter models.}
An alternative perspective is to consider models for anomalous transport that have their origins in soft-matter physics \cite{rubinstein2003polymer}. This is a rich area of research and has the advantage that anomalous dynamics are directly related to molecular architectures. Thus Zimm, Rouse and semi-flexible modes are expected for the sub-diffusive internal dynamics of polymer chains \cite{granek1997semi}. Membrane dynamics have their own characteristic sub-diffusive spectra. The collective dynamics of simple colloidal materials are also rich in phenomena e.g. in dense colloidal suspensions cage hopping is observed with a spatially dependent diffusion coefficient \cite{pusey1991liquids}. However, in molecular and cellular biology the variety of interacting molecules is often so vast, complex, and inaccurately characterised that the motion of particles cannot be directly related to their molecular structures. Thus statistical approaches that are agnostic with respect to the exact molecular details are preferable e.g. HAT. Furthermore, HAT emphasizes the statistics of the processes involved, which are often neglected in standard soft-matter models, e.g., standard Zimm, Rouse and semi-flexible models of polymers assume Gaussian fluctuations which are often poor approximations when HAT occurs. Thus, HAT models can be more accurate in describing the underlying statistical physics of the biological processes considered.

Anomalous transport leads to the motion of molecules in cells and these often react with one another. Reaction-diffusion equations form the mathematical framework to describe these phenomena and they have been important to understand pattern formation in biology, e.g. Alan Turing's seminal work on morphogenesis \cite{turing1990chemical,pismen2021active}. Similar to equation (1) the mean square position of a wavefront, $\langle R^2 (t) \rangle$, can be considered as a function of time (\emph{t}) and regimes of anomalous transport can be defined,
\begin{equation}
\label{Wavefront}
\langle R^2 (t) \rangle
\sim R_c ^2 + D_{\beta} t^{\beta}
\end{equation}
where $R_c$ is the critical radius for nucleation of the wavefront, $D_\beta$ is a generalized diffusion coefficient and $\beta$ is the anomalous scaling exponent. Similar to $\alpha$, when $\beta<1$ the wavefront moves sub-diffusively, $\beta=1$ is diffusive, $\beta>1$ is super-diffusive, $\beta=2$ is ballistic and $\beta>2$ is super-ballistic. Historically, the non-ballistic motion of wavefronts was considered analytically by perturbative calculations from perfect ballistic motion by invoking the idea of pushed and pulled wavefronts \cite{dieterle2021diffusive}. However, these models struggle to handle the motion of strongly anomalous wavefronts. Thus, anomalous reaction-diffusion equations have been developed to understand particles with anomalous motility \cite{mendez2010reaction} e.g. waves of activity that spread across cells. How heterogeneity affects the motion of these waves \cite{xin2000front} has not been quantified in detail in experiments. Agent based models for bacterial biofilms are able to predict super-diffusive wavefronts based on a fire-diffusive-fire formalism across a heterogeneous biofilm (with varying bacterial concentration) \cite{blee2019spatial}. A subtlety is seen in how to define the position of the wavefronts in such systems. With bacterial signalling the best choice is the threshold concentration for excitation of the cells, not the peak concentration of the diffusing particles (which is strongly affected by the dimensionality and shape of the biofilms). 

An extensive range of theoretical models were developed for transport in porous heterogeneous materials driven by technological applications e.g. oil exploration in which the oil needs to be extracted from porous rock. Some of these models could be applied to biology, although they tend to stress heterogeneity in \emph{space} and ignore that in \emph{time} \cite{lanoiselee2018diffusion,lanoiselee2018model}.

\section{Modelling heterogeneous diffusion in heterogeneous media}

Diffusion in heterogeneous media is intimately related to heterogeneous anomalous transport.    
Diffusion in heterogeneous media itself shows many anomalous features. A classical example is the diffusion on percolation clusters \cite{ben2000diffusion}. Relatively new developments include anomalous diffusion caused by ensemble heterogeneity and randomly changing diffusivity.  

\subsection{Diffusion in quenched random environments}

The study of diffusion in disordered materials has a long history \cite{bouchaud1990anomalous} with many applications e.g. the motion of oil in porous materials for extraction by the petroleum industry or water in glassy carbohydrates which is important in food spoiling and production.
Heterogeneous diffusion in porous materials has been considered in detail by the pulsed NMR community
\cite{callaghan2011translational}. Models were developed based on the analogy of an ant crawling diffusively in a maze. Simple models relate the rate of diffusion at long times to the porosity of the maze \cite{callaghan2011translational}.
Diffusion on percolation clusters, named by de Gennes ‘an ant in a labyrinth’  \cite{de1976percolation}, is the simplest system to study how diffusion is affected by the presence of microscopic random irregularities.  Anomalous diffusion can result in the long-time behaviour of the ant if the maze has a fractal structure with fractal dimension $d_w$, $\mbox{MSD(t)} 
\sim D t^{2/d_w}$. For continuum percolation (the Swiss cheese
model) in three dimensions, 
$d_w \simeq 3.2$ corresponds to subdiffusion \cite{bunde2005diffusion} and a non-Gaussian propagator \cite{metzler2004restaurant}. Due to strong disorder, the generalized diffusion coefficients for single realizations of clusters, $D=\mbox{MSD(t)}/t^{2/d_w}$, fluctuate randomly in time but their distributions at fixed times show universal fluctuations \cite{pacheco2022universal}. Furthermore, a universal L{\'e}vy distribution of single-particle diffusivity was found in a quenched trap model (a random walk on a quenched random energy landscape) with diverging mean trapping time  \cite{akimoto2016universal}. 
 
\subsection{Diffusing diffusivity models}

For annealed disorder, a particle diffusing in a heterogeneous environment can be characterized by the time-dependent local diffusion coefficients randomly changing in time, $D(t)=\mbox{MSD(t)}/t=2dk_BT/\gamma(t)$,  
following a stochastic differential equation   \cite{chubynsky2014diffusing,chechkin2017brownian,metzler2020superstatistics,metzner2015superstatistical}. Chubynsky and Slater considered the position of a Brownian particle, $x$, described by Langevin equation with random time-dependent diffusivity that was governed by Ornstein–Uhlenbeck process via an auxiliary variable $Y$: 
\begin{equation}
\label{DD}
\frac{d}{dt} x(t) = \sqrt{2D(t)} \zeta(t),
\end{equation}
\begin{equation*}
\label{DD}
D(t) = Y^2, \; \; \frac{d}{dt} Y(t) = -\frac{Y(t)}{\tau} + \sigma \nu(t),
\end{equation*}
where $\zeta$ and $\nu$ are independent Gaussian processes, $\tau$ is the correlation time of the Ornstein-Uhlenbeck process, and $\sigma$ is the amplitude of the fluctuations of $Y$. The squared dependence, $D(t) = Y^2$, guarantees the positivity of the diffusion coefficient.
The MSD, in this case, grows linearly with time but the displacement distribution of particles has a non-Gaussian form at short times scale (hence the name “Brownian 
yet non-Gaussian”  diffusion) and transitions to a Gaussian distribution at longer times \cite{wang2012brownian}. Evidence for this type of motion was found for submicron tracers moving along linear phospholipid bilayer tubes and in entangled actin networks \cite{wang2009anomalous}, for tracer dynamics in hard-sphere colloidal suspensions, nanoparticles adsorbed at fluid interfaces  and membranes, inside colloidal suspensions and the motion of nematodes (see \cite{metzler2020superstatistics} for a recent review and references therein).

Diffusing diffusivity phenomenon is also expected in heterogeneous diffusion of single molecules in dilute solutions due to the flexibility and reflects internal dynamic modes. Examples include flexible polymers (Rouse or Zimm chains), semi-flexible polymers, and membranes \cite{granek1997semi}.
Other phenomena, such as reptation, the snake-like motion of entangled polymers in concentrated solutions, depend on one-dimensional diffusion along a tube and it is also affected by heterogeneous diffusion. Heterogeneous diffusion in reptation will be driven by both chain flexibility and the interaction with neighbouring flexible chains in the tubes.

\subsection{Superstatistical models: anomalous diffusion caused by ensemble heterogeneity}

Often non-Gaussian probability distributions of displacements can emerge as a consequence of polydispersity in the particles' masses and radii \cite{gheorghiu2004heterogeneity}
or particles moving in patches of different local properties attaining different diffusion constants, $D$ \cite{wang2012brownian}. In contrast to diffusing diffusivity models, $D$ in superstatistical models are random variables, but are assigned constant values for each trajectory. The probability distributions of displacements $P(x,t)$ of such  systems are given by (according to the \emph{super-statistical approach}) the superposition of the probability distribution $G(x,t\vert D)$ to find a particle at $x$ at time $t$ given its diffusion coefficient $D$ with the probability distribution of diffusion coefficients $p(D)$ \cite{chechkin2017brownian,metzler2020superstatistics,beck2003superstatistics,beck2006superstatistical},
\begin{equation}
\label{superstat}
P(x,t) = \int_{0}^{\infty} G(x,t\vert D) p(D) dD.
\end{equation}
For a large class of HAT systems (containing \emph{anomalous yet Brownian diffusion} \cite{chubynsky2014diffusing,metzler2020superstatistics,wang2012brownian}), $G(x,t\vert D)$ is given by a Gaussian  distribution and thus non-Gaussian tails for $P(x,t)$ are due to the distribution of diffusion coefficients, $P(D)$.  Homogeneous diffusion with a single value for the diffusion coefficient, $D_0$, corresponds to a Dirac delta function probability distribution for $D$, $p(D)=\delta(D - D_0)$. If the probability distribution of diffusion coefficients is exponential ($D_0$ is now a characteristic value for the distribution of diffusion coefficients), $p(D)=\exp(-D/D_0)/D_0$, or follows the Weibull density function $p(D) = c (D^{c-1}/D_0^{c}) \exp(-(D/D_0)^c)$ ($c>0$), the resulting probability distribution of displacements has a two-sided exponential (Laplacian) form, $P(x,t) \simeq \exp(-\vert x \vert/\sqrt{D_0 t})/\sqrt{4 D_0 t}$ \cite{korabel2022hemocytes,chubynsky2014diffusing,wang2012brownian,lampo2017cytoplasmic}. The Gamma probability distribution of diffusion coefficients ($\gamma$ and $\beta$ are constants), $p(D) \sim D^{\beta-1} \exp(-\gamma D)$, produces Laplacian tails in the $P(x,t)$ \cite{hapca2009anomalous}. On the contrary, power-law distributed diffusion coefficients, $p(D) \sim D^{-1-\gamma}$,  lead to a power-law density of displacements, $P(\zeta) \sim \vert\zeta\vert^{-1-2\gamma}$, where $\zeta=x/\left<x^2\right>^{1/2}$ is the scaled displacement \cite{korabel2021local,chechkin2017brownian,sadoon2018anomalous}. 
As discussed in \cite{wang2012brownian}, a constant probability distribution of polydisperse diffusion coefficients is not the only interpretation of (\ref{superstat}). 
Instead, a particle can intermittently change gears to attain different diffusion coefficients from the distribution $p(D)$.

The superstatistical approach to random diffusivity can also be formulated via stochastic processes. The superstatistical model for FBM of a variable $Y(t)$ is defined as $Y(t) = X B_H(t)$ where $X$ is a random (but constant) diffusivity drawn from a distribution of diffusivities \cite{mura2008non,mura2009class,mura2008characterizations,molina2016fractional,sposini2018random,mackala2019statistical} and $B_H$ is FBM with zero mean and unit variance. A generalised Langevin equation,  which is closely related to FBM, describing non-Gaussian
viscoelastic anomalous diffusion, was studied with a superstatistical approach via a random parameterisation of the
stochastic force \cite{slkezak2018superstatistical}. If the scale parameter is given by $X=\sqrt{S_{\alpha}}$ with $S_{\alpha}$ drawn from the Mittag-Leffler distribution, the process is called generalized gray Brownian motion (ggBM) \cite{mura2008characterizations,mura2009class}. At short times ggBM is equivalent to the diffusing diffusivity model but behaves differently in the long time limit e.g., the probability distribution functions of ggBM remain non-Gaussian, whereas they tend to a Gaussian form in the diffusing diffusivity model.

\section{Measuring HAT: experimental methods and measurables}
State of the art experiments to probe heterogeneous transport in molecular or cellular biology  are mostly based on optical microscopy \cite{mertz2019introduction}. These experiments allow the discrimination of individual particles (e.g. cells, single molecules, organelles, nanoparticles etc) and allow intrinsic heterogeneity in both time and space to be probed simultaneously without significant ensemble averaging. There have been many modern developments in optical microscopy that are ideal for studies of HAT in cellular biology. Specific examples include selective plane illumination microscopy with embryos
\cite{korabel2022hemocytes}, confocal microscopy with bacterial biofilms
\cite{drescher2016architectural}, and super-resolution fluorescence microscopy (specifically, stochastic optical reconstruction microscopy, STORM) with bacterial capsular rafts on the membranes of live \emph{E. coli} \cite{phanphak2019super}. 

In our experience, the most robust platform for unambiguously and routinely measuring HAT over a wide range of time and length scales has been to use absorption contrast of particles in an inverted optical microscope with a bright LED and ultra-fast CMOS camera ($10^5$ fps) \cite{waigh2016advances}. The apparatus is identical to that extensively used by our group for particle tracking microrheology. Vibration damping is crucial in the design of the apparatus to isolate measurement noise from the intrinsic stochastic processes in soft biological materials. Thus a 'princess and the pea' architecture is used for our apparatus, with the microscope placed on a slab of acoustic isolation foam, which is placed on an active noise cancellation table, which in turn is placed on a large (high inertia) floated optical table. This provided us with routine access to particle tracks with sub $10^{-4}$ s time and 10 nm spatial resolution e.g. for endosomes containing gold nanoparticles inside human cancer cells \cite{kenwright2012first}.

 Fluorescence correlation spectroscopy (FCS) is normally microscope based and it has the advantage that faster motility can be explored (down to microseconds) than standard microscope-based methods that use fluorescence, which is useful for rapidly photobleached fluorophores, but length scale information tends to be lost in FCS (this can be ameliorated  using multiple experiments with different measurement volumes, but this is often impractical with cells \cite{stolle2019anomalous} and only intensity correlation functions can be conveniently extracted. 
 
A wide range of other experimental techniques support the existence of HAT. Inelastic and quasi-elastic scattering techniques demonstrate that anomalous transport exists ubiquitously in soft materials (synthetic polymers, membranes, globular proteins, etc.) at time scales down to picoseconds and length scales down to Angstroms. Such methods have been invaluable in establishing the field of anomalous transport (e.g. quasi-elastic neutron scattering, inelastic neutron scattering \cite{roosen2011protein}, X-ray photon correlation spectroscopy \cite{madsen2010beyond} and dynamic light scattering \cite{berne2000dynamic}, but studies are predominantly constrained to ensemble-averaged measurements with a relatively poor spatial resolution (precluding single cell measurements). NMR modalities also provide strong evidence for HAT in ensemble-averaged data sets \cite{callaghan2011translational}, but again their spatial resolution precludes single-cell measurements. Recently developed optically detected NMR techniques with nitrogen defects in diamonds provide improved resolution \cite{holzgrafe2020nanoscale}, but the measurements are then again optical microscope based (similar to standard instruments for HAT) and have not yet been specifically used for fluorescence-based studies of HAT.

Super-resolution fluorescence techniques based on stimulated emission-depletion (STED) microscopy have recently demonstrated with 1.7 nm resolution per millisecond for tracks of kinesins in live cells \cite{wolff2023minflux,deguchi2023direct}. Specifically, the MINIFLUX version of STED was implemented in which minimum intensities of overlapping fluorescence emissions from fluorophores excited with a series of laser pulses were measured. Such world-leading performance will be invaluable for future studies of HAT in molecular molecular biology e.g. to model the transition from subdiffusion to superdiffusion for transport with motor proteins \cite{kenwright2012first}.

\subsection{Heterogeneous microrheology}

Rheology is defined as the study of the flows of materials. Practically it can be thought of as a form of generalized fluid mechanics in which all materials that flow are considered, not just those that are purely fluids. Thus, viscoelasticity (in which the responses of materials to flow are intermediate between solids and liquids) is of central importance in rheological studies. 

Microrheology is an extension of classical rheological techniques to consider the flow of materials as a function of length scale; typically distances on the order of microns or smaller are considered \cite{waigh2005microrheology, waigh2016advances}. There are a range of microrheology techniques, but the majority measure the response of nm-micron-sized probes to known stresses (either thermal stresses in \emph{passive microrheology} or externally applied stresses in \emph{active microrheology}). With structured complex fluids at the nm-micron scale, the responses of the probes is invariably heterogeneous and they experience heterogeneous anomalous diffusion. The expression for the viscoelastic compliance (\ref{Compliance}) shows that heterogeneous anomalous diffusion of the probe spheres necessarily implies heterogeneous linear viscoelasticity in equilibrium systems. In practice this is true in both active and non-active systems, although (\ref{Compliance}) is not necessarily an accurate approximation \cite{mizuno2008active}. 

There is a large range of evidence for heterogeneous microviscoelasticity in polymeric, colloidal and surfactant suspensions \cite{waigh2005microrheology, waigh2016advances}. It is 
a universal phenomenon appearing in all gels, structured complex fluids, and semi-dilute polymeric solutions. Specific examples of heterogeneous compliance (and thus heterogeneous MSDs) in polymer gels include cytoskeletal filaments \cite{tassieri2008dynamics}, mucins \cite{georgiades2014particle} and aggrecans \cite{papagiannopoulos2008viscoelasticity}. Eukaryotic and prokaryotic cells both demonstrate heterogeneous intracellular microrheology \cite{rogers2008intracellular}.

In addition to heterogeneous viscoelasticity, the simpler field of heterogeneous viscosity is also actively being studied. For example, in suspensions of marine phytoplankton cells gradients of viscosity are observed in the vicinity of the cells e.g. 40 times the viscosity of seawater at 30 \( \mu \)m away from the cells \cite{guadayol2021microrheology}. Non-equilibrium concentration gradients of secreted molecules are expected to occur around the majority of cells (not just those embedded in biofilms or the extracellular matrix) and they will perturb the microrheology of their surroundings, which in turn will affect their motility.

\subsection{Measurables to probe heterogeneous anomalous transport}
Different statistical measures tend to emphasize different aspects of a stochastic data set. The choice of measure is dictated by the ability to robustly calculate them in experiments, their information content, and the ease with which they can be modelled, e.g. via analytic theory or stochastic simulations. In dealing with experimental HAT data, it is important to use as many complementary statistical tools as possible to fully describe the behaviour over as broad a range of time and length scales as possible. 

\subsection{Ensemble and time mean squared displacements}
The ensemble-averaged mean squared displacement (eMSD) of 3D trajectories with coordinates $r_i=\{x_i, y_i, z_i\}$ is defined as
\begin{equation}
\mbox{eMSD}(t) = \int_{-\infty}^{\infty} r^2 P(r,t) dr = \frac{1}{l^2} \left< (r_i(t)-r_i(0))^2 \right> = 6 D_{\alpha} \left( \frac{t}{\tau} \right)^{\alpha},
\label{emsd}
\end{equation}
where $P(r,t)$ is the probability density function to find the tracked particle at time $t$ and position $r$. The angled brackets denote averaging over an ensemble of trajectories, $\left< A \right>=\sum_{i=1}^{N(t)} A_i/N(t)$, where $N(t)$ is the number of trajectories in the ensemble available at time $t$.  
Plotting the eMSD versus the time is a first check for anomalous diffusion defined via (\ref{MSD}). The MSD can be fitted to a power law function to extract the anomalous exponent $\alpha$ and the generalized diffusion coefficient $D_{\alpha}$. Notice that $\alpha$ and $D_{\alpha}$ are constants that characterize averaged transport properties of the ensemble of trajectories. To make the generalized diffusion coefficient $D_{\alpha}$ dimensionless, the time and length scales are commonly set to $\tau=1$ sec and the length scale $l=1$ $\mu$m. This facilitates the comparison of generalized diffusion coefficients between experiments, otherwise, they have varying fractional units which makes life awkward e.g. plotting them on a single axis of a graph is challenging. 

A standard procedure for the estimation of the ensemble of the anomalous exponents and the generalized diffusion coefficients from the ensemble-averaged eMSD is to fit a straight line on a log-log scale using the least squares method to extract the slope and the intercept. The optimal number of data points for the best fit as well as accurate handling of the static and dynamic errors
were derived for Brownian diffusion \cite{michalet2010mean,vestergaard2014optimal} and for fractional Brownian motion \cite{backlund2015chromosomal,lanoiselee2018optimal}. Some guidelines were also given to optimally fit mean square displacements of single particles in the general case of anomalous diffusion \cite{kepten2015guidelines}. However, it is assumed the ensemble anomalous exponents represent all the trajectories at all times and, therefore, cannot accurately describe HAT. However, the eMSD is useful to determine whether ergodicity occurs as we discuss in Section 5.13.

\subsection{Time-dependent anomalous exponents and generalized diffusion coefficients}

The time dependent ensemble anomalous exponent $\alpha(t)$ can be calculated by the division of the logarithm of $\mbox{eMSD}(t)$ by the logarithm of time,
\begin{equation}
\alpha(t) = \frac{\log \mbox{eMSD}(t)}{\log t}.
\label{alpha_naive}
\end{equation}
This method is not very accurate and leads to large errors due to poor statistics at longer times, especially when trajectories in the ensemble have different lengths i.e. different tracks in the ensemble have different durations. Fewer data points are available for averaging at larger times, which makes the eMSD noisy. 

For Brownian diffusion in 3D, the instantaneous diffusion coefficient is calculated as
\begin{equation}
D_{inst}(t) = \frac{1}{6} \frac{\partial \mbox{eMSD}(t)}{\partial t},
\label{D_inst}
\end{equation}
and the equilibrium diffusion coefficient is $D_{eq}=\lim_{t \rightarrow \infty} D_{inst}(t)$. In the case of anomalous transport, the MSD increases as a power law (\ref{emsd}) and $D_{inst}(t)$ become coupled to the anomalous exponent $\alpha$. Therefore, a fractional derivative in (\ref{D_inst}) or other definitions of diffusion coefficient must be used. A two-parameter fit of (\ref{emsd}) is typically used to calculate the generalized diffusion coefficient $D_{\alpha}$. 

The time-averaged mean squared displacement (tMSD) of an individual trajectory is one of the most commonly used tools to probe diffusive properties of single-particle trajectories \cite{gal2013particle},
\begin{equation}
\mbox{tMSD}_i(t) = \frac{1}{l^2} \frac{1}{T-t} \int_{0}^{T-t} \left( r_i(t'+t)-r_i(t') \right)^2 dt', 
\label{tmsd}
\end{equation}
where $l$ is the characteristic length scale, $T$ is the duration of the track, and $i$ is the index of the track. For example, $l$ is often set as $l=1$ $\mu$m in experiments with cellular motion. The time averaging is used to improve the signal-to-noise ratios and typically the resulting short time data on the tMSD is much less noisy than the long time data due to the relative number of MSD samples that are averaged over.

\subsection{Velocity auto-correlation function}

The time-averaged velocity auto-correlation function (VACF) of an individual trajectory is defined as 
\begin{equation}
\mbox{VACF}(t) = \frac{ \int_{0}^{T-t - \delta} \vec{v}(t'+t) \vec{v}(t') dt'}{T-t - \delta}.
\label{tvacf}
\end{equation}
where $\vec{v}=(\vec{r}(t+\delta)-\vec{r}(t))/\delta$ , $\delta$ is the time increment, $r(t)$ is the particle displacement at time ($t$) and $T$ is the duration of the track. To improve statistics, the VACF can be further averaged over the ensemble of all the trajectories. For Brownian motion, the VACFs exponentially decay to zero, as observed in many cases of single-cell motility e.g. isolated human dermal keratinocyte cells \cite{selmeczi2008cell}. For anomalous diffusion processes, the decay of the VACF is non-exponential and very sensitive to the nature of the anomalous diffusion processes \cite{burov2011single,weber2012analytical}. Positive valued VACFs can be attributed to inertial effects or intermittent persistent movement \cite{korabel2021local}. Negative valued VACFs can be a sign of the viscoelasticity of the medium which induces antipersistent behavior in a particle’s trajectory \cite{weber2012analytical}.

\subsection{Local MSD, local anomalous exponent, and generalized diffusion coefficient}

The local tMSD (L-tMSD), defined as the tMSD$_i$ in a small sliding time window $(t-W/2,t+W/2)$ \cite{han2020deciphering,arcizet2008temporal,nandi2012distributions} (window duration \emph{W}), can be used to describe heterogeneous diffusion processes which switch between different states along a track. By fitting small numbers (e.g. $10$ data points) of L-tMSD data points to power law functions, the time-dependent local anomalous exponents $\alpha_L(t)$ and the time-dependent local generalized diffusion coefficients $D_{\alpha_L}(t)$ can be extracted,
\begin{equation}
\mbox{L-tMSD}(t) = 6 D_{\alpha_L} \left( \frac{t}{\tau} \right)^{\alpha_L}.
\label{ltmsdfit}
\end{equation}
The time scale $\tau=1$ sec 
is introduced in order to make the local generalized diffusion coefficients dimensionless and facilitate comparison. The behaviour of the MSD determines the character of the movement of the ensemble of particles with normal diffusion for $\alpha _L= 1$, sub-diffusion for $\alpha _L< 1$, and super-diffusion for $\alpha _L > 1$. Similarly, the time dependence of the L-tMSD determines the character of the local movement of a particle from its trajectory. 

\subsection{Distributions of anomalous exponents and diffusion coefficients}

Since HAT is defined by the time and space dependence of anomalous exponents and diffusion coefficients, one can characterise them by calculating their probability distributions. The probability distribution of local anomalous exponents $\alpha_L$ and generalized diffusion coefficients $D_L$ were used to describe HAT in the dynamics of endosomes within eukaryotic cells \cite{korabel2023ensemble} and the movement of single cells in Drosophila embryos \cite{korabel2022hemocytes}. However, one should keep in mind that limited statistics in a sliding window to calculate L-tMSD introduces errors in $\alpha_L$ and $D_L$ (see section 5.6 above) and distorts their true distributions (it tends to broaden them). For example, for FBM with constant anomalous exponent $\alpha_L=2H$ and constant diffusion coefficient, one would get a normal distribution of $\alpha_L$ and log-normal distribution of $D_L$ which will approach the expected Dirac delta distributions upon increasing the window size. Similar distortion was also found for individual anomalous exponents and diffusion coefficients due to the limiting statistics in L-tMSDs \cite{speckner2021single}. Another artifact of the calculation of MSDs in overlapping windows is the occurrence of spurious positive correlations between $\alpha_L$ and $D_L$, $D_L \sim \exp(\alpha_L)$. Interestingly, $\alpha_L$ and $D_L$ of hemocyte cells in Drosophila embryos were found to be anticorrelated \cite{korabel2022hemocytes}. One way to overcome these distortions is to avoid calculations of L-tMSDs altogether, for example by estimating anomalous diffusion and diffusion coefficient via machine learning techniques (see sections 5.14 and 5.15). Another possibility to obtain the probability distribution of diffusion coefficients for Gaussian processes (e.g. Brownian diffusion or FBM) is to use the Richardson-Lucy non-parametric method which is routinely used to deconvolve images that have been blurred by a Gaussian point spread function \cite{lucy1974iterative}. The method computes the probability distribution of diffusion coefficients, $f(D)$, iteratively by inverting the probability distribution of displacements $P(x,t)$ \cite{wang2012brownian}:
\begin{equation}
f_{i+1}(D) =  f_{i}(D) \int \frac{P(x,t)}{P_n(x,t)} G(x\vert D) dx,
\label{RLucy}
\end{equation}
where $P_n=\int_{0}^{D_0} f_n(D) G(x\vert D) dD$ is an approximation of the displacement distribution, $G(x\vert D)$ is the Gaussian density, and the initial guess could be chosen as exponential PDF, $f_1(D)=\exp(-\hat{D}/D)/\hat{D}$. The probability distribution $f(D)$ is normalized $\int_{0}^{D_0} f(D) dD=1$. In Sections 6 and 7 we will discuss various  probability distributions of local anomalous exponents and instantaneous diffusion coefficients measured in experiments which also depends on how the anomalous exponents and diffusion coefficients were defined.   

\subsection{First passage probability}

The first passage probability (FFP) can be the key statistical property in a physical problem e.g. the times for chemical reactions, the exit times of a particle from a maze, or the chain retraction times for reptation of branched polymers \cite{redner2001guide}. In general, the FPP is the probability distribution of the times for a process to reach a threshold value for the first time. For a set of tracks \(\{R_{n}(T)\}\), the first passage probability is calculated by finding the smallest positive time $t$ that satisfies \(|R_{n}(T+t)-R_{n}(T)|=L\) at each starting time point \emph{T} for each track (indexed by \emph{n}) that travels a length \emph{L} \cite{rogers2010first}. In the analysis of HAT, the FPP has the advantage that probability distributions can be considered as a function of transit length (\emph{L}) e.g. to ask the question of whether long endosomal transits travel faster than short transits \cite{kenwright2012first}. The mean of the FPP (the MFPP) provides an alternative to the MSD to characterise anomalous transport \cite{kenwright2012first}. Different scaling regimes of a FPP distribution on time necessitate that there will be different scaling regimes for any reaction rates (which can be anomalous), since all reaction rates are to a degree controlled by transport processes (chemical reactivity can also be important).

\subsection{Survival analysis}

Another fundamental statistical process to describe time-varying data is described using \emph{survival analysis}. This was originally developed in medicine and describes the fraction of patients alive in a trial after a certain time in terms of the survival probability $\Psi(t)$. Useful algorithms have been developed to correct for censuring of the data, e.g. when people leave a trial before they die, and the empirical survival probability can be estimated by using the non-parametric Kaplan-Meier estimator \cite{aalen2008survival}.
Furthermore, a \emph{hazard function} can be defined that facilitates the analysis of complex survival functions $\gamma(t) = - \Psi'(t)/\Psi(t)$ where $\Psi'(t)=d\Psi(t)/dt$, e.g. to characterise non-Poisson distributions in which the hazard rate varies with time. The survival function $\Psi(t)$ can be written in terms of the probability
distribution of the first passage time $F(t)$ as $\Psi(t)=1-F(t)$. 
Survival analysis was used to understand the switching between persistent and antipersistent motions of endosomes in human cells and hemocyte cells in drosophila embryos
\cite{korabel2022hemocytes,han2020deciphering,fedotov2018memory}.

\subsection{Directional persistence}

Directional persistence is a key parameter of particle tracks, which is invisible to a mean square displacement analysis since MSDs are only sensitive to amplitudes of motion, not directions \cite{harrison2013modes}. One method is to consider the average cosine of the angles between successive displacements ($r_i, r_j$) of the particle as a function of time interval ($\tau$). This can be conveniently calculated using vector calculus,
\begin{equation}
\langle \cos\theta (\tau) \rangle=\Biggl \langle\frac{r_i \cdot r_j}{|r_i| |r_j|} \Biggl \rangle.
\label{costheta}
\end{equation}
$\langle \cos\theta (\tau) \rangle$ is bounded between 1 and -1 which corresponds to persistent and anti-persistent motion respectively. The values of the cosine can be averaged over time and considered as a function of time interval ($\tau$), similar to a time-averaged MSD. Directional persistence was used to analyse tracks from endosomes in human cells and anti-persistence was induced by the motor protein tethers of the endosomal cargoes \cite{harrison2013modes}. 
 
\subsection{Intensity correlation function analysis}

An experimental technique introduced to measure the mobility of molecules is called \emph{fluorescence correlation spectroscopy} (FCS) \cite{elson201340}. The technique measures the time correlation functions of the intensity fluctuations of fluorescence emitted by fluorophores in a sample that are created by changes of the local concentrations. Often FCS can explore the dynamics of molecules at 2-3 orders of magnitude faster time scales than an equivalent particle tracking experiment using fluorophores due to the reduced requirements on its detector (point detectors are used for FCS as opposed to CMOS pixel arrays for microscopy imaging). Single molecules can also be examined with FCS by calculating the auto-correlation curve of the fluorescence intensity $F(t)$
\begin{equation}
C(\tau) = \frac{\left< \delta F(t) \delta F(t+\tau)\right>}{\left< \delta F(t)\right>^2},
\label{FCorr}
\end{equation}
where $\delta F(t) = F(t) - \left< F(t) \right>$ and $\left< ... \right>$ is the time averaging.
Fitting the
autocorrelation function, C($\tau$), to the function \cite{banks2005anomalous,weiss2003anomalous,lubelski2009fluorescence}
\begin{equation}
C(\tau) = \frac{A}{1+D_{\alpha}\tau^{\alpha}/B},
\label{FCS}
\end{equation}
where A and B are constants, one can determine the generalized diffusion coefficient $D_{\alpha}$ and  the anomalous exponent $\alpha$. The main weakness of the technique is that the correlation functions mean that data is analysed less directly, and both time and spatial averaging occur. There are therefore more sources of ambiguity in elucidating the fundamental causes of HAT in a FCS experiment than in an equivalent microscopy tracking experiment.

\subsection{Distribution of displacements and increments}

The probability density function (PDF), $P(r,t)$, to find a particle at displacement $r(t)$ at time $t$ assuming it begins from the origin $r(0)=0$, is an important statistical tool to describe the transport properties of particles. The ensemble mean-squared displacement is the second moment of the displacement PDF, which provides a connection to equation (\ref{emsd}). Particles moving via Brownian diffusion have Gaussian probability distributions of displacements. A departure from the Gaussian form is a hallmark of anomalous transport. In low-resolution data the departure from a Gaussian can be quantified using the fourth moment of the distribution, the \emph{kurtosis}, which is more robust to smaller numbers of samples than attempting to fit the full probability distribution. The third moment of the PDF is called the \emph{skew} and often indicates driven transport e.g. the activity of motor proteins or convective transport that breaks the symmetry \cite{perkins2021network}. However, skew is not indicative of anomalous transport, since driven motion with a significantly skewed PDF can occur in both Brownian and anomalous systems.

Non-Gaussian distributions of displacements were experimentally observed for a subdiffusive motion of cytoplasmic RNA-protein particles in live {\it Escherichia coli} and {\it Saccharomyces cerevisiae} cells \cite{lampo2017cytoplasmic} and for the acetylcholine receptors diffusing on live muscle-cell membranes \cite{he2016dynamic}.
Furthermore, the distribution of increments of $150$ nm diameter nanoparticles passively moving in {\it Dictyostelium discoideum} cells was found to have a parabolic Gaussian-like region for
small increments and an exponential tail for larger increments \cite{witzel2019heterogeneities}. 
In living cells, non-Gaussian transport features are thought to be due to 1) \emph{sample-based variability}, 2) \emph{rarely
occurring events} with large amplitude motions, 3) \emph{aging} and 4) \emph{spatiotemporal heterogeneities} of the medium.

Another important tool that can be used to test HAT is the distribution of increments $\Delta r$, $P(\Delta r,\tau)$, a particle makes during the time interval $\tau$. Brownian diffusion and fractional Brownian motion are characterized by  Gaussian $P(\Delta r,\tau)$. Non-Gaussian distributions of increments can be a signature of HAT.

\subsection{Ergodicity breaking and aging} 

For ergodic motion, such as Brownian diffusion, the \emph{Birkhoff ergodic theorem} allows the diffusive properties to be estimated using a single long trajectory via the 
equivalence of ensemble averages and
long-time averages
\cite{metzler2014anomalous,feller2008introduction},
\begin{equation}
\lim_{t \rightarrow \infty} \mbox{tMSD}_i(t) = \mbox{eMSD}(t).
\label{ergod}
\end{equation}
In experiments, frequently only short trajectories are available and the long time limit of $\mbox{tMSD}_i(t)$ is replaced by additional averaging over the ensemble of trajectories to improve statistics,
\begin{equation}
\mbox{tMSD}(t) = \left< \mbox{tMSD}_i(t) \right> = \mbox{eMSD}(t).
\label{etmsd}
\end{equation} 
Checking for the equivalence between the time-averaged and ensemble-averaged MSDs can be used to discriminate between different models of anomalous diffusion i.e. testing for ergodicity \cite{barkai2012single,metzler2014anomalous}. 
Several ergodicity breaking (EB) parameters were introduced to probe the ergodicity of anomalous diffusion \cite{metzler2014anomalous,he2008random}. An EB parameter based on the fourth moment of the particle displacement, 
\begin{equation}
\mbox{EB}(t) = \left< \xi^2(t)\right>-1,
\label{eb}
\end{equation} 
with $\xi(t) = \mbox{tMSD}_i(t)/\mbox{tMSD}(t)$, works well when all particle trajectories have the same duration. For heterogeneous trajectories which have different durations, the ergodicity breaking parameter is defined using the ratio between the ensemble and the time-averaged MSDs,
\begin{equation}
\mbox{EB}(t) = \frac{\mbox{tMSD}(t)}{\mbox{eMSD}(t)}-1,
\label{eb1}
\end{equation} 
works better. For ergodic processes, the EB parameter converges to zero.

Many anomalous transport processes in living cells demonstrate ergodicity breaking \cite{golding2006physical,jeon2011vivo}, i.e. the non-equivalence of time and ensemble averages, $\left< \mbox{tMSD}_i(t) \right> \ne \mbox{eMSD}(t).$ The time-averaged MSDs for single trajectories, $tMSD_i(t)$, remain random functions and can be described by probability distributions that depend on the anomalous dynamics. For example, with mono-fractional CTRW-type motion, this distribution takes the form of a one-sided L\'evy stable probability distribution \cite{he2008random} and approaches the expected delta-function as the anomalous exponent goes to 1. Ergodicity breaking could stem from the non-stationary nature of the process when the probability distribution of the times associated with immobilization events has a divergent mean \cite{barkai2012single,metzler2014anomalous}. Several experiments have shown that nonergodic behavior is a consequence of interactions occurring in heterogeneous cellular environments \cite{manzo2015weak,jeon2011vivo}.
Ergodicity breaking could also emerge as a consequence of the heterogeneity of the system, in the absence of immobilization \cite{molina2016fractional}. In some systems ergodic and non-ergodic processes coexist, e.g. in plasma membranes  \cite{weigel2011ergodic} and the intracellular transport of insulin granules \cite{tabei2013intracellular}.

Aging is probed by delaying the beginning of the measurement over the aging time $t_a$ from the initiation of the process at $t=0$ or equivalently aging the system in $(-t_a,0)$ and starting the measurement at $t=0$. Statistical aging is defined as the dependence of quantities on both the measurement time $t$ and the aging time $t_a$ \cite{monthus1996models}. Aging is related to ergodicity breaking \cite{bouchaud1992weak}.
The ensemble MSD for aging is \cite{barkai2003aging,burov2010aging}
\begin{equation}
\mbox{eMSD}(t,t_a) = \int_{-\infty}^{\infty} r^2 P(r,t,t_a) dr = \frac{1}{l^2} \left< (r_i(t+t_a)-r_i(t_a))^2 \right>.
\label{emsd_aging}
\end{equation}
Similarly, the aging tMSD of a single trajectory is defined as
\begin{equation}
\mbox{tMSD}_i(\Delta,t_a) = \frac{1}{l^2} \frac{1}{T'
-\Delta} \int_{t_a}^{t_a+T'-\Delta} \left( r_i(t'+\Delta)-r_i(t'))^2 \right) dt'.
\label{tmsd_aging}
\end{equation}
Notice that for experimental trajectories, the duration of a trajectory, $T'$, must be adjusted using $T'=T-t_a$, where $T$ is the duration of a non-aged trajectory. The adjustment can be omitted when dealing with simulated trajectories which can be made as long as required, but it is necessary in typical experiments. 

\subsection{Machine learning of single trajectories}

Hidden Markov Models (HMMs), a variety of machine learning, have been used to analyze heterogeneity in biological experiments based on the analysis of classical diffusion coefficients \cite{pinholt2021single}. 
Unfortunately, most of the current models for anomalous transport required to motivate (\ref{MSD}) have a memory (in agreement with experimental measures, such as the survival time) and are thus non-Markovian. This presents a serious flaw of HMMs \cite{han2020deciphering} and they are expected to perform badly in the majority of cases in intracellular and cellular biology as a result. An alternative method is to use deep learning neural networks (see the section below) trained on non-Markovian models that can better handle the subtle effects involved. Thus, calculating a spectrum of FBM exponents for endosomal motility with a neural network was found to outperform an equivalent HMM \cite{han2020deciphering} which failed to accurately describe changes in the persistence of the motility. 

A \emph{Bayesian approach} for anomalous dynamics in many cases relies on Markovian approximations and inference with HMM models. It was used for least-squares fitting of empirical MSD curves of particle motion in live cells \cite{monnier2012bayesian}, to discriminate between FBM and GLE models \cite{lysy2016model}, for model selection and parameter inference for L\'evy walk \cite{park2021bayesian}, FBM \cite{krog2018bayesian}, diffusing diffusivity \cite{cherstvy2019non, thapa2018bayesian}, and time-dependent random walk \cite{metzner2015superstatistical}.
A \emph{variational Bayesian} framework for reaction-diffusion models \cite{persson2013extracting}  
based on a hidden Markov model identified  transitions between several diffusing states from short trajectories. 

\emph{Random forests} are another machine learning method that was used to classify the underlying diffusion mechanism of anomalous diffusion trajectories and estimate the anomalous exponent \cite{munoz2020single}.

\subsection{Neural networks for time series analysis}

A wide range of machine learning techniques are showing promise for the analysis of time series data e.g. data from electrocardiograms (ECGs), stock market returns, and particle tracks \cite{nielsen2019practical}. In tracking experiments, neural networks (NN) can provide an order of magnitude improvement in sensitivity and robust analysis of highly non-linear phenomena that are hard to handle analytically, allowing heterogeneity to be robustly quantified in space and time \cite{munoz2021objective,han2020deciphering,granik2019single}. NNs and other more old-fashioned machine learning techniques provide a major impetus for research in heterogeneous anomalous transport. 
Important tasks performed by NNs with HAT include the estimation of the anomalous exponent, the classification of the anomalous diffusion model, and the automatic segmentation of trajectories. Here NNs are being used to emulate generalized non-linear functions, which is permitted by the \emph{universal approximation theorem}
\cite{aggarwal2018neural}.

Different NN architectures have been used to analyze HAT. Here we overview several NN architectures with an emphasis on those which were used for inference of the anomalous exponents and diffusion coefficients. We will discuss their morphology, and training in order to better understand their pros and cons. The benchmarking of different NNs is an open problem that has been attempted during the AnDi challenge where different software to analyse anomalous transport were compared based on common datasets \cite{munoz2021objective}. Due to the rich variety of possible parameters to quantify the algorithms' performance, it is concluded that the software chosen needs to be matched to the applications. 

In addition to the NN architecture chosen, how the training data is presented to a NN can make a big difference e.g. whether a network is trained on particle positions or particle displacements, normalized or unnormalized data is used or 3D data is reduced to 1D (e.g. using the absolute magnitude of the displacement). This process of \emph{feature engineering} is an art form during the use of neural networks \cite{zheng2018feature}. Intuitive choices for feature engineering tend to focus on the reduction of ambiguity in data sets (reducing the variability in data sets due to parameters that are irrelevant to the models used) to maximize their information content.

A simple \emph{feedforward neural network} (FFNN) performed well when calculating instantaneous time-dependent anomalous exponents \cite{han2020deciphering}. To achieve this, a window of size $M$ was fed with the displacements scaled by their range and shifted along the particle trajectory. The input layer was equal to the length of the window and a single output neuron was used. Multiple NN architectures were tested (anti-triangular, rectangular, and triangular deep learning NN) with different numbers of hidden layers. The NN had no regularization and was trained on FBM trajectories without noise. The results showed no improvement for NNs with more than 3 layers and were insensitive to the NN architecture. The accuracy of NN assessed by the mean absolute error, $\sigma_H=\sum_{n=1}^{N} \left|H_{n}^{sim} - H_{n}^{est}\right|/N$, was $\sigma_H \simeq 0.05$ for trajectory lengths of 50 data points which is much lower than other existing methods. The main advantage of the NN is its high accuracy and simplicity. The drawback of this NN is that it only works for data sets that are consistent with FBM motion due to its training, therefore a preparatory statistical analysis is necessary to ensure this. Secondly, the NN in the present architecture only predicts the time-dependent anomalous exponent. Further work is needed to modify it  to predict both the anomalous exponent and the generalized diffusion coefficient. This NN was not a part of the AnDi challenge \cite{munoz2021objective}, however, it has been made available for benchmarking \cite{han2020deciphering}. Previously FFNN with backpropagation was applied to differentiate between Brownian, confined, and directed modes of motion and detect their transitions in time \cite{dosset2016automatic}. Combined with feature engineering based on classical statistics, FFNN was used to identify the anomalous diffusion model, infer the anomalous exponent as well as segment trajectories \cite{gentili2021characterization,manzo2021extreme}.

\emph{Recurrent neural networks} (RNN) have been used for inference of the anomalous diffusion exponent and for the classification of anomalous diffusion models \cite{bo2019measurement,argun2021classification,garibo2021efficient,szarek2021neural} and changes between diffusion models \cite{arts2019particle}. Although questions exist with regard to their accuracy in handling long-time correlations. The architecture of the NN called RANDI was chosen using two layers of RNNs with long-short-term memory (LSTM), with dimensions of 250 and 50. 
The input trajectory was fed into the first  LSTM layer. The NN was trained on increments of trajectories ($\Delta x_i = x_{i+1}  - x_i$) generated similar to tests of the AnDi datasets which consisted of a mixture of trajectories from several anomalous diffusion models \cite{munoz2021objective}. The increments were normalized to have zero mean $E[\Delta x] = 0$ and unit variance $\sigma^2(\Delta x) = 1$ for each trajectory. The NN was trained on trajectories of a fixed length
when tested on trajectories of different lengths. To make a prediction on a trajectory of a certain length, the predictions made on the closest lengths were combined. The mean absolute error ranged from $0.34$ for trajectories containing $32$ points to $0.1$ for trajectories containing $1024$ points. The authors made RANDI freely available as a Python package \cite{argun2021classification}. The disadvantage of RNN models is that it can take a long time to train them since they are typically updated for each training example. 

\emph{Bayesian Neural Networks (BNN) or Deep Learning Bayesian NN (BDLNN)} use an ensemble of neural networks to model the distribution of outputs \cite{mackay1992practical}. In \cite{seckler2022bayesian} a recurrent
LSTM neural network architecture was chosen to obtain anomalous exponent inference, and anomalous diffusion model selection together with the corresponding error estimates. Trajectories from AnDi data sets \cite{munoz2021objective} used for training were converted to increments $\Delta x_t = x_{t+1} - x_t$ 
and normalised to a unit standard deviation. The NN was implemented using the PyTorch library and the code was made freely available online \cite{seckler2022bayesian}.

A \emph{convolutional neural network} (CNN) architecture was used for classification to discriminate between FBM, CTRW and Brownian diffusion models, the inference of anomalous exponents from FBM trajectories, and for the inference of the diffusion coefficient of Brownian motion \cite{granik2019single}. 
The network architecture consisted of four 
sets of convolution blocks with different filter sizes which operated in parallel and were implemented in Python using TensorFlow.  The network was trained on $300,000$ simulated trajectories with added normally distributed localization error and received the velocity autocorrelation function as the input, equ. (17) and was tested with the anomalous exponents of experimentally obtained trajectories of fluorescent beads diffusing in entangled F-actin network gels
with various mesh sizes estimated via time-averaged MSDs. Both experimental trajectories and neural networks were made available \cite{granik2019single}.

A CNN with sparse long-distance connections (WaveNet) is expected to provide state-of-the-art performance for the analysis of anomalous transport \cite{li2021wavenet} due to the accurate handling of long-time correlations. The NN consisted of the input, WaveNet encoder, recurrent neural network (3 stacked long short-term memory LSTM units), and output multilayer perceptron (MLP). The NN was implemented using the PyTorch library and trained on  trajectories from the AnDi datasets  \cite{munoz2021objective}.
Before training or inference, trajectories were normalized to ensure 
zero average and unit standard deviation of the positions. The addition of the convolutional unit improved the NN performance compared to recurrent NN, however, it increased the complexity during design and training. The code is freely available online \cite{li2021wavenet}.

\emph{Graph neural networks} (GNN) use supervised learning to infer the physical properties of anomalous random walks by representing trajectories in the form of graphs and associating a vector of features with each observed position
\cite{verdier2021learning}. Trajectories were normalized either by the standard deviation of step sizes or position or by using the mean step size. The GNN architecture consisted of an encoder and  task-specific multi-layer perceptrons  estimating the anomalous exponent and the random walk model. 
GNNs are expected to perform poorly if long-range correlations are present, therefore special care must be taken for 
classifying various random walks and inferring their anomalous exponents \cite{verdier2021learning}. 

\emph{Transformer neural networks} are a recent development in time series analysis \cite{drori2022science} (they famously occur in ChatGPT) and they may be efficient to describe anomalous transport since their attention mechanism could be matched to the memory of the non-Markovian transport process \cite{firbas2023characterization}. Specifically, an attention mechanism is included in the transformer architectures and this can provide longer terms memories in comparison with other architectures e.g. the long short-term memory (LSTM) constraint on RNNs \cite{pml1Book}. 
In \cite{firbas2023characterization} a bi-layered CNN architecture was combined with a transformer (the Convolutional Transformer) to extract features from trajectories by CNNs and then fed them to  transformer encoding blocks to infer
anomalous exponent or to perform anomalous diffusion model classification. The model was trained in Python with Pytorch framework using the AnDi datasets \cite{munoz2021objective}.

\emph{Generative adversarial neural networks} (GANs) have not yet been extensively used with anomalous transport (they are a key algorithm in the creation of deep fakes), but they could be used to improve statistical measurements with relatively limited training data (\cite{goodfellow2014generative}) through the creation of virtual tracks.  In general, GANs are often used to leverage relatively small amounts of training data during data analysis.

\emph{Variational autoencoders} (VA) are closely related to GANS and can provide non-linear separation of features (a non-linear generalization of classical linear principal component analysis for spectral decomposition \cite{murphy2022probabilistic} i.e. dimensionality reduction of data sets which is often used in classification problems. VAs could be used to analyze tracking data in which non-linear dependencies are expected e.g. distributions of run times on subsequent rests. An unsupervised convolutional autoencoder architecture was used to classify anomalous diffusion models and detect mixed diffusion models \cite{munoz2021unsupervised}.

In the next section, we discuss further perspectives on the use of neural networks for time series analysis.

\subsection{Perspectives on the use of NNs}

An advantage of many NN models in the analysis of tracks is their sensitivity, so a small length of the trajectory (window size) can allow robust classification of the anomalous exponent. Thus HAT can often be explored as a function of time and space, without any additional experimental requirements on track length or data acquisition rate, if NNs are used to analyse the data. Furthermore, it is possible to calculate anomalous exponents over a range of time intervals without retraining a NN by regularly down-sampling data sets. This allows the calculation of coherence times over which there are distinct changes between the anomalous exponent probability distribution functions extracted using the NNs e.g. from peaks at $\alpha=2$ to $\alpha=1$ for a run and tumble model of a bacterium as longer time intervals are considered. Thus, the multi-fractal properties of the data sets over a range of time intervals can be probed.

Skeptics have questioned whether time series data is being overfitted with the new NN models. Assuming the data is well described by a particular anomalous transport model (including HAT) then clearly the answer is no. The fact that FBM has an obvious geometrical interpretation implies that it will correctly model changes in the persistence of motility. Thus as a form of non-linear spectral analysis with strong geometrical foundations, FBM using NNs will provide useful information in most cases. Classical models typically neglect directionality and with NNs it is thought to be a key contributor to their increased sensitivity e.g. with NNs trained on FBM it makes them sensitive to transitions between persistent and anti-persistent motion. Alternatives to HAT calculated with NNs normally involve underfitting with a bad classical model e.g. using heterogeneous classical diffusion for motility in high-concentration systems, such as cells.

It is expected that neural networks can be used to analyse time series data in microrheology. For example, tracks of probe spheres in a complex fluid with time- or space-varying viscoelasticity could be measured. The NN could then be used to characterise the viscoelasticity of the fluids with time or space e.g. the process of gelation with antibodies in response to a pH switch could be followed or the large spatial heterogeneity in the response of formulated complex fluids (e.g. lamellar gels in hair conditioner) and, in general, it would be reasonably straightforward to probe a wide range of power-law fluids that can be modelled with fractional Brownian motion. Thus NNs should provide an order of magnitude improvement in sensitivity in probing heterogeneous viscoelasticity in both space and time. 

Scaling laws are very common in physics e.g. to calculate the conformations of polymer chains \cite{rubinstein2003polymer}. It is expected that analogous neural network techniques to those used for tracking experiments \cite{han2020deciphering} (e.g. considering polymer chains in space rather than tracks in time) can be applied to a broad range of scaling problems in soft-matter physics to provide a robust description of heterogeneity e.g. to better describe boundary effects on flexible polymer chains, non-Markovian effects on polymers (semi-flexible polymers will have longer memories and thus higher fractional exponents than flexible polymers to describe the spatial persistence), heteropolymer conformations, the conformations of polymer chains inside gels in the presence of quenched disorder, block copolymers and flexible polyelectrolytes. Scaling models for polymers are often \emph{ad hoc} multi-fractals e.g. de Gennes introduced the idea of blobs to understand flexible polymers. The blob length can be found by equating the thermal energy (\emph{kT}) to the free energy of the fraction of the chain involved and the chain conformations are multi-fractals, as a result, i.e. the scaling exponent for the chain size as a function of the degree of polymerisation has a sudden change in behaviour at the blob size. The use of a NN to analyze the positions of monomers along a fluorescently labelled polymer chain or the chain conformations from molecular dynamics simulations of polymers could allow rigorous quantification of the blob idea e.g. monodisperse blobs may be a convenient mathematical fiction and flexible polymeric chains are more accurately described by a continuous spectrum of scaling exponents as a function of length scale (or equivalently the number of monomers) \cite{han2020deciphering}. 

In addition to the analysis of tracks, NNs can be useful to create tracks from microscopy experiments. Convolutional neural networks can be used to segment particles in images from a movie, which are subsequently linked together to form tracks \cite{newby2018convolutional}. CNNs tend to require less user optimisation than Gaussian trackers and they can function well with less symmetric particles e.g. bacterial cells. However, care is needed to physically constrain NNs for their optimal performance and under some conditions classical tracking techniques are preferable e.g. Gaussian trackers used with movies of diffraction-limited particles that are well described by Gaussian functions can outperform unconstrained CNNs.

\subsection{Computer simulations}

Simulations play a crucial role in testing the underlying physical mechanisms of anomalous transport. They are also indispensable for the creation of training sets for neural networks. Several anomalous diffusion models e.g. FBM \cite{mandelbrot1968fractional}, CTRW \cite{montroll1965random,metzler2000random}, L{\'e}vy flights/walks \cite{zaburdaev2015levy}, scaled Brownian motion \cite{jeon2014scaled} and diffusion processes with space-dependent diffusivity \cite{cherstvy2013population} have become popular. Some standard algorithms for simulating FBM are the exact methods due to Hosking, Cholesky-Levinson, Davies-Harte, and Wood-Chan. Although the Hosking and Choleski-Levinson methods are exact, they are slow, since their computational complexity is of order $O(N^2)$ and $O(N^2log N)$, respectively. 
The Wood-Chan method is exact for simulating  fractional Gaussian noise (increments of FBM), and is faster $O(N log(N))$. Several approximate methods also exist to simulate FBM e.g. the stochastic representation method \cite{mandelbrot1968fractional} and wavelet-based simulations \cite{coeurjolly2000simulation,dieker2004simulation}.
Some models have been extended to heterogeneous FBM \cite{korabel2021local,han2020deciphering,janczura2021identifying,balcerek2022fractional}, multifractional FBM with deterministic and random exponents \cite{ayache2018new,chan1998simulation}, heterogeneous CTRW \cite{grebenkov2018heterogeneous}, heterogeneous L{\'e}vy flights/walks, and subordinated combinations to simulate HAT.

\section{Case studies in molecular biology}

There are a large number of examples of HAT in soft matter physics. Here we will focus on examples found to date in molecular biology experiments, but many more are expected to occur due to the rich variety of molecules available e.g. carbohydrates, polyphenols, proteoglycans, etc. will all experience HAT. We thus expect to find HAT in the majority of  biological systems \emph{in vivo}.

\subsection{DNA}

DNA is unusual compared with most synthetic polymers in that it can be completely monodisperse i.e. all DNA chains have the same length when sourced from a specific chromosome of a single organism. 
However, the 3D spatial organization of the chromatin 
(DNA with associated bound proteins) in the nucleus of eukaryotic cells is hierarchical and highly heterogeneous \cite{misteli2020self}. 
Chromosomes occupy specific regions of a nucleus, called chromosome territories and each consists of megabase A compartments which contain a large amount of the bases for expressed genes, and B compartments with less transcriptionally active bases \cite{lieberman2009comprehensive}. Compartments are further spatially structured into multiple self-interacting topologically associating domains (TADs). 

Unsurprisingly, due to its polymeric nature, the dynamics of DNA is mostly anomalous on a short time scale (e.g., \cite{cabal2006saga,weber2010bacterial} and many others). A simple Rouse polymer model predicts sub-diffusive dynamics of the chains with the anomalous exponent $\alpha=0.5$ \cite{doi1988theory}, the inclusion of the hydrodynamic interaction between neighbouring sections of the chain gives $\alpha=2/3$ (Zimm dynamics) and inclusion of both semi-flexibility and hydrodynamic interactions gives $\alpha=0.75$. The heterogeneous structure of chromatin, interactions with various proteins (e.g. cohesin and RNA polymerase II), different geometrical constraints, and ATP-driven active processes lead to the heterogeneous dynamics of the DNA (see \cite{ashwin2020heterogeneous} for a recent review). Polymeric modeling shows that the formation of TADs is a key driver of heterogeneity in the motion of chromatin \cite{salari2022spatial}. Various values of anomalous exponent $\alpha$ were reported depending on the position in the genome, cell type, and cell condition \cite{javer2013short}. 
A wide range of  spot-to-spot variability in
diffusion dynamics was found in a mouse ES cell line carrying TetO arrays inserted at random genomic locations \cite{oliveira2021precise} quantified using $D_{\alpha}$s and $\alpha$s. Heterochromatin-rich regions showed less movement, while the activity of transcriptional machinery increased the amplitude of the dynamics \cite{nozaki2017dynamic}.

Telomere (regions at the ends of chromosomes) dynamics in living mammalian cells is highly heterogeneous \cite{stadler2017non,bronstein2009transient,bronshtein2015loss}. Analysis of single telomeres' trajectories revealed exponential distributions of displacement increments \cite{stadler2017non}. There was a crossover of tMSDs from sub-diffusion $\alpha<1$ for time lags $\tau<10$ seconds to normal diffusion $\alpha \simeq 1$ for large $\tau$ \cite{stadler2017non,bronstein2009transient}. Although there could be several reasons for such a  crossover, it was argued that  the heterogeneity of telomeres' motions  stems from non-equilibrium cytoskeletal forces that drive their anomalous diffusion \cite{stadler2017non}. 

\begin{figure}
    \centering
    \includegraphics[width=1\textwidth]{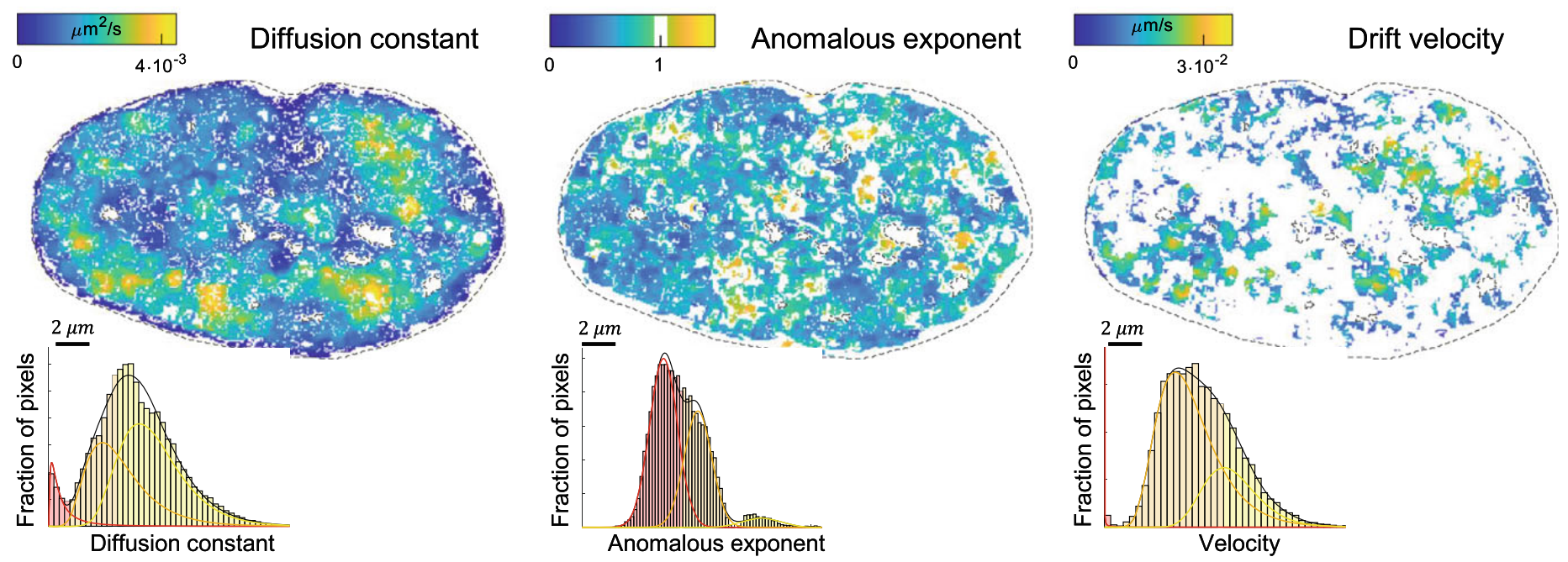}
    \caption{{\bf Spatially resolved mapping of genome dynamic properties at nanoscale resolution in living U2OS human bone cancer cells.} DNA dynamics in the nuclear interior are spatially partitioned into $0.3–3$-$\mu$m domains in a mosaic-like pattern. Maps of biophysical parameters ($D_{\alpha}$, $\alpha$ and drift velocity $V$) reveal the local dynamic behavior of DNA in large domains. The distribution is deconvolved using a generalized mixture model. Reprinted from \cite{shaban2020hi}.}
    \label{ShabanHiD}
\end{figure}

Spatially resolved mapping of genome dynamics was organized into domains with a mosaic pattern (figure \ref{ShabanHiD}) \cite{nozaki2017dynamic,shaban2018formation, shaban2020hi,barth2020coupling}.
Such correlated motion might be caused by active mechanisms \cite{saintillan2018extensile}, such as transcription \cite{barth2020coupling}. 
Long-range directional movement of interphase chromatin was also found which was super-diffusive \cite{chuang2006long,levi2005chromatin,zidovska2013micron}. Similarly, coherent movements over 1 $\mu$m distances by human RNA polymerase II molecules were observed over the entire nucleus in human cancer cells \cite{barth2022spatially}. 

Many DNA-binding proteins (Rad4-Rad23 in yeast or telomeric sequence binding proteins TRF1, TRF2, and SA1) have anomalous dynamics, which were reviewed recently in the context of facilitated diffusion \cite{park2021mini}. 
\begin{figure}
    \centering
    \includegraphics[width=1\textwidth]{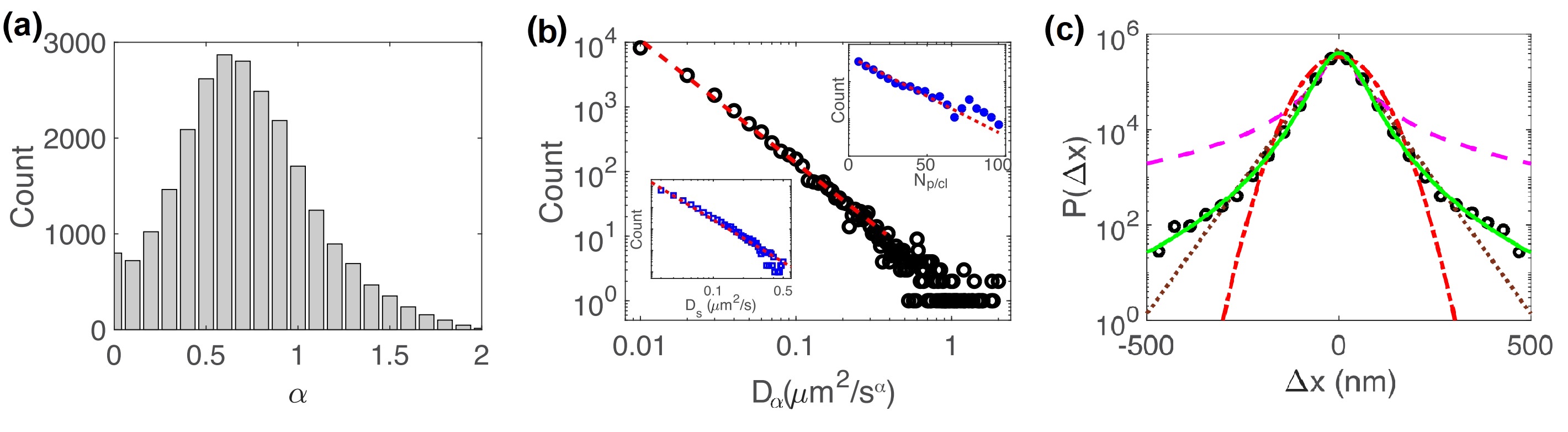}
    \caption{{\bf Heterogeneous anomalous dynamics of individual histone-like nucleoid-structuring (H-NS) proteins  in live \emph{Escherichia coli} bacteria.} {\bf (a)} Histogram of the anomalous exponents $\alpha$. {\bf (b)} Histogram of the generalized diffusion coefficients, $D_{\alpha}$, fitted with $P(D_{\alpha}) \simeq D_{\alpha}^{-(\beta+1)}$ with $\beta = 0.97 \pm 0.07$ (red dashed line). {\bf (c)}  Probability distributions of displacements ($P(\Delta x)$, black circles) fitted with the power-law-tailed (Pearson VII) distribution (green solid line) fits the data very well. Reprinted from \cite{sadoon2018anomalous}
    }
    \label{SaddonWang}
\end{figure}
A broad probability distribution of anomalous exponents and a power-law probability distribution for the generalized diffusion coefficients of individual histone-like nucleoid-structuring (H-NS) proteins were found for their anomalous dynamics (figure \ref{SaddonWang}) in \emph{Escherichia coli} \cite{sadoon2018anomalous}. Furthermore, non-Gaussian power-law-type (Pearson VII) probability distributions of protein displacements were observed. The H-NS proteins interact with both other proteins and DNA simultaneously. The broad probability distribution of anomalous exponents was due to the active motion of a sub-population of the H-NS proteins, while the power-law probability distribution of generalized diffusion coefficients, $P(D_{\alpha}) \simeq D_{\alpha}^{-(\beta+1)}$, where $\beta$ is a constant, was suggested to stem from protein polymerization. 
Transient non-specific interactions of DNA-binding proteins with DNA were shown to lead to highly heterogeneous dynamics with a broad probability distribution of diffusion coefficients \cite{stracy2021transient} and a broad (power-law) survival probability distribution for nonspecific (TetR-DNA) interactions \cite{normanno2015probing}. These results indicate the current paradigm of facilitated classical diffusion for the dynamics of DNA-binding proteins needs to be extended. 

\subsection{RNA}

RNA-protein particles (MS2-mRNA complexes) in the cytoplasm of the bacterium \emph{E. coli} and the eukaryotic microorganism \emph{S. cerevisiae} (budding yeast) exhibit heterogeneous subdiffusive behavior of 
viscoelastic origin \cite{lampo2017cytoplasmic}. Probability distributions of the displacements of the complexes, $\Delta x = x(t+\delta) - x(t)$, where $\delta$ is a time increment, were found to be non-Gaussian at all observed timescales and were well
described by a Laplace distribution,
\begin{equation}
\mbox{P}(\Delta x) = \frac{1}{\sqrt{2} \sigma} \exp \left( - \frac{\sqrt{2} \left| \Delta x \right| }{\sigma} \right)
\label{LaplacePDF}
\end{equation}
where $\sigma$ describes the breadth of the distribution.
Subdiffusive dynamics of MS2-mRNA complexes which had Laplace distributions of displacements were also reported  \cite{stylianidou2014cytoplasmic}. In \cite{lampo2017cytoplasmic}, the probability distributions of generalized diffusion coefficients extracted from the time-averaged MSD of
individual RNA-particle trajectories were well approximated by an exponential distribution,
\begin{equation}
\mbox{P}(D) = \frac{1}{\left< D \right>} \exp \left( - \frac{D}{\left< D \right>} \right),
\label{ExpPDF}
\end{equation}
where $\langle D \rangle$ describes the breadth of the distribution.
This non-Gaussian behavior is
a consequence of significant heterogeneity between trajectories and dynamic heterogeneity along single trajectories.

\subsection{Membranes}

Membranes fulfill many crucial functions in cellular biology as external barriers that maintain the integrity of cells and as barriers that define specialized organelles. 
The cell membrane was long known to be highly heterogeneous \cite{kusumi2014tracking} with patches with strongly varying diffusivity \cite{serge2008dynamic} e.g. the \emph{lipid raft} hypothesis. 
Various mechanisms underlying anomalous diffusion in plasma membranes have been reviewed \cite{metzler2016non,krapf2015mechanisms}. 
For example, anomalous subdiffusion can be caused by transient immobilization and spatial heterogeneity
\cite{saxton1997single} or specific interactions between molecules which temporarily modify their diffusive behavior \cite{masson2014mapping,torreno2016uncovering,charalambous2017nonergodic,sanz2023broadband}.

\begin{figure}
    \centering
    \includegraphics[width=1\textwidth]{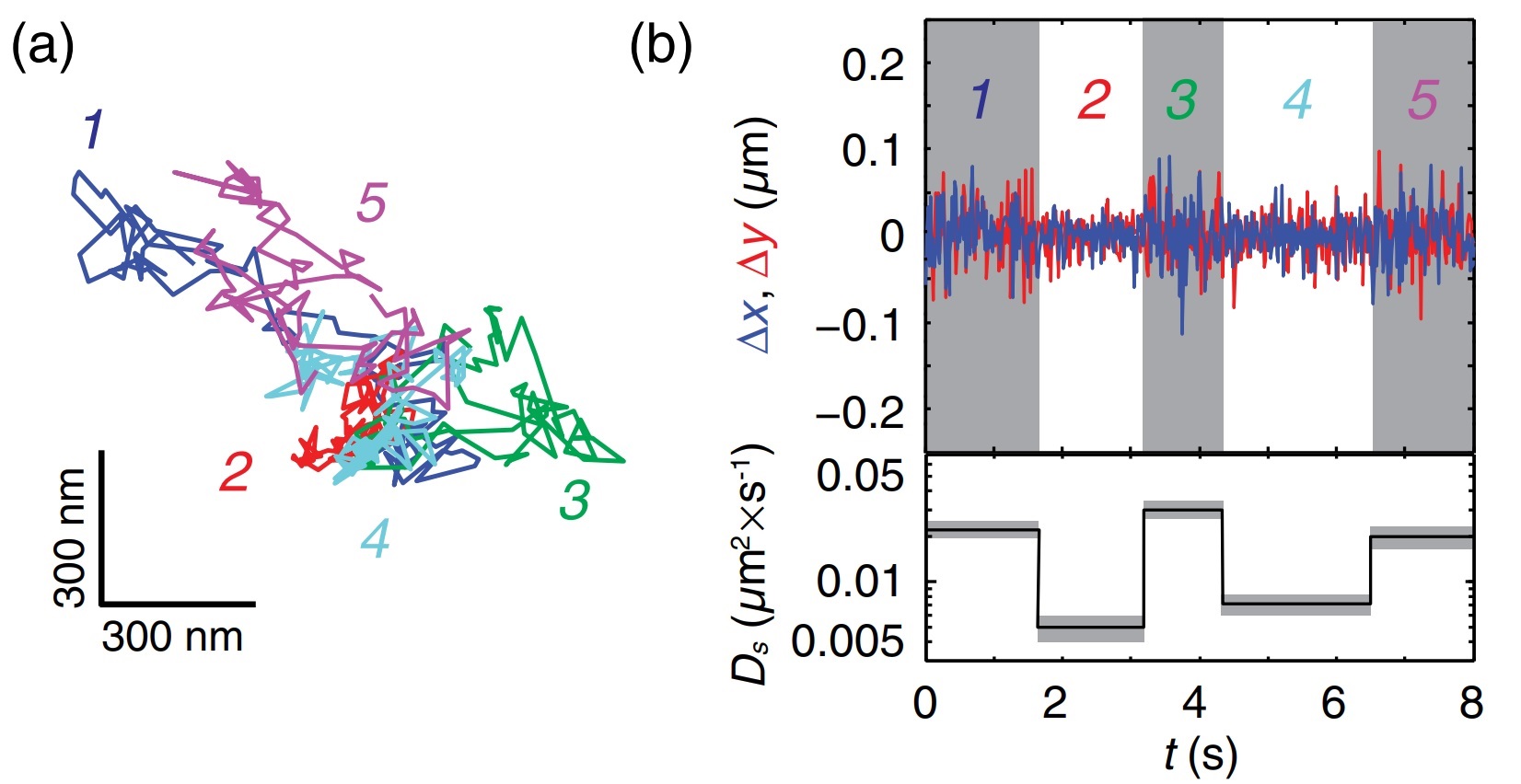}
    \caption{{\bf  Dendritic cell-specific intercellular adhesion molecule 3-grabbing nonintegrin (DC-SIGN) motion experience changes in diffusivity on living-cell membranes.} {\bf (a)} Representative  trajectory displaying changes of diffusivity. The change-point analysis provides evidence for five different regions that are represented with different colors. {\bf (b)} Plot of the x (blue) and y (red) displacements for the trajectory in (a) as a function of time. The lower panel displays the corresponding short-time diffusion coefficient $D_s$ as obtained from a linear fit of the time-averaged MSD for the five different regions. Reprinted from \cite{manzo2015weak}.
    }
    \label{Manzo}
\end{figure}

The spatially heterogeneous dynamics of two membrane
proteins transfected into COS-7 cells was measured using photoactivated localization microscopy (PALM) combined
with live-cell single-particle tracking \cite{manley2008high}. Some trajectories showed sub-diffusion typical for motion in a membrane. Broad probability distributions of short-time diffusion coefficients revealed dynamic heterogeneities in the cell membrane. Hyperspectral microscopy allowed single particle tracking of quantum dot labels at $27$ frames per second and could resolve the time-dependent instantaneous diffusion coefficient of membrane proteins interacting with membrane-associated structures \cite{cutler2013multi}. 
The motion of dendritic cell-specific intercellular adhesion molecules on cell membranes revealed nonergodic subdiffusion, which was interpreted as due to changes of diffusivity (figure \ref{Manzo}) \cite{manzo2015weak}. 
Single receptor trajectories displayed Brownian motion with relatively constant diffusivity over intervals of varying length but changed significantly between intervals. The probability distribution of diffusion coefficients, $D$, could be described by a Gamma distribution, 
\begin{equation}
P(D) = \frac{D^{\sigma-1}}{b^{\sigma}\Gamma(\sigma)} \exp \left(-D/b\right), 
\end{equation}
with $\sigma>0$ and \emph{b} is a constant.
Anomalous dynamics of Kv2.1 potassium channels 
that form stable clusters in the plasma membrane of transfected human embryonic kidney cells and native neurons have both ergodic and nonergodic components with very broadly distributed anomalous exponents \cite{weigel2011ergodic,weron2017ergodicity}. A probable reason for the HAT behaviour is the binding of the ion channel clusters to clathrin-coated pits with a heavy-tailed distribution of immobilization times. 
Molecular dynamics simulations suggested that interactions between membrane-binding proteins and lipids can lead to fluctuating diffusivity \cite{yamamoto2017dynamic}.

Numerical simulations predict multi-fractal, non-Gaussian, and spatiotemporally heterogeneous anomalous lateral diffusion for the motions of lipids and membrane proteins \cite{jeon2016protein}. Protein crowding induces spatiotemporal heterogeneity in the lateral diffusion of lipids that stochastically transition between high and low diffusivity states. The probability distribution of generalized diffusivities $D_{\alpha}$ has a double-Gaussian form. 

Self-assembled lipid vesicles created in solution are popular as a model for cell membranes. Heterogeneous fluctuations 
of the environment lead to non-Gaussian diffusion of fluorescently tagged unilamellar lipid vesicles (liposomes) diffusing in nematic solutions of aligned F-actin filaments with a  spectrum of diffusivities. The MSDs of liposomes grew linearly with time, but the PDFs of the displacements showed a crossover from a Laplace (double-exponential) form at small times to a Gaussian form at later times \cite{wang2012brownian}. 

Dynamical heterogeneities were recently reported for the diffusion of transmembrane AChRs receptors in Xenopus embryo muscle cells  \cite{he2016dynamic} which moved subdiffusively with a broad spectrum of anomalous exponents $\alpha$ on time scales of up to $1$ second and performed Brownian diffusion at larger times. The transition from sub-diffusion to Brownian diffusion and non-Gaussian distributions of displacements with exponential tails suggest spatiotemporal fluctuations of diffusivity, which is also supported by the observation that
the mobile trajectories have some immobile segments of varying  durations \cite{he2016dynamic}. These transient confinements of AChRs could be caused by the transient binding of the AChRs to the cortical actin network.

Lipid microdomains (lipid rafts)
attached to capsular carbohydrates were found to experience HAT on the surfaces of \emph{E. coli} in super-resolution fluorescence microscopy experiments \cite{phanphak2019super}. 
The motion of the labeled rafts was subdiffusive with the average anomalous exponent $\alpha = 0.33 \pm 0.23$ at 1 h and $\alpha = 0.36 \pm 0.20$ at 1.5 h. The rafts eventually merge together at long time scales to sterically stabilise the bacteria with giant carbohydrate brushes ($>$200 nm in height).

Finally, the superdiffusive motion of molecules that target C2 domains in membranes alternated between phases of two-dimensional motility on the membrane and 
three-dimensional diffusive excursions
\cite{campagnola2015superdiffusive}. Such intermittent dynamics was characterized by a probability distribution of displacements with both Gaussian and Cauchy components and a broad probability distribution of diffusion coefficients. The field of non-Brownian interfacial anomalous diffusion was recently reviewed \cite{wang2020non}. 

\subsection{Endosomal transport}

\emph{Endosomes} are sophisticated membrane-bound packages that are transmitted intracellularly (extracellular transport is also possible via exosomes). Endosomes constitute a complex targeted postal system inside eukaryotic cells (e.g. membrane-bound proteins, such as Rab, can label endosomal packages that direct processes of sorting) and are used for the active transport (in contrast to passive diffusion) of molecules. Early studies of active transport considered engulfed microspheres moving along microtubules in living eukaryotic cells \cite{caspi2000enhanced}. The mean square displacement of microspheres was superdiffusive at short times, with a clear crossover to subdiffusive or ordinary diffusion scaling at longer times. However, for large spheres this is a process of phagocytosis and the biology is different to endocytosis that creates standard endosomes (there is a size filter that differentiates between the two possibilities, with smaller particles experiencing endocytosis). In general, the biological phenomena associated with endocytosis of nanoparticles are very rich 
\cite{rennick2021key}. 

\begin{figure}
    \centering
    \includegraphics[width=0.9\textwidth]{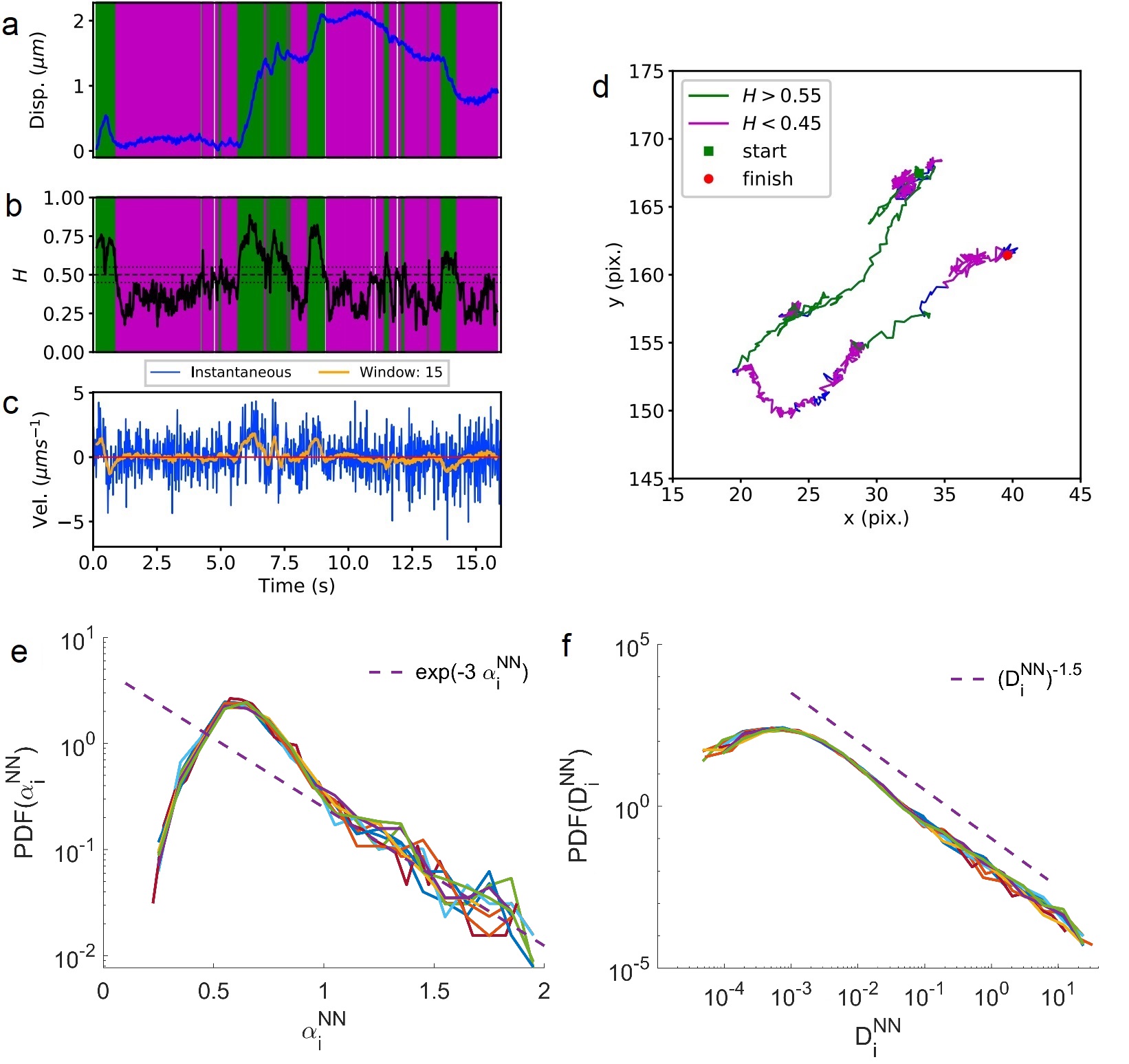}
    \caption{{\bf  Neural network (NN) analysis of a GFP-Rab5 endosome trajectory reveals its heterogeneous dynamics.} (a) Plot of the displacement of a single trajectory in an MRC-5 human lung fibroblast cell (blue). Shaded areas show persistent ($0.55 < H < 1$ in green) and anti-persistent ($0 < H < 0.45$ in magenta) behaviour. (b) A $15$ point moving window NN exponent estimate for the trajectory (black) with a line (dashed) marking diffusion $H = 0.5$ and two lines (dotted) marking confidence bounds for estimation marking $H = 0.45$ and $0.55$. (c) The plot of instantaneous and moving ($15$ point) window velocity. (d) Plot of the endosome trajectory with start and finish positions indicated. Persistent (green) and anti-persistent (magenta) segments are shown. Sections that were $0.45 < H < 0.55$, were not classified as persistent or anti-persistent and are depicted in blue. (e, f) Probability density functions of local anomalous exponents $\alpha_i^{NN} = 2H$ estimated with the NN. (e) PDFs of generalized diffusion coefficients $D_{\alpha_i}^{NN}$. Different
curves in d and e correspond to PDFs estimated at $t = 0.2,0.4,0.8,1,2,4,8,10,12,14$ s.
    Reprinted from \cite{han2020deciphering}.
    }
    \label{Han}
\end{figure}

In a series of publications over the last 10 years, our group has considered a number of the issues involved in heterogeneous endosomal transport. Our early microscopy experiments suffered from issues with signal-to-noise ratios and temporal sensitivity. Robust statistics (e.g. MSDs) could only be extracted from data sets that were averaged over both the time and the ensemble.  The process of endosomal transport was found to be anomalous and super-diffusive \cite{flores2011roles,fedotov2018memory,korabel2018non}. This was followed by the use of an ultrafast camera with gold-containing endosomes (plasmon-based contrast using bright field microscopy). These experiments allowed the heterogeneous anomalous transport of single endosomes to be probed at very fast (sub-millisecond) time scales. The anomalous transport was still found in single particle tracks and at fast time scales. A crossover was observed from sub-diffusive motion at fast time scales ($<$0.01 s) to super-diffusive motion at intermediate time scales i.e. clear evidence for HAT as a function of time interval was observed. This was followed by slower fluorescence microscopy experiments that labelled specific types of endosomes (Rab5 early endosomes). These experiments allowed us to collect a large ensemble (more than $10^4$) of trajectories and track them over much longer time scales. More recently, based on observations pointing to FBM-like dynamics, we developed a neural network (NN) to probe the heterogeneities in the anomalous transport of endosomes \cite{han2020deciphering}. The NN successfully learned patterns in endosomal dynamics in terms of persistence (a tendency to keep or reverse the direction of motion quantified by the Hurst exponent $H$) and elucidated switching from sub-diffusive to super-diffusive motion in a single trajectory (figure \ref{Han}) without using any arbitrary thresholds \cite{fedotov2018memory,chen2015memoryless}. 
Instead of normal and log-normal probability distributions of anomalous exponents and generalized diffusion coefficients respectively, expected for homogeneous systems, an exponential probability distribution of anomalous exponents and a broad power-law probability distribution of generalized diffusion coefficients were found \cite{korabel2021local} (figure \ref{Han}e, f), similar to heterogeneous dynamics of histone-like nucleoid-structuring proteins in live \emph{Escherichia coli} \cite{sadoon2018anomalous}. For endosomes, we also found power-law probability distributions of displacements and increments \cite{korabel2021local}. Power-law probability distributions were previously reported for trajectories of individual nucleoid-structuring H-NS proteins \cite{sadoon2018anomalous}.
Endosomal fusion and fission events were recently studied 
using Smoluchowski-like coagulation equations to describe the power-law probability distributions of endocytosed low-density lipoprotein \cite{foret2012general}, endosome maturation \cite{castro2021fusion} and clustering of nanoparticles \cite{alexandrov2022dynamics}. A power-law probability distribution of cluster
sizes naturally leads to a power-law probability distribution of diffusion coefficients \cite{vicsek1985scaling}.

The dynamics of endosomes share many similarities with other organelles, e.g., lysosomes. Lysosomal trajectories studied using wavelet-based analysis were found to be super-diffusive and composed of alternating bursts and pauses in their active transport  \cite{polev2022large}. 

\subsection{Endoplasmic reticulum}
The endoplasmic reticulum (ER) is the largest organelle in eukaryotic cells and can span the entire width of cells. Membranes are the principle component of the ER and they adopt a wide variety of different structures due to the interplay of lipids with other surface active molecules e.g. membrane-bound proteins. Many of the structures are non-equilibrium and continuous motor activity is needed to maintain them \cite{perkins2021intertwined}. A previous model from our group for the ER was that it acts as a stirred reaction vessel to increase the chemical kinetics, with motor proteins doing the stirring. The ER is a molecular factory that performs a wide range of functions including the synthesis of proteins and lipids. It is thought that ribosomes (protein factories) associate with the membrane of the ER to reduce the dimensionality of the reactions responsible for making new proteins and thereby increase their rate of synthesis.
\begin{figure}
    \centering
    \includegraphics[width=1\textwidth]{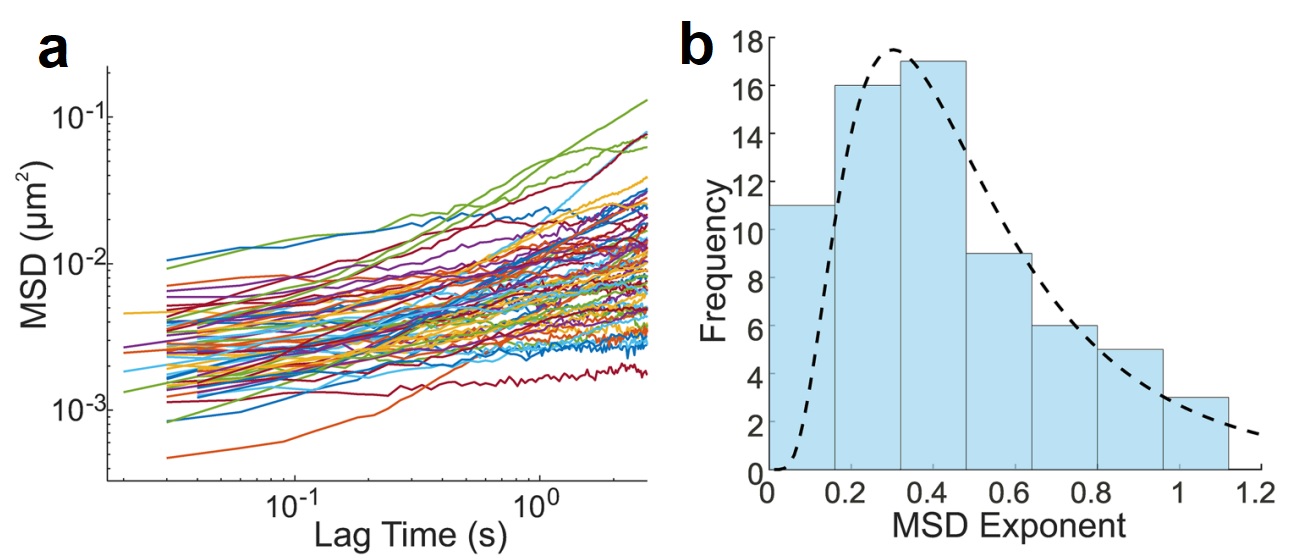}
    \caption{{\bf  Mean squared displacements of the transverse displacements of endoplasmic reticulum (ER) tubules in live African green monkey kidney Vero cells.} 
    {\bf (a)} MSDs of measured transverse tubule 
    fluctuations as a function of lag time. Traces begin at different time points due to slight differences in video frame rates. {\bf (b)} Histogram of the exponents of the power law fit to the MSDs in (a). The dotted line shows a lognormal fit with mean exponent $0.54\pm0.04$. 
Reprinted from \cite{perkins2021network}.
    }
    \label{ER}
\end{figure}
The ER is divided into rough and smooth varieties. The rough ER is constructed from a network of membranous tubules and a practical constraint for tracking experiments is that they resemble linear objects rather than point objects. Thus, they need to be analyzed in images using snakes algorithms and the dynamics resembles that of semi-flexible polymers \cite{perkins2021network,georgiades2017flexibility}. 

The statistics of the driven transverse fluctuations of ER tubules were quantified by our group \cite{perkins2021network,georgiades2017flexibility}. HAT was observed for these fluctuations with the majority of anomalous exponents around $0.5$ (the predicted value for stressed semi-flexible filaments), but a substantial minority had super-diffusive exponents driven by motor proteins (figure \ref{ER}). These results are similar to the dynamics of three-way ER junctions in the cell periphery of HeLa cells    \cite{speckner2018anomalous}. ER junctions show heterogeneous anomalous diffusion with a broad probability distribution of anomalous exponents around
a mean $\alpha = 0.49$ which was interpreted as ER driven by microtubule-dependent active noise. 

Heterogeneous dynamics was also found for individual quantum dots loaded into the cytoplasm of HeLa cells \cite{sabri2020elucidating, etoc2018non}. It was suggested that the observed heterogeneous subdiffusion stemmed from nonspecific binding of quantum dots to the cytoskeleton and the ER network causing broad probability distributions of anomalous exponents, diffusion coefficients, and exponential decays for the PDFs of the normalized increments.  

The ER interacts with a huge number of molecules and organelles in the cell, so there is a huge range of possible sources of dynamic heterogeneity in both time and space. Correctly quantifying the source of the dynamic heterogeneity is an important future goal in understanding the activity of the ER and is expected to relate to disease states e.g. spastic mutations \cite{montenegro2012mutations}. Machine learning techniques were used to find patterns in the ER's HAT e.g. how the anomalous transport varied with the distance to the cell's nucleus \cite{perkins2021network}.

\subsection{Globular proteins}

Anomalous diffusion of proteins due to molecular crowding has been studied for decades \cite{banks2005anomalous}. However, anomalous diffusion in the cytoplasm of living cells can be caused by different mechanisms in addition to macromolecule crowding effects. Dilute biomacromolecules, such as globular proteins, are thought to demonstrate heterogeneous diffusion due to internal flexibility \cite{yamamoto2021universal}. This will inevitably lead to HAT in more concentrated suspensions e.g. a flexible protein moving in a heterogeneous polymeric mesh \cite{gupta2016protein}. The process of time-fluctuating instantaneous diffusivity has been observed in molecular dynamics simulations and predicted for simple analytic models of polymers e.g. Rouse, Zimm, etc., and for reptation in concentrated solutions. To examine the fluctuations of the diffusivity, the magnitude, and orientation, correlation functions of the diffusivity were introduced to distinguish between different polymer models \cite{miyaguchi2017elucidating}. 

Caging phenomena lead to heterogeneous diffusion in concentrated suspensions of colloids \cite{cipelletti2011glassy}. Globular proteins are examples of colloids, so caging phenomena and jamming are also expected in these systems.
The viscoelasticity of concentrated antibodies is important for injections in anti-cancer immunotherapies. Classical liquid state theories typically predict a heterogeneous diffusion coefficient that depends on length scale via the momentum transfer $q$, $D(q)$. Historically, inelastic scattering techniques were the primary technique used to measure $D(q)$, since they operate in Fourier space (hence the use of $q$). A multi-fractal model for HAT is expected to be superior to $D(q)$, since it is more flexible e.g. it can describe the directional phenomena involved in both the antipersistence of caged particle motions and more persistent hopping phenomena between cages. Systematic comparison has not to our knowledge been completed for the multi-fractal analysis of caging in colloids. 

An important practical issue is to monitor the size of antibodies during their production in bioprocessing plants e.g. monoclonal antibodies extracted from Chinese hamster cells. Antibodies have many crucial applications in current biotechnology such as pregnancy kits, COVID tests, and immunotherapy in cancer treatments. Currently, the top ten best-selling pharmaceuticals are cancer treatments based on antibodies. Monomeric forms of antibodies are clinically preferred and dimerisation or more extensive aggregation can lead to inactivity and less favorable outcomes
\cite{roberts2014therapeutic}. Thus, differentiating diffusing diffusion from aggregation is an important issue in this context. Furthermore, therapeutic antibody formulations are required at high concentrations to increase dosages (e.g. to decrease the probability of resistance to treatment with cancers) and reduce the frequency at which they are administered. Thus studies of the HAT of high-concentration antibodies and how it relates to viscoelasticity are important e.g. how the antibody formulations can be injected through syringes.

Sub-diffusive motion has been established for the internal dynamics of peptides based on standard molecular dynamics simulations (GROMACS) \cite{xia2020origin,hassani2022multiscale}. Power law and logarithmic relaxation of hydrated proteins were also found in a molecular dynamics simulation with elastin-like polypeptides \cite{kampf2012power}. Evidence for fractional Brownian motion was found in quasi-elastic neutron scattering from myoglobin and C-phycocyanine \cite{kneller2005quasielastic}. A two-state model for the folding of globular proteins with anomalous dynamics was presented and scattering functions were calculated 
\cite{dieball2022scattering}. A generalized elastic model yielded a fractional Langevin equation \cite{taloni2010generalized} which could be used to describe a broad range of phenomena related to anomalous dynamics in soft-condensed matter including proteins. 

NMR evidence for fractional dynamics was found for methyl side-chain dynamics in proteins \cite{calligari2012toward}. 
Fractional dynamics was also seen in NMR spectroscopy from lysozyme and was confirmed with molecular dynamics simulations
\cite{calandrini2010fractional}.

\subsection{Amyloids}
Amyloid diseases include Alzheimer's, Parkinson's, and bovine spongiform encephalopathy (BSE). Progression of the diseases is driven by the aggregation of misfolded proteins into giant plaques and such aggregates have thus experienced intensive research over recent years. The self-assembly of amyloids is fairly easy to recreate \emph{in vitro} in a laboratory. Misfolded proteins commonly adopt beta sheets, which are thought to be their most energetically favourable structures. Less obviously, many standard globular proteins (e.g., lysozyme \cite{booth1997instability}) experience self-assembly of the beta sheets into giant aggregates that resemble the amyloids observed in the disease. Synthetic peptides also can create giant aggregates and show promise as designer functional materials that can be created at relatively low costs. The transverse displacements of self-assembled peptide fibres measured in dynamic light scattering experiments follow the predictions for semi-flexible polymers i.e. they have anomalous exponents for their transverse displacement fluctuations with $\alpha$=3/4 \cite{carrick2005internal}. Extensive research has been done on the self-assembly of peptide surfactants \cite{hu2020recent}. We were able to measure entanglement tubes at the single molecule level within peptide gels using single-molecule fluorescence microscopy \cite{cox2018single}. Pre-stresses in the gel fibres due to quenched anomalous disorder provided a novel source of heterogeneous anomalous diffusion \cite{cox2019active}.

\subsection{Mucins}
Heterogeneous anomalous diffusion in mucin is a crucial step during many forms of drug delivery (e.g. to deliver drugs through the intestines, lungs, eye, mouth etc \cite{mcguckin2011mucin}) and for the transport of infectious organisms (e.g. viruses and bacteria). In general, mesoscopic interactions are thought to play an important role in modulating the diffusive transport of nanoparticles in hydrogels \cite{zhang2015particle}. Two broad heterogeneous dynamic modes are seen in experiments and the dynamics of synthetic probe spheres have been interpreted in terms of polyelectrolyte scaling theories \cite{georgiades2014particle}. In \cite{lysy2016model} heterogeneity displayed by trajectories of 1 $\mu$m diameter tracer particles in human lung mucus were described by a hierarchical FBM equivalent to a GLE model with a distribution of parameters which induce ergodicity breaking. More modern analysis of particles in mucin gels finds evidence for ergodicity breaking (as expected for most polymeric gels), non-Gaussian pdfs and non-Fickian diffusion \cite{wagner2017rheological,cherstvy2019non}. 

\section{Case studies in cellular biology}

\subsection{Embryonic cells}

Stem cell-aided repair is one of the principal goals of tissue engineering in medicine and stem cells are found in embryos. Cells in live embryos also provide important insights into morphogenesis, which impacts all of the physiology of the resultant organism. Reaction-diffusion equations can describe pattern formation in embryos e.g. how the leopard got its spots. HAT will therefore affect such pattern formation and a formalism for reaction-anomalous diffusion is needed.

Temporal heterogeneity within  trajectories of haematopoietic stem cells was found in the form of alternating periods of a confined random walk followed by processive motion \cite{christodoulou2020live}.
Statistical analysis of 
trajectories of persistently migrating cells in
2D and 3D culture environments revealed that the cells display non-Gaussian super-diffusion with 
a spectrum of anomalous exponents and diffusivities \cite{luzhansky2018anomalously}. In contrast to normal diffusion, cells during super-diffusive motion are covering new areas more quickly. The heterogeneity of single-cell trajectories could originate from both the phenotypic and functional heterogeneity of stem cells, with cells moving between two or more metastable states \cite{graf2008heterogeneity}. {\em In vitro}, mouse fibroblast cells on a two-dimensional substrate were also found to move super-diffusively due to heterogeneous noise during the runs, leading to large fluctuations in speed and a run-and-tumble behaviour, where cells move in a straight line for a while before rapidly changing direction \cite{passucci2019identifying}. However, the displacement probability distribution was found to have the classical Gaussian form. 

\subsection{Cancer cells}

Anomalous heterogeneous diffusion of cancerous cells emerges during the migration of tumor cells, metastasis formation, and the analysis of crawling gaits. One origin of heterogeneity is the migration and proliferation dichotomy in tumour-cell invasion \cite{giese1996dichotomy} (also known as the \emph{Go-or-Grow mechanism}) which could be described using a continuous time random walk model \cite{fedotov2007migration}. On 2D substrates and in 3D collagen matrices, cell heterogeneity results in the super-diffusive migration of individual cancer cells, with an exponential probability distribution of cell displacements and a non-Gaussian velocity probability distribution  \cite{wu2014three}. 
Non-Gaussian double exponential probability distributions of displacements and super-diffusive ensemble averaged MSDs were found for the motility of breast cancer carcinoma cells 
\cite{metzner2015superstatistical}.  
The interpretation of the anomalous diffusion of cancerous cells is based on models that range from L{\'e}vy walks  
\cite{huda2018levy} to the fractional Klein–Kramers equation \cite{dieterich2008anomalous}. Higher resolution experimental data is needed to rigorously differentiate between the different possibilities.

Bayesian inference of time-dependent persistence and local activity (noise
amplitude) parameters revealed heterogeneity in measured cell trajectories with large variations of cell behaviour, both with time and between individual cancer cells. 
Anomalous diffusion is heterogeneous even within  trajectories of individual tracer particles in the cytoplasm of cancerous cells. In Hela cells, analyses of the movements of quantum dots revealed heterogeneous sub-diffusion 
with antipersistent increments \cite{sabri2020elucidating}. 
It was suggested that quantum dots move with single constant anomalous exponents but stochastically switch between different mobility states, most likely due to transient associations with the cytoskeleton-shaken endoplasmic reticulum network. Previously, the emergence
of heterogeneous subdiffusion in HeLa cells and the values of the anomalous exponent were shown to depend both on particle size and non-specific interactions with the cytoplasmic interior \cite{etoc2018non}. 

\subsection{Leukocytes/hemocytes}
Diffusion of T cells (a type of leukocyte) is essential for the immune response of mammals. \emph{In vivo}, T cell motility is orchestrated by a complex combination of chemokines, interactions with antigen-presenting cells, and multiple guidance cues leading to the persistent directional movement of cells for several minutes following short periods of rest, broadly similar to the run-and-tumble motion of bacteria. There is a lot of evidence that T cell movement is super-diffusive and highly heterogeneous \cite{krummel2016t}. In na\"{i}ve CD4$^{+}$ T cells in the inguinal lymph nodes of anesthetized mice, T cells were found to move via anomalous diffusion, cycling between states of low and high motility roughly every 2 min  \cite{miller2003autonomous}. 
\begin{figure}
    \centering
    \includegraphics[width=1\textwidth]{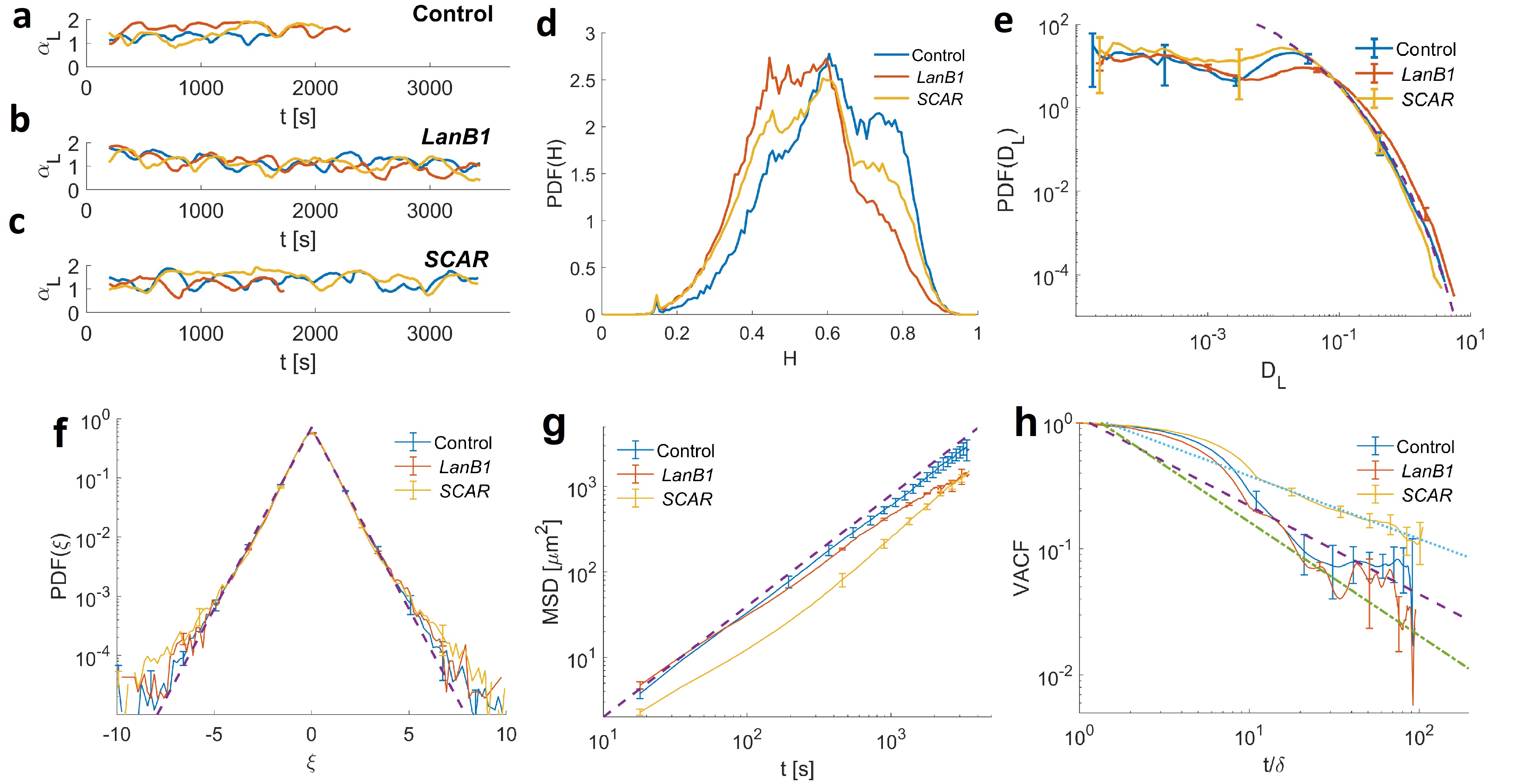}
    \caption{{\bf  Characterization of hemocytes’ heterogeneous anomalous diffusion in control and mutant (LanB1 and SCAR) \emph{Drosophila melanogaster}
embryos.} 
    {\bf (a), (b), (c)} Local anomalous exponents $\alpha_L(t)$ for representative hemocyte trajectories in a control, {\it LanB1} and {\it SCAR} embryo. $\alpha_L(t)$ display oscillatory behaviour. 
    {\bf (d)} Probability distributions of time dependent Hurst exponents $H(t)$ estimated using the neural network. {\bf (e)} Probability distributions of local generalized diffusion coefficients $D_{\alpha_L}$. The dashed line represents a fit of the distribution tail with the Weibull density function, $PDF(\eta) \sim c (\eta^{c-1}/\eta_0^{c}) \exp(-(\eta/\eta_0)^c)$ with $c=0.39$ and $\eta_0=0.008$. {\bf (f)} Probability distributions of displacements, $dx=x(t+\tau)-x(t)$, calculated at $\tau=17.6$ s are non-Gaussian and follow the Laplace distribution (the dashed line), $PDF(\xi)=\exp \left( - |\xi|/b \right)/2b$ with $\xi=dx/\sigma_x$ and $b=0.71$. {\bf (g)}  MSDs and TMSDs as a function of time interval grow super-diffusively with anomalous exponents $\alpha=1.3 \pm 0.1$ in the  control and $\alpha=1.1 \pm 0.1$ for {\it LanB1} embryo. {\it LanB1} and \emph{SCAR} embryos have a progressively more anti-persistent motion on the shortest time scales while their long-time behaviour remains superdiffusive. At longer time scales the MSD of SCAR increases with the  anomalous exponent $\alpha=1.5 \pm 0.1$. The dashed line represents the power-law function $t^{1.3}$ for the control as a guide. {\bf (h)} Velocity auto-correlation functions as a function of time interval are positive and decay to zero as power-laws, $(t/\delta)^{\alpha-2}$, with $\alpha$ determined from the MSDs. Statistical quantities in this figure were averaged over $5$ data sets of control, $2$ {\it LanB1}, $4$ {\it SCAR} embryos. The error bars correspond to the standard error of the mean.
    }
    \label{HEMOCYTES}
\end{figure}
In the brains of chronically infected mice, T cells were also moving super-diffusively on time scales of up to several minutes with non-Gaussian probability distributions of displacements  \cite{harris2012generalized}.  
The probability distributions of velocities suggested different types of movement of na\"{i}ve T cells, with slow moving cells showing a heavy-tailed probability distribution while faster moving cells were more Brownian \cite{fricke2016persistence}. Super-diffusion of T cells was also observed in larval zebrafish \cite{jerison2020heterogeneous} with the variability in motility amongst the cells attributed to heterogeneity in speed and turning behavior amongst the cells. The heterogeneity of motility emerged from a positive feedback
loop between actin flows and the maintenance of cell polarity in motile cells which results in a higher persistence of motility for faster cells \cite{jerison2020heterogeneous,maiuri2015actin}. Run-and-stop motion in the three-dimensional migration of T cells with periodic shape oscillations was observed \cite{cavanagh2022t}.

 
Hemocytes are phagocytes of invertebrates that plays a crucial role in their immune system. Recently, our group discovered heterogeneous anomalous diffusion of hemocytes during their migration in live \emph{Drosophila melanogaster} embryos (figure \ref{HEMOCYTES}). Hemocytes were moving super-diffusively in the control and in two mutant embryos (\emph{LanB1} and \emph{SCAR}) with positive velocity auto-correlation functions which decay to zero as power-laws, $t^{\alpha-2}$, with $\alpha$ in agreement with values calculated from the MSDs. The probability distributions of incremental displacements  $dx=x(t+\tau)-x(t)$, $dy=y(t+\tau)-y(t)$ and $dz=z(t+\tau)-z(t)$ follow Laplace distributions instead of Gaussians. The non-Gaussian form of the probability distributions of the displacements suggests that the motility of hemocytes is heterogeneous. Using a deep learning neural network \cite{han2020deciphering}, a broad spectrum of Hurst exponents $H$ (related to anomalous exponents as $H=\alpha/2$) and generalized diffusion coefficients were found \cite{korabel2022hemocytes}. The probability distributions of Hurst exponents were characteristic of the type of mutant considered. The hemocytes are switching between different levels of persistent motility, rather than distinct subpopulations of cells having altered dynamics. 
Interestingly, local (time resolved) anomalous exponents (and also local generalized diffusion coefficients) of individual hemocytes were found to exhibit oscillatory behaviour which could be a manifestation of a positive feedback loop between actin flows and maintenance of cell polarity   \cite{jerison2020heterogeneous,maiuri2015actin} and/or similar periodic shape oscillations to those observed during the run-and-stop motions of T cells \cite{cavanagh2022t}. Upon their contact, we found that the hemocyte motion was synchronised, although their clear sense of directional motion was lost \cite{korabel2022hemocytes}. Contact of the cells inhibits their motion (a process of contact inhibitory locomotion) and the anomalous transport of cells continues to be synchronised in terms of the anomalous exponents, generalized diffusion coefficients, displacements, and velocities.  An anomalous tango is thus observed between the hemocyte cells when they touch. 

\subsection{Bacterial cells}
The simplest model for bacterial motion is Poisson distributed stretches of ballistic motion followed by tumbles that randomise the direction \cite{lovely1975statistical}. It is equivalent to the model for the conformations of freely jointed polymers in which space is replaced with time. However, it has been known that bacterial motion is to some degree anomalous, since it was first quantified with tracking experiments e.g. the runs clearly have some curvature in the images of H. Berg (e.g. on the front cover of the book "Random Walks in Biology" \cite{berg2018random}) and thus the runs are subballistic with a reduced persistence. Therefore the motion of the bacteria during the runs is not fully persistent as expected for ballistic transport. 

\begin{figure}
    \centering
    \includegraphics[width=1\textwidth]{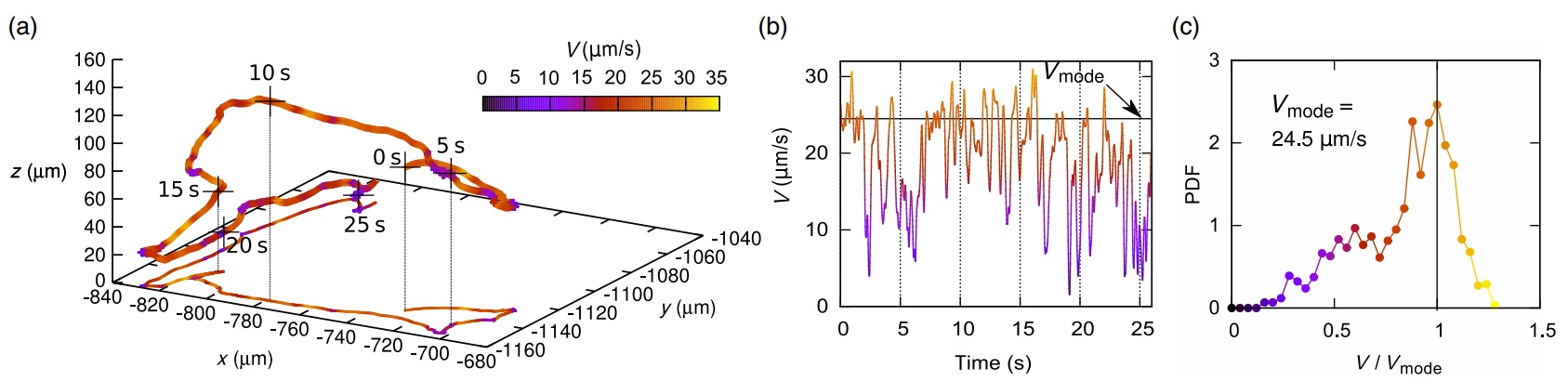}
    \caption{{\bf  A typical heterogeneous trajectory of a RP437 wild-type \emph{E. coli} bacterial cell} . 
    {\bf (a)} A 3D trajectory and its projection on the x-y plane, {\bf (b)} the velocity as a function of time and {\bf (c)} the velocity probability distribution. The marks every 5 s in the 3D track are references for comparison with (b). Reprinted from \cite{figueroa20203d}.
    }
    \label{BACTERIA}
\end{figure}

\begin{figure}
    \centering
    \includegraphics[width=1\textwidth]{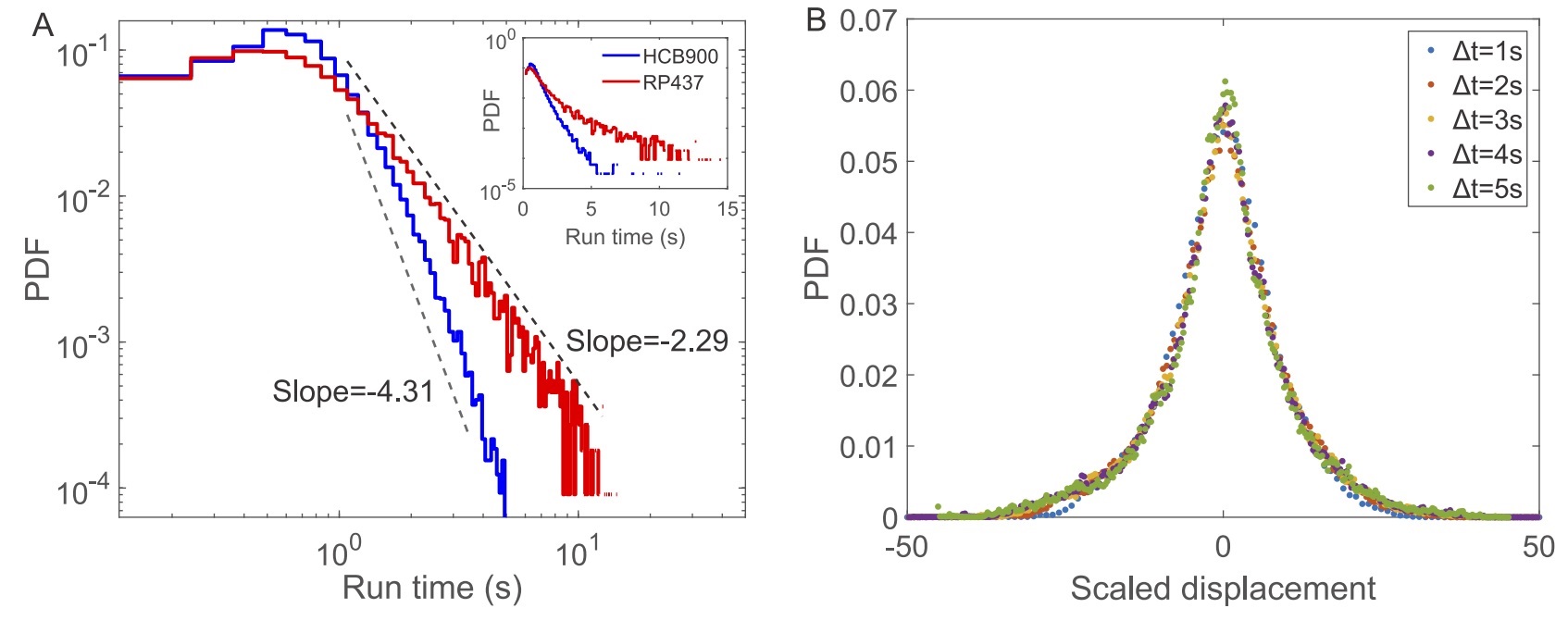}
    \caption{{\bf  The trajectories of RP437 wild-type \emph{E. coli} in dilute suspension are consistent with Lévy walks.} 
    {\bf (A)} The probability distributions of run times for the two strains (red line, RP437; blue line, HCB900). The inset shows a semi-log plot of the probability density function (PDF). {\bf (B)} The probability distribution of the displacements during fixed time intervals $\Delta t$, scaled by $\Delta t^{\gamma}$ where $\gamma=1/(3-\alpha)$. Reprinted from \cite{figueroa20203d}.
    }
    \label{HUO}
\end{figure}

Super-diffusive 
bacterial motion was probed by micron-scale beads in a freely suspended soap film \cite{wu2000particle} that contained a dense swarming bacterial suspension \cite{ariel2015swarming}. Efforts have been made to understand the anomalous transport of bacteria using fractional Brownian motion \cite{zonia2009swimming}. A multi-fractal model has also been introduced to describe the swimming of bacteria \cite{koorehdavoudi2017multi}.
A more sophisticated form of HAT with bacteria is related to their memory due to sensory proteins \cite{figueroa20203d}. Large behavioral variability of wild-type \emph{E. coli} was revealed in their three-dimensional trajectories (figure \ref{BACTERIA}) with a broad probability distribution of persistence times within a monoclonal population of bacteria.
HAT for the internal motion of proteins inside bacteria has been examined using a super-statistical approach. A q-Gaussian was found for the displacement probability distribution \cite{itto2021superstatistical}. 
L{\'e}vy walks have been successful in describing anomalous transport in concentrated suspensions of bacteria \cite{ariel2015swarming}. More recently there is evidence for the applicability of L{\'e}vy walks in quiescent dilute solutions of \emph{E. coli} (figure \ref{HUO}) \cite{huo2021swimming} and the dynamics of gold nanorods in bacterial swarms \cite{feng2019single}. These models better describe the distribution of run times (moving beyond simple Poisson models with exponential distributions), but still fail to describe the reduced persistence of runs observed in experiments. Correlated truncated L{\'e}vy processes were used to describe mixing in dilute suspensions of algae and bacteria \cite{zaid2011levy}.

\subsection{Bacterial biofilms}
Most bacteria spend most of their time in surface-attached aggregates. There are numerous advantages for them to follow this communal style of living and the main one is protection from antibiotics.

Bacterial biofilms are rheologically heterogeneous on the micron scale and can become more homogeneous as a function of height as they
mature. The motion of \emph{P. aeruginosa} (a rod-shaped swimming bacterium) is a combination of active and thermal motion, whereas with non-motile \emph{S. aureus} the motion is only thermal. A wide range of anomalous transport in biofilms was found with both varieties of bacteria \cite{rogers2008microrheology,hart2019microrheology}.
The biofilms were locally heterogeneous and the MSDs of neighboring bacteria differed by up to a factor of 10. Heterogeneity on individual trajectories was also observed during the biofilm formation. Trajectories of individual \emph{B. subtilis} cells can be composed of super- and subdiffusive regimes as cells continuously enter or leave aggregates \cite{worlitzer2022biophysical}, figure 9. Elastic mechanical waves have been observed to propagate across bacterial biofilms due to their motile activity \cite{xu2022autonomous}. 
\begin{figure}
    \centering
    \includegraphics[width=0.8\textwidth]{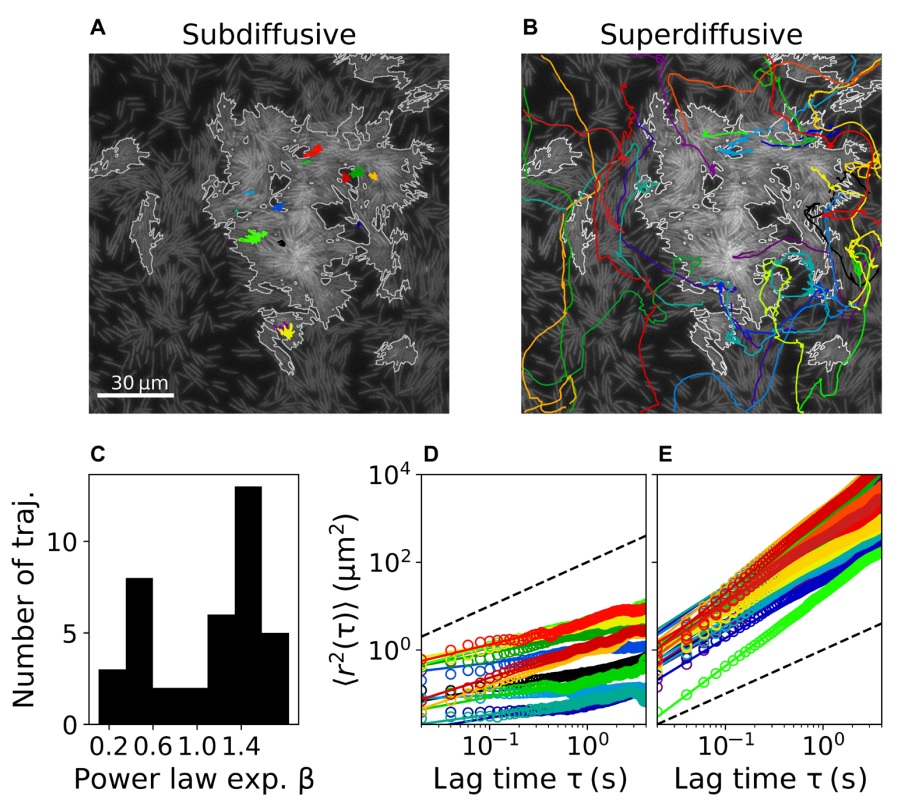}
    \caption{{\bf Heterogeneous trajectories of \emph{B. subtilis} bacteria in a growing biofilm.} 
    {\bf (A)} Subdiffusive and {\bf (B)} superdiffusive trajectories superimposed on the microscopic image at $t = 0$, where stationary aggregates are outlined. {\bf (C)} Histogram of power law exponents obtained from nontransitive trajectories. Mean square displacements (open circles) as a function of lag time for {\bf (D)} subdiffusive trajectories [color matches the trajectories of (A)] and {\bf (E)} superdiffusive trajectories [color matches the trajectories of (B)]. Fits are provided as solid lines, while the expected behaviour for normal diffusion is indicated as a dashed line. Reprinted from \cite{worlitzer2022biophysical}.
    }
    \label{BIOFILM}
\end{figure}

Bacterial biofilms have heterogeneous structures on the micron and nanoscale.  Furthermore, in general, both \emph{polymeric} and \emph{colloidal gels} exhibit heterogeneous viscoelasticity (even simple one-component gels with a solvent are heterogeneous), so both properties may contribute to the heterogeneous viscoelasticity of biofilms (both predominantly colloidal, i.e. very little polymer, and polymeric, i.e. very few cells, biofilms can be made).
The average spatial heterogeneity of bacterial cells in biofilms can be quantified separately using the Ripley K function \cite{hart2019microrheology}, but pair correlation functions can also be used.

Recent studies indicate that the motion of some antibiotics in biofilms is subdiffusive \cite{kosztolowicz2020modelling} 
and biofilm/sputum composites have also been considered. Transport in the composites was studied using a system consisting of two different media \cite{korabel2011boundary,kosztolowicz2020diffusion}. Thus, the effective action of biofilms against antibiotics could be due to the introduction of slow subdiffusive transport, reducing the rate at which high concentrations of antibiotics arrive at the bacteria. Other factors are also known to be important, such as phenotypic changes of the bacteria in the biofilms, which adopt very low metabolic states e.g. persister cells.

\subsection{Eukaryotic microorganisms}


Eukaryotic microorganisms are known to generate non-equilibrium fluctuations in their surroundings, which have been studied using passive tracer particles immersed in suspensions of eukaryotic swimmers. The long-time dynamics of the tracers was described with a normal (Brownian) yet non-Gaussian diffusion model \cite{leptos2009dynamics}. Theoretically it was predicted that hydrodynamic interactions between the tracer and the swimmers lead to the occurrence of a L\'evy flight regime in the tracer dynamics and non-Gaussian distributions of displacements \cite{kanazawa2020loopy}. L{\'e}vy flight models are nonphysical since they imply instantaneous transport of particles during jumps, so these results should be revisited with a L{\'e}vy walk model. Interestingly, the shape asymmetry of tracers is predicted to induce ratchet effects that alter fluctuations in cell motion and lead to asymptotic  super-diffusion \cite{granek2022anomalous}. 

The  two-dimensional motion of the social amoeba \emph{Dictyostelium discoideum} (a species of soil-dwelling amoeba) was found to be super-diffusive with 
broadly distributed 
diffusivities 
\cite{li2008persistent} (probably due to cell-to-cell variability) and was characterized using an exponential probability distribution of displacements and generalised gamma probability distribution of cell speeds \cite{cherstvy2018non}. The diffusion of the parasitic soil nematode \emph{Phasmarhabditis hermaphrodita} was found to be normal but with a significant variation of the diffusion coefficients among the
the population that followed a gamma probability distribution \cite{hapca2009anomalous}.
Analysis of the motion of four flagellated Protozoa species revealed super-diffusion with non-Gaussian probability distributions of displacements together with a generalised gamma probability distribution for their speeds and a broad probability distribution of Hurst exponents characterizing the self-similarity of the cells' velocity time series \cite{alves2016transient}.

\section*{Conclusions}
The questions involved in heterogeneous anomalous transport are of central importance to quantitative studies of molecular and cellular biology. The main challenge is to decipher the origin of the underlying distributions of anomalous exponents and generalized diffusion coefficients. In specific cases quantitative treatments are possible e.g. for diffusing particles undergoing fission and fusion, a power-law probability distribution of cluster size was predicted by Smoluchowski’s coagulation theory, which naturally leads to polydispersity with diffusion coefficients that scale with the cluster size as a power law \cite{vicsek1985scaling}. Alternative theoretical methodologies to consider the motility of molecules and cells include focusing on the small system thermodynamics of trajectories \cite{peliti2021stochastic} or soft matter models \cite{de1992soft}. 
Here we prioritised the experimental measurement of the statistics of motion and the development of agnostic models which can directly describe the anomalous transport observed. The extreme complexity of high-concentration biological systems in living cells  often precludes more fundamental stochastic thermodynamic or soft matter modelling. The hope is that heterogeneous anomalous transport will allow the gap to be bridged to more fundamental molecular models. Furthermore, many classical soft matter models were based on Gaussian fluctuations and still need to be extended to describe the heterogeneous anomalous statistics observed experimentally. Machine learning techniques (e.g. deep learning neural networks) combined with high-resolution microscope-based tracking allow some of the major challenges in the characterisation of HAT to be overcome. Continued development of anomalous agent-based models will allow a fundamental physical framework to understand collective anomalous motility to be developed e.g. models for the collective motion of cells in tissue based on HAT.

So, what if there are more complicated models based on HAT that are required to fit the stochastic motion of particles in biology? There are a number of advantages to this approach:

$\bullet$ HAT has \emph{diagnostic potential} in medicine. Models based on anomalous transport provide more detailed information to differentiate mutants or diseased states of cells \cite{korabel2022hemocytes}. This could be used for diagnostic purposes e.g. cancer cells often have altered motile phenotypes in leukemia that can drive metastasis and could be diagnosed from a blood sample \cite{olson2010linking}. Although genetic analysis of the cancer cells would also be a priority for diagnosis, connecting genotype to phenotype is a crucial step, and making the connection necessitates motility assays and robust methods to quantify them. Furthermore, gait analysis (how cells swim or crawl) could be used to classify varieties of microorganisms, which again has diagnostic potential
\cite{lauga2020fluid}. Also, diseased states of eukaryotic cells (e.g. in humans or livestock) are often associated with impaired intracellular motility e.g. in motor neuron disease where the motility of neurotransmitters in endosomes can be compromised \cite{hirokawa2010molecular}. Information could be used from HAT analysis for gene therapy targeted to treat the phenotypes of impaired intracellular motility and to understand the success of treatments.

$\bullet$ HAT provides \emph{fundamental insights} into biology. The machinery inside cells is finely optimised by evolution to provide efficient solutions to biological challenges. How do cells and molecules make use of anomalous stochastic phenomena \cite{hofling2013anomalous,bressloff2014stochastic,woringer2020anomalous}?

$\bullet$ Anomalous transport presents a \emph{fundamental barrier} in a diverse range of problems in quantitative cellular and molecular biology e.g. understanding the action of enzymes inside cells where the kinetics can be dramatically affected 
\cite{hellmann2012enhancing}, how cells move in developing embryos \cite{korabel2022hemocytes}, the transcription of information from DNA \cite{golding2006physical,sumner2021random}, how reaction-diffusion processes lead to robust patterning during morphogenesis \cite{pismen2021active} and the finely orchestrated movement of endosomes inside cells \cite{korabel2021local,han2020deciphering}. Without a thorough quantitative understanding of anomalous transport, we lack tools to handle a large number of crucial biological problems.

$\bullet$ Numerous \emph{new physical phenomena} are still left to be discovered that are related to heterogeneous anomalous transport e.g. the motility of molecular motors in active matter \cite{pismen2021active}, collective cellular motility in embryos \cite{korabel2022hemocytes}, the motion of wavefronts of electrical signalling in bacterial biofilms \cite{blee2019spatial}
and how bacteria swim \cite{lauga2016bacterial}. The field provides an interesting perspective to the subject of soft matter physics, focusing on the statistics of motility.

$\bullet$ The \emph{applied mathematics} of HAT is novel and interesting in its own right e.g. fractional partial differential equations \cite{podlubny1999fractional}, the generalized central limit theorem, Ito calcululus \cite{gardiner2009stochastic} and multi-fractals \cite{falconer2004fractal}.

\section*{Acknowledgements}
We would like to thank Viki Allan, Sergei Fedotov, Eli Barkai, Victor Martorelli, Salman Rogers, Ian Roberts, Tom Millard, Jian Lu, Daniel Han, Jack Hart, Sorasak Phanphak, Joanna Blee, David Kenwright and Andrew Harrison for their invaluable help over the years.

\section*{References}

\bibliographystyle{iopart-num} 
\bibliography{Rbib}      

\providecommand{\newblock}{}
\begin{thebibliography}{100}
\expandafter\ifx\csname url\endcsname\relax
  \def\url#1{{\tt #1}}\fi
\expandafter\ifx\csname urlprefix\endcsname\relax\def\urlprefix{URL }\fi
\providecommand{\eprint}[2][]{\url{#2}}

\bibitem{bouchaud1990anomalous}
Bouchaud J~P and Georges A 1990 {\em Physics reports\/} {\bf 195} 127--293

\bibitem{metzler2000random}
Metzler R and Klafter J 2000 {\em Physics reports\/} {\bf 339} 1--77

\bibitem{metzler2004restaurant}
Metzler R and Klafter J 2004 {\em Journal of Physics A: Mathematical and
  General\/} {\bf 37} R161

\bibitem{klages2008anomalous}
Klages R, Radons G and Sokolov I~M 2008 {\em Anomalous transport\/} (Wiley
  Online Library)

\bibitem{barkai2012single}
Barkai E, Garini Y and Metzler R 2012 {\em Phys. Today\/} {\bf 65} 29

\bibitem{hofling2013anomalous}
H{\"o}fling F and Franosch T 2013 {\em Reports on Progress in Physics\/} {\bf
  76} 046602

\bibitem{bressloff2014stochastic}
Bressloff P~C 2014 {\em Stochastic processes in cell biology\/} vol~41
  (Springer)

\bibitem{metzler2014anomalous}
Metzler R, Jeon J~H, Cherstvy A~G and Barkai E 2014 {\em Physical Chemistry
  Chemical Physics\/} {\bf 16} 24128--24164

\bibitem{banks2005anomalous}
Banks D~S and Fradin C 2005 {\em Biophysical journal\/} {\bf 89} 2960--2971

\bibitem{de1979scaling}
De~Gennes P~G and Gennes P~G 1979 {\em Scaling concepts in polymer physics\/}
  (Cornell university press)

\bibitem{rubinstein2003polymer}
Rubinstein M, Colby R~H {\em et~al.\/} 2003 {\em Polymer physics\/} vol~23
  (Oxford university press New York)

\bibitem{feynman2011six}
Feynman R~P, Leighton R~B and Sands M 2011 {\em Six easy pieces: Essentials of
  physics explained by its most brilliant teacher\/} (Basic Books)

\bibitem{Mandelbrot1983}
Mandelbrot B~B 1983 {\em The fractal geometry of nature\/} 3rd ed (New York: W.
  H. Freeman and Comp.)

\bibitem{munoz2021objective}
Mu{\~n}oz-Gil G, Volpe G, Garcia-March M~A, Aghion E, Argun A, Hong C~B, Bland
  T, Bo S, Conejero J~A, Firbas N {\em et~al.\/} 2021 {\em Nature
  communications\/} {\bf 12} 1--16

\bibitem{meroz2015toolbox}
Meroz Y and Sokolov I~M 2015 {\em Physics Reports\/} {\bf 573} 1--29

\bibitem{sokolov2005diffusion}
Sokolov I~M and Klafter J 2005 {\em Chaos: An Interdisciplinary Journal of
  Nonlinear Science\/} {\bf 15} 026103

\bibitem{sokolov2012models}
Sokolov I~M 2012 {\em Soft Matter\/} {\bf 8} 9043--9052

\bibitem{zaburdaev2015levy}
Zaburdaev V, Denisov S and Klafter J 2015 {\em Reviews of Modern Physics\/}
  {\bf 87} 483

\bibitem{saxton1997single}
Saxton M~J and Jacobson K 1997 {\em Annual review of biophysics and
  biomolecular structure\/} {\bf 26} 373--399

\bibitem{kusumi2014tracking}
Kusumi A, Tsunoyama T~A, Hirosawa K~M, Kasai R~S and Fujiwara T~K 2014 {\em
  Nature chemical biology\/} {\bf 10} 524--532

\bibitem{manzo2015weak}
Manzo C, Torreno-Pina J~A, Massignan P, Lapeyre~Jr G~J, Lewenstein M and Parajo
  M~F~G 2015 {\em Physical Review X\/} {\bf 5} 011021

\bibitem{stolle2019anomalous}
Stolle M~D and Fradin C 2019 {\em Biophysical journal\/} {\bf 116} 791--806

\bibitem{weiss2003anomalous}
Weiss M, Hashimoto H and Nilsson T 2003 {\em Biophysical journal\/} {\bf 84}
  4043--4052

\bibitem{berne2000dynamic}
Berne B~J and Pecora R 2000 {\em Dynamic light scattering: with applications to
  chemistry, biology, and physics\/} (Courier Corporation)

\bibitem{giavazzi2009scattering}
Giavazzi F, Brogioli D, Trappe V, Bellini T and Cerbino R 2009 {\em Physical
  Review E\/} {\bf 80} 031403

\bibitem{kneller2005quasielastic}
Kneller G~R 2005 {\em Physical Chemistry Chemical Physics\/} {\bf 7} 2641--2655

\bibitem{calandrini2010fractional}
Calandrini V, Abergel D and Kneller G~R 2010 {\em The Journal of chemical
  physics\/} {\bf 133} 10B604

\bibitem{yamamoto2021universal}
Yamamoto E, Akimoto T, Mitsutake A and Metzler R 2021 {\em Physical review
  letters\/} {\bf 126} 128101

\bibitem{miyaguchi2017elucidating}
Miyaguchi T 2017 {\em Physical Review E\/} {\bf 96} 042501

\bibitem{waigh2005microrheology}
Waigh T~A 2005 {\em Reports on progress in physics\/} {\bf 68} 685

\bibitem{waigh2016advances}
Waigh T~A 2016 {\em Reports on Progress in Physics\/} {\bf 79} 074601

\bibitem{mason1995optical}
Mason T~G and Weitz D~A 1995 {\em Physical review letters\/} {\bf 74} 1250

\bibitem{xu1998compliance}
Xu J, Viasnoff V and Wirtz D 1998 {\em Rheologica acta\/} {\bf 37} 387--398

\bibitem{bonfanti2020fractional}
Bonfanti A, Kaplan J~L, Charras G and Kabla A 2020 {\em Soft Matter\/} {\bf 16}
  6002--6020

\bibitem{hunter2011tracking}
Hunter G~L, Edmond K~V, Elsesser M~T and Weeks E~R 2011 {\em Optics express\/}
  {\bf 19} 17189--17202

\bibitem{cheng2003rotational}
Cheng Z and Mason T 2003 {\em Physical review letters\/} {\bf 90} 018304

\bibitem{jain2017diffusing}
Jain R and Sebastian K 2017 {\em The Journal of chemical physics\/} {\bf 146}
  214102

\bibitem{frisch1995turbulence}
Frisch U 1995 {\em Turbulence: the legacy of AN Kolmogorov\/} (Cambridge
  university press)

\bibitem{malm2017elastic}
Malm A and Waigh T 2017 {\em Scientific reports\/} {\bf 7} 1--13

\bibitem{dunkel2013fluid}
Dunkel J, Heidenreich S, Drescher K, Wensink H~H, B{\"a}r M and Goldstein R~E
  2013 {\em Physical review letters\/} {\bf 110} 228102

\bibitem{shlesinger1993strange}
Shlesinger M~F, Zaslavsky G~M and Klafter J 1993 {\em Nature\/} {\bf 363}
  31--37

\bibitem{shlesinger1987levy}
Shlesinger M~F, West B and Klafter J 1987 {\em Physical Review Letters\/} {\bf
  58} 1100

\bibitem{meneveau1991multifractal}
Meneveau C and Sreenivasan K 1991 {\em Journal of Fluid Mechanics\/} {\bf 224}
  429--484

\bibitem{brunton2020machine}
Brunton S~L, Noack B~R and Koumoutsakos P 2020 {\em Annual review of fluid
  mechanics\/} {\bf 52} 477--508

\bibitem{mendez2010reaction}
Mendez V, Fedotov S and Horsthemke W 2010 {\em Reaction-transport systems:
  mesoscopic foundations, fronts, and spatial instabilities\/} (Springer
  Science \& Business Media)

\bibitem{keener2009mathematical}
Keener J and Sneyd J 2009 {\em Mathematical physiology: II: Systems
  physiology\/} (Springer)

\bibitem{volpert2013fronts}
Volpert V, Nec Y and Nepomnyashchy A 2013 {\em Philosophical Transactions of
  the Royal Society A: Mathematical, Physical and Engineering Sciences\/} {\bf
  371} 20120179

\bibitem{henry2006anomalous}
Henry B, Langlands T and Wearne S 2006 {\em Physical Review E\/} {\bf 74}
  031116

\bibitem{erban2020stochastic}
Erban R and Chapman S~J 2020 {\em Stochastic modelling of reaction--diffusion
  processes\/} vol~60 (Cambridge University Press)

\bibitem{blee2019spatial}
Blee J, Roberts I and Waigh T 2019 {\em Physical Review E\/} {\bf 100} 052401

\bibitem{mainardi2007levy}
Mainardi F 2007 {\em Lecture Notes on Mathematical Physics\/}

\bibitem{lim2002self}
Lim S~C and Muniandy S~V 2002 {\em Physical Review E\/} {\bf 66} 021114

\bibitem{jeon2014scaled}
Jeon J~H, Chechkin A~V and Metzler R 2014 {\em Physical Chemistry Chemical
  Physics\/} {\bf 16} 15811--15817

\bibitem{chechkin2017brownian}
Chechkin A~V, Seno F, Metzler R and Sokolov I~M 2017 {\em Physical Review X\/}
  {\bf 7} 021002

\bibitem{fulinski2011anomalous}
Fuli{\'n}ski A 2011 {\em Physical Review E\/} {\bf 83} 061140

\bibitem{cherstvy2015ergodicity}
Cherstvy A~G and Metzler R 2015 {\em Journal of Statistical Mechanics: Theory
  and Experiment\/} {\bf 2015} P05010

\bibitem{wang2020fractional}
Wang W, Cherstvy A~G, Chechkin A~V, Thapa S, Seno F, Liu X and Metzler R 2020
  {\em Journal of Physics A: Mathematical and Theoretical\/} {\bf 53} 474001

\bibitem{wang2020unexpected}
Wang W, Seno F, Sokolov I~M, Chechkin A~V and Metzler R 2020 {\em New Journal
  of Physics\/} {\bf 22} 083041

\bibitem{mackala2019statistical}
Ma{\'c}ka{\l}a A and Magdziarz M 2019 {\em Physical Review E\/} {\bf 99} 012143

\bibitem{peltier1995multifractional}
Peltier R~F and V{\'e}hel J~L 1995 {\em Multifractional Brownian motion:
  definition and preliminary results\/} Ph.D. thesis INRIA

\bibitem{benassi1997elliptic}
Benassi A, Roux D and Jaffard S 1997 {\em Revista matem{\'a}tica
  iberoamericana\/} {\bf 13} 19--90

\bibitem{adler2010geometry}
Adler R~J 2010 {\em The geometry of random fields\/} (SIAM)

\bibitem{benzi1984multifractal}
Benzi R, Paladin G, Parisi G and Vulpiani A 1984 {\em Journal of Physics A:
  Mathematical and General\/} {\bf 17} 3521

\bibitem{falconer2004fractal}
Falconer K 2004 {\em Fractal geometry: mathematical foundations and
  applications\/} (John Wiley \& Sons)

\bibitem{harte2001multifractals}
Harte D 2001 {\em Multifractals: theory and applications\/} (Chapman and
  Hall/CRC)

\bibitem{mandelbrot2013multifractals}
Mandelbrot B~B 2013 {\em Multifractals and 1/ƒ noise: Wild self-affinity in
  physics (1963--1976)\/} (Springer)

\bibitem{mura2008characterizations}
Mura A and Pagnini G 2008 {\em Journal of Physics A: Mathematical and
  Theoretical\/} {\bf 41} 285003

\bibitem{mura2008non}
Mura A, Taqqu M~S and Mainardi F 2008 {\em Physica A: Statistical Mechanics and
  its Applications\/} {\bf 387} 5033--5064

\bibitem{ayache2005multifractional}
Ayache A and Taqqu M~S 2005 {\em Publicacions matematiques\/}  459--486

\bibitem{molina2016fractional}
Molina-Garc{\'\i}a D, Pham T~M, Paradisi P, Manzo C and Pagnini G 2016 {\em
  Physical Review E\/} {\bf 94} 052147

\bibitem{han2020deciphering}
Han D, Korabel N, Chen R, Johnston M, Gavrilova A, Allan V~J, Fedotov S and
  Waigh T~A 2020 {\em Elife\/} {\bf 9} e52224

\bibitem{korabel2021local}
Korabel N, Han D, Taloni A, Pagnini G, Fedotov S, Allan V and Waigh T~A 2021
  {\em Entropy\/} {\bf 23} 958

\bibitem{korabel2023ensemble}
Korabel N, Taloni A, Pagnini G, Allan V, Fedotov S and Waigh T~A 2023 {\em
  Scientific Reports\/} {\bf 13} 8789

\bibitem{korabel2022hemocytes}
Korabel N, Clemente G~D, Han D, Feldman F, Millard T~H and Waigh T~A 2022 {\em
  Communications physics\/} {\bf 5} 269

\bibitem{cherstvy2016anomalous}
Cherstvy A~G and Metzler R 2016 {\em Physical Chemistry Chemical Physics\/}
  {\bf 18} 23840--23852

\bibitem{cherstvy2013population}
Cherstvy A~G and Metzler R 2013 {\em Physical Chemistry Chemical Physics\/}
  {\bf 15} 20220--20235

\bibitem{montroll1965random}
Montroll E~W and Weiss G~H 1965 {\em Journal of Mathematical Physics\/} {\bf 6}
  167--181

\bibitem{klafter2011first}
Klafter J and Sokolov I~M 2011 {\em First steps in random walks: from tools to
  applications\/} (OUP Oxford)

\bibitem{wong2004anomalous}
Wong I, Gardel M, Reichman D, Weeks E~R, Valentine M, Bausch A and Weitz D~A
  2004 {\em Physical review letters\/} {\bf 92} 178101

\bibitem{xu2011subdiffusion}
Xu Q, Feng L, Sha R, Seeman N and Chaikin P 2011 {\em Physical review
  letters\/} {\bf 106} 228102

\bibitem{weigel2011ergodic}
Weigel A~V, Simon B, Tamkun M~M and Krapf D 2011 {\em Proceedings of the
  National Academy of Sciences\/} {\bf 108} 6438--6443

\bibitem{chechkin2005fractional}
Chechkin A~V, Gorenflo R and Sokolov I~M 2005 {\em Journal of Physics A:
  Mathematical and General\/} {\bf 38} L679

\bibitem{korabel2010paradoxes}
Korabel N and Barkai E 2010 {\em Physical review letters\/} {\bf 104} 170603

\bibitem{korabel2011anomalous}
Korabel N and Barkai E 2011 {\em Journal of Statistical Mechanics: Theory and
  Experiment\/} {\bf 2011} P05022

\bibitem{roth2020inhomogeneous}
Roth P and Sokolov I~M 2020 {\em Physical Review E\/} {\bf 102} 012133

\bibitem{dentz2012diffusion}
Dentz M, Gouze P, Russian A, Dweik J and Delay F 2012 {\em Advances in Water
  Resources\/} {\bf 49} 13--22

\bibitem{grebenkov2018heterogeneous}
Grebenkov D~S and Tupikina L 2018 {\em Physical Review E\/} {\bf 97} 012148

\bibitem{barkai2020packets}
Barkai E and Burov S 2020 {\em Physical review letters\/} {\bf 124} 060603

\bibitem{korabel2011boundary}
Korabel N and Barkai E 2011 {\em Physical Review E\/} {\bf 83} 051113

\bibitem{kosztolowicz2012subdiffusion}
Koszto{\l}owicz T, Dworecki K and Lewandowska K~D 2012 {\em Physical Review
  E\/} {\bf 86} 021123

\bibitem{zaburdaev2008random}
Zaburdaev V, Schmiedeberg M and Stark H 2008 {\em Physical Review E\/} {\bf 78}
  011119

\bibitem{fedotov2017emergence}
Fedotov S and Korabel N 2017 {\em Physical Review E\/} {\bf 95} 030107

\bibitem{sliusarenko2019finite}
Sliusarenko O~Y, Vitali S, Sposini V, Paradisi P, Chechkin A, Castellani G and
  Pagnini G 2019 {\em Journal of Physics A: Mathematical and Theoretical\/}
  {\bf 52} 095601

\bibitem{tabei2013intracellular}
Tabei S~A, Burov S, Kim H~Y, Kuznetsov A, Huynh T, Jureller J, Philipson L~H,
  Dinner A~R and Scherer N~F 2013 {\em Proceedings of the National Academy of
  Sciences\/} {\bf 110} 4911--4916

\bibitem{kozubowski2006fractional}
Kozubowski T~J, Meerschaert M~M and Podgorski K 2006 {\em Advances in applied
  probability\/} {\bf 38} 451--464

\bibitem{fox2021aging}
Fox Z~R, Barkai E and Krapf D 2021 {\em Nature communications\/} {\bf 12} 1--9

\bibitem{chechkin2003distributed}
Chechkin A~V, Gorenflo R, Sokolov I~M and Gonchar V~Y 2003 {\em Fractional
  Calculus and Applied Analysis\/} {\bf 6} 259--280

\bibitem{chechkin2012natural}
Chechkin A, Sokolov I~M and Klafter J 2012 Natural and modified forms of
  distributed-order fractional diffusion equations {\em Fractional Dynamics:
  Recent Advances\/} (World Scientific) pp 107--127

\bibitem{sandev2015distributed}
Sandev T, Chechkin A~V, Korabel N, Kantz H, Sokolov I~M and Metzler R 2015 {\em
  Physical Review E\/} {\bf 92} 042117

\bibitem{fedotov2019asymptotic}
Fedotov S and Han D 2019 {\em Physical review letters\/} {\bf 123} 050602

\bibitem{granek1997semi}
Granek R 1997 {\em Journal de Physique II\/} {\bf 7} 1761--1788

\bibitem{pusey1991liquids}
Pusey P 1991 Liquids, freezing and the glass transition

\bibitem{turing1990chemical}
Turing A~M 1990 {\em Bulletin of mathematical biology\/} {\bf 52} 153--197

\bibitem{pismen2021active}
Pismen L 2021 {\em Active Matter Within and Around Us: From Self-Propelled
  Particles to Flocks and Living Forms\/} (Springer Nature)

\bibitem{dieterle2021diffusive}
Dieterle P~B and Amir A 2021 {\em Physical Review E\/} {\bf 104} 014406

\bibitem{xin2000front}
Xin J 2000 {\em SIAM review\/} {\bf 42} 161--230

\bibitem{lanoiselee2018diffusion}
Lanoisel{\'e}e Y, Moutal N and Grebenkov D~S 2018 {\em Nature communications\/}
  {\bf 9} 1--16

\bibitem{lanoiselee2018model}
Lanoisel{\'e}e Y and Grebenkov D~S 2018 {\em Journal of Physics A: Mathematical
  and Theoretical\/} {\bf 51} 145602

\bibitem{ben2000diffusion}
Ben-Avraham D and Havlin S 2000 {\em Diffusion and reactions in fractals and
  disordered systems\/} (Cambridge university press)

\bibitem{callaghan2011translational}
Callaghan P~T 2011 {\em Translational dynamics and magnetic resonance:
  principles of pulsed gradient spin echo NMR\/} (Oxford University Press)

\bibitem{de1976percolation}
de~Gennes P~G {\em et~al.\/} 1976 {\em La recherche\/} {\bf 7} 919--927

\bibitem{bunde2005diffusion}
Bunde A and Kantelhardt J~W 2005 Diffusion and conduction in percolation
  systems {\em Diffusion in Condensed Matter: Methods, Materials, Models\/}
  (Springer) pp 895--914

\bibitem{pacheco2022universal}
Pacheco-Pozo A and Sokolov I~M 2022 {\em Journal of Physics A: Mathematical and
  Theoretical\/} {\bf 55} 345001

\bibitem{akimoto2016universal}
Akimoto T, Barkai E and Saito K 2016 {\em Physical review letters\/} {\bf 117}
  180602

\bibitem{chubynsky2014diffusing}
Chubynsky M~V and Slater G~W 2014 {\em Physical review letters\/} {\bf 113}
  098302

\bibitem{metzler2020superstatistics}
Metzler R 2020 {\em The European Physical Journal Special Topics\/} {\bf 229}
  711--728

\bibitem{metzner2015superstatistical}
Metzner C, Mark C, Steinwachs J, Lautscham L, Stadler F and Fabry B 2015 {\em
  Nature communications\/} {\bf 6} 1--8

\bibitem{wang2012brownian}
Wang B, Kuo J, Bae S~C and Granick S 2012 {\em Nature materials\/} {\bf 11}
  481--485

\bibitem{wang2009anomalous}
Wang B, Anthony S~M, Bae S~C and Granick S 2009 {\em Proceedings of the
  National Academy of Sciences\/} {\bf 106} 15160--15164

\bibitem{gheorghiu2004heterogeneity}
Gheorghiu S and Coppens M~O 2004 {\em Proceedings of the National Academy of
  Sciences\/} {\bf 101} 15852--15856

\bibitem{beck2003superstatistics}
Beck C and Cohen E~G 2003 {\em Physica A: Statistical mechanics and its
  applications\/} {\bf 322} 267--275

\bibitem{beck2006superstatistical}
Beck C 2006 {\em Progress of Theoretical Physics Supplement\/} {\bf 162} 29--36

\bibitem{lampo2017cytoplasmic}
Lampo T~J, Stylianidou S, Backlund M~P, Wiggins P~A and Spakowitz A~J 2017 {\em
  Biophysical journal\/} {\bf 112} 532--542

\bibitem{hapca2009anomalous}
Hapca S, Crawford J~W and Young I~M 2009 {\em Journal of the Royal Society
  Interface\/} {\bf 6} 111--122

\bibitem{sadoon2018anomalous}
Sadoon A~A and Wang Y 2018 {\em Physical Review E\/} {\bf 98} 042411

\bibitem{mura2009class}
Mura A and Mainardi F 2009 {\em Integral Transforms and Special Functions\/}
  {\bf 20} 185--198

\bibitem{sposini2018random}
Sposini V, Chechkin A~V, Seno F, Pagnini G and Metzler R 2018 {\em New Journal
  of Physics\/} {\bf 20} 043044

\bibitem{slkezak2018superstatistical}
\'Slezak J, Metzler R and Magdziarz M 2018 {\em New Journal of Physics\/} {\bf
  20} 023026

\bibitem{mertz2019introduction}
Mertz J 2019 {\em Introduction to optical microscopy\/} (Cambridge University
  Press)

\bibitem{drescher2016architectural}
Drescher K, Dunkel J, Nadell C~D, Van~Teeffelen S, Grnja I, Wingreen N~S, Stone
  H~A and Bassler B~L 2016 {\em Proceedings of the National Academy of
  Sciences\/} {\bf 113} E2066--E2072

\bibitem{phanphak2019super}
Phanphak S, Georgiades P, Li R, King J, Roberts I~S and Waigh T~A 2019 {\em
  Langmuir\/} {\bf 35} 5635--5646

\bibitem{kenwright2012first}
Kenwright D~A, Harrison A~W, Waigh T~A, Woodman P~G and Allan V~J 2012 {\em
  Physical Review E\/} {\bf 86} 031910

\bibitem{roosen2011protein}
Roosen-Runge F, Hennig M, Zhang F, Jacobs R~M, Sztucki M, Schober H, Seydel T
  and Schreiber F 2011 {\em Proceedings of the National Academy of Sciences\/}
  {\bf 108} 11815--11820

\bibitem{madsen2010beyond}
Madsen A, Leheny R~L, Guo H, Sprung M and Czakkel O 2010 {\em New Journal of
  Physics\/} {\bf 12} 055001

\bibitem{holzgrafe2020nanoscale}
Holzgrafe J, Gu Q, Beitner J, Kara D~M, Knowles H~S and Atat{\"u}re M 2020 {\em
  Physical Review Applied\/} {\bf 13} 044004

\bibitem{wolff2023minflux}
Wolff J~O, Scheiderer L, Engelhardt T, Engelhardt J, Matthias J and Hell S~W
  2023 {\em Science\/} {\bf 379} 1004--1010

\bibitem{deguchi2023direct}
Deguchi T, Iwanski M~K, Schentarra E~M, Heidebrecht C, Schmidt L, Heck J, Weihs
  T, Schnorrenberg S, Hoess P, Liu S {\em et~al.\/} 2023 {\em Science\/} {\bf
  379} 1010--1015

\bibitem{mizuno2008active}
Mizuno D, Head D, MacKintosh F and Schmidt C 2008 {\em Macromolecules\/} {\bf
  41} 7194--7202

\bibitem{tassieri2008dynamics}
Tassieri M, Evans R, Barbu-Tudoran L, Khaname G~N, Trinick J and Waigh T~A 2008
  {\em Physical Review Letters\/} {\bf 101} 198301

\bibitem{georgiades2014particle}
Georgiades P, Pudney P~D, Thornton D~J and Waigh T~A 2014 {\em Biopolymers\/}
  {\bf 101} 366--377

\bibitem{papagiannopoulos2008viscoelasticity}
Papagiannopoulos A, Waigh T and Hardingham T 2008 {\em Faraday Discussions\/}
  {\bf 139} 337--357

\bibitem{rogers2008intracellular}
Rogers S~S, Waigh T~A and Lu J~R 2008 {\em Biophysical journal\/} {\bf 94}
  3313--3322

\bibitem{guadayol2021microrheology}
Guadayol {\`O}, Mendonca T, Segura-Noguera M, Wright A~J, Tassieri M and
  Humphries S 2021 {\em Proceedings of the National Academy of Sciences\/} {\bf
  118} e2011389118

\bibitem{michalet2010mean}
Michalet X 2010 {\em Physical Review E\/} {\bf 82} 041914

\bibitem{vestergaard2014optimal}
Vestergaard C~L, Blainey P~C and Flyvbjerg H 2014 {\em Physical Review E\/}
  {\bf 89} 022726

\bibitem{backlund2015chromosomal}
Backlund M~P, Joyner R and Moerner W~E 2015 {\em Physical Review E\/} {\bf 91}
  062716

\bibitem{lanoiselee2018optimal}
Lanoisel{\'e}e Y, Sikora G, Grzesiek A, Grebenkov D~S and Wy{\l}oma{\'n}ska A
  2018 {\em Physical review E\/} {\bf 98} 062139

\bibitem{kepten2015guidelines}
Kepten E, Weron A, Sikora G, Burnecki K and Garini Y 2015 {\em PLoS One\/} {\bf
  10} e0117722

\bibitem{gal2013particle}
Gal N, Lechtman-Goldstein D and Weihs D 2013 {\em Rheologica Acta\/} {\bf 52}
  425--443

\bibitem{selmeczi2008cell}
Selmeczi D, Li L, Pedersen L~I, Nrrelykke S, Hagedorn P~H, Mosler S, Larsen
  N~B, Cox E~C and Flyvbjerg H 2008 {\em The European Physical Journal Special
  Topics\/} {\bf 157} 1--15

\bibitem{burov2011single}
Burov S, Jeon J~H, Metzler R and Barkai E 2011 {\em Physical Chemistry Chemical
  Physics\/} {\bf 13} 1800--1812

\bibitem{weber2012analytical}
Weber S~C, Thompson M~A, Moerner W~E, Spakowitz A~J and Theriot J~A 2012 {\em
  Biophysical journal\/} {\bf 102} 2443--2450

\bibitem{arcizet2008temporal}
Arcizet D, Meier B, Sackmann E, R{\"a}dler J~O and Heinrich D 2008 {\em
  Physical review letters\/} {\bf 101} 248103

\bibitem{nandi2012distributions}
Nandi A, Heinrich D and Lindner B 2012 {\em Physical Review E\/} {\bf 86}
  021926

\bibitem{speckner2021single}
Speckner K and Weiss M 2021 {\em Entropy\/} {\bf 23} 892

\bibitem{lucy1974iterative}
Lucy L~B 1974 {\em The astronomical journal\/} {\bf 79} 745

\bibitem{redner2001guide}
Redner S 2001 {\em A guide to first-passage processes\/} (Cambridge university
  press)

\bibitem{rogers2010first}
Rogers S~S, Flores-Rodriguez N, Allan V~J, Woodman P~G and Waigh T~A 2010 {\em
  Physical Chemistry Chemical Physics\/} {\bf 12} 3753--3761

\bibitem{aalen2008survival}
Aalen O, Borgan O and Gjessing H 2008 {\em Survival and event history analysis:
  a process point of view\/} (Springer Science \& Business Media)

\bibitem{fedotov2018memory}
Fedotov S, Korabel N, Waigh T~A, Han D and Allan V~J 2018 {\em Physical Review
  E\/} {\bf 98} 042136

\bibitem{harrison2013modes}
Harrison A~W, Kenwright D~A, Waigh T~A, Woodman P~G and Allan V~J 2013 {\em
  Physical biology\/} {\bf 10} 036002

\bibitem{elson201340}
Elson E~L 2013 40 years of fcs: how it all began {\em Methods in enzymology\/}
  vol 518 (Elsevier) pp 1--10

\bibitem{lubelski2009fluorescence}
Lubelski A and Klafter J 2009 {\em Biophysical Journal\/} {\bf 96} 2055--2063

\bibitem{perkins2021network}
Perkins H~T, Allan V~J and Waigh T~A 2021 {\em Scientific Reports\/} {\bf 11}
  1--13

\bibitem{he2016dynamic}
He W, Song H, Su Y, Geng L, Ackerson B~J, Peng H and Tong P 2016 {\em Nature
  communications\/} {\bf 7} 1--8

\bibitem{witzel2019heterogeneities}
Witzel P, G{\"o}tz M, Lanoisel{\'e}e Y, Franosch T, Grebenkov D~S and Heinrich
  D 2019 {\em Biophysical journal\/} {\bf 117} 203--213

\bibitem{feller2008introduction}
Feller W 2008 {\em An introduction to probability theory and its applications,
  vol 1\/} (John Wiley \& Sons)

\bibitem{he2008random}
He Y, Burov S, Metzler R and Barkai E 2008 {\em Physical Review Letters\/} {\bf
  101} 058101

\bibitem{golding2006physical}
Golding I and Cox E~C 2006 {\em Physical review letters\/} {\bf 96} 098102

\bibitem{jeon2011vivo}
Jeon J~H, Tejedor V, Burov S, Barkai E, Selhuber-Unkel C, Berg-S{\o}rensen K,
  Oddershede L and Metzler R 2011 {\em Physical review letters\/} {\bf 106}
  048103

\bibitem{monthus1996models}
Monthus C and Bouchaud J~P 1996 {\em Journal of Physics A: Mathematical and
  General\/} {\bf 29} 3847

\bibitem{bouchaud1992weak}
Bouchaud J~P 1992 {\em Journal de Physique I\/} {\bf 2} 1705--1713

\bibitem{barkai2003aging}
Barkai E 2003 {\em Physical review letters\/} {\bf 90} 104101

\bibitem{burov2010aging}
Burov S, Metzler R and Barkai E 2010 {\em Proceedings of the National Academy
  of Sciences\/} {\bf 107} 13228--13233

\bibitem{pinholt2021single}
Pinholt H~D, Bohr S~S~R, Iversen J~F, Boomsma W and Hatzakis N~S 2021 {\em
  Proceedings of the National Academy of Sciences\/} {\bf 118} e2104624118

\bibitem{monnier2012bayesian}
Monnier N, Guo S~M, Mori M, He J, L{\'e}n{\'a}rt P and Bathe M 2012 {\em
  Biophysical journal\/} {\bf 103} 616--626

\bibitem{lysy2016model}
Lysy M, Pillai N~S, Hill D~B, Forest M~G, Mellnik J~W, Vasquez P~A and McKinley
  S~A 2016 {\em Journal of the American Statistical Association\/} {\bf 111}
  1413--1426

\bibitem{park2021bayesian}
Park S, Thapa S, Kim Y, Lomholt M~A and Jeon J~H 2021 {\em Journal of Physics
  A: Mathematical and Theoretical\/} {\bf 54} 484001

\bibitem{krog2018bayesian}
Krog J, Jacobsen L~H, Lund F~W, W{\"u}stner D and Lomholt M~A 2018 {\em Journal
  of Statistical Mechanics: Theory and Experiment\/} {\bf 2018} 093501

\bibitem{cherstvy2019non}
Cherstvy A~G, Thapa S, Wagner C~E and Metzler R 2019 {\em Soft Matter\/} {\bf
  15} 2526--2551

\bibitem{thapa2018bayesian}
Thapa S, Lomholt M~A, Krog J, Cherstvy A~G and Metzler R 2018 {\em Physical
  Chemistry Chemical Physics\/} {\bf 20} 29018--29037

\bibitem{persson2013extracting}
Persson F, Lind{\'e}n M, Unoson C and Elf J 2013 {\em Nature methods\/} {\bf
  10} 265--269

\bibitem{munoz2020single}
Mu{\~n}oz-Gil G, Garcia-March M~A, Manzo C, Mart{\'\i}n-Guerrero J~D and
  Lewenstein M 2020 {\em New Journal of Physics\/} {\bf 22} 013010

\bibitem{nielsen2019practical}
Nielsen A 2019 {\em Practical time series analysis: Prediction with statistics
  and machine learning\/} (O'Reilly Media)

\bibitem{granik2019single}
Granik N, Weiss L~E, Nehme E, Levin M, Chein M, Perlson E, Roichman Y and
  Shechtman Y 2019 {\em Biophysical journal\/} {\bf 117} 185--192

\bibitem{aggarwal2018neural}
Aggarwal C~C {\em et~al.\/} 2018 {\em Springer\/} {\bf 10} 978--3

\bibitem{zheng2018feature}
Zheng A and Casari A 2018 {\em Feature engineering for machine learning:
  principles and techniques for data scientists\/} (" O'Reilly Media, Inc.")

\bibitem{dosset2016automatic}
Dosset P, Rassam P, Fernandez L, Espenel C, Rubinstein E, Margeat E and Milhiet
  P~E 2016 {\em BMC bioinformatics\/} {\bf 17} 1--12

\bibitem{gentili2021characterization}
Gentili A and Volpe G 2021 {\em Journal of Physics A: Mathematical and
  Theoretical\/} {\bf 54} 314003

\bibitem{manzo2021extreme}
Manzo C 2021 {\em Journal of Physics A: Mathematical and Theoretical\/} {\bf
  54} 334002

\bibitem{bo2019measurement}
Bo S, Schmidt F, Eichhorn R and Volpe G 2019 {\em Physical Review E\/} {\bf
  100} 010102

\bibitem{argun2021classification}
Argun A, Volpe G and Bo S 2021 {\em Journal of Physics A: Mathematical and
  Theoretical\/} {\bf 54} 294003

\bibitem{garibo2021efficient}
Garibo-i Orts {\`O}, Baeza-Bosca A, Garcia-March M~A and Conejero J~A 2021 {\em
  Journal of Physics A: Mathematical and Theoretical\/} {\bf 54} 504002

\bibitem{szarek2021neural}
Szarek D 2021 {\em International Journal of Advances in Engineering Sciences
  and Applied Mathematics\/} {\bf 13} 257--269

\bibitem{arts2019particle}
Arts M, Smal I, Paul M~W, Wyman C and Meijering E 2019 {\em Scientific
  reports\/} {\bf 9} 17160

\bibitem{mackay1992practical}
MacKay D~J 1992 {\em Neural computation\/} {\bf 4} 448--472

\bibitem{seckler2022bayesian}
Seckler H and Metzler R 2022 {\em Nature Communications\/} {\bf 13} 6717

\bibitem{li2021wavenet}
Li D, Yao Q and Huang Z 2021 {\em Journal of Physics A: Mathematical and
  Theoretical\/} {\bf 54} 404003

\bibitem{verdier2021learning}
Verdier H, Duval M, Laurent F, Cass{\'e} A, Vestergaard C~L and Masson J~B 2021
  {\em Journal of Physics A: Mathematical and Theoretical\/} {\bf 54} 234001

\bibitem{drori2022science}
Drori I 2022 {\em The Science of Deep Learning\/} (Cambridge University Press)

\bibitem{firbas2023characterization}
Firbas N, Garibo-i Orts {\`O}, Garcia-March M~{\'A} and Conejero J~A 2023 {\em
  Journal of Physics A: Mathematical and Theoretical\/} {\bf 56} 014001

\bibitem{pml1Book}
Murphy K~P 2022 {\em Probabilistic Machine Learning: An introduction\/} (MIT
  Press) \urlprefix\url{probml.ai}

\bibitem{goodfellow2014generative}
Goodfellow I, Pouget-Abadie J, Mirza M, Xu B, Warde-Farley D, Ozair S,
  Courville A and Bengio Y 2014 {\em Advances in neural information processing
  systems\/} {\bf 27}

\bibitem{murphy2022probabilistic}
Murphy K~P 2022 {\em Probabilistic machine learning: an introduction\/} (MIT
  press)

\bibitem{munoz2021unsupervised}
Mu{\~n}oz-Gil G, i~Corominas G~G and Lewenstein M 2021 {\em Journal of Physics
  A: Mathematical and Theoretical\/} {\bf 54} 504001

\bibitem{newby2018convolutional}
Newby J~M, Schaefer A~M, Lee P~T, Forest M~G and Lai S~K 2018 {\em Proceedings
  of the National Academy of Sciences\/} {\bf 115} 9026--9031

\bibitem{mandelbrot1968fractional}
Mandelbrot B~B and Van~Ness J~W 1968 {\em SIAM review\/} {\bf 10} 422--437

\bibitem{coeurjolly2000simulation}
Coeurjolly J~F 2000 {\em Journal of statistical software\/} {\bf 5} 1--53

\bibitem{dieker2004simulation}
Dieker T 2004 {\em Simulation of fractional Brownian motion\/} Ph.D. thesis
  Masters Thesis, Department of Mathematical Sciences, University of Twente~…

\bibitem{janczura2021identifying}
Janczura J, Balcerek M, Burnecki K, Sabri A, Weiss M and Krapf D 2021 {\em New
  Journal of Physics\/} {\bf 23} 053018

\bibitem{balcerek2022fractional}
Balcerek M, Burnecki K, Thapa S, Wy{\l}oma{\'n}ska A and Chechkin A 2022 {\em
  arXiv preprint arXiv:2206.03818\/}

\bibitem{ayache2018new}
Ayache A, Esser C and Hamonier J 2018 {\em Risk and Decision Analysis\/} {\bf
  7} 5--29

\bibitem{chan1998simulation}
Chan G and Wood A~T 1998 Simulation of multifractional brownian motion {\em
  COMPSTAT\/} (Springer) pp 233--238

\bibitem{misteli2020self}
Misteli T 2020 {\em Cell\/} {\bf 183} 28--45

\bibitem{lieberman2009comprehensive}
Lieberman-Aiden E, Van~Berkum N~L, Williams L, Imakaev M, Ragoczy T, Telling A,
  Amit I, Lajoie B~R, Sabo P~J, Dorschner M~O {\em et~al.\/} 2009 {\em
  Science\/} {\bf 326} 289--293

\bibitem{cabal2006saga}
Cabal G~G, Genovesio A, Rodriguez-Navarro S, Zimmer C, Gadal O, Lesne A, Buc H,
  Feuerbach-Fournier F, Olivo-Marin J~C, Hurt E~C {\em et~al.\/} 2006 {\em
  Nature\/} {\bf 441} 770--773

\bibitem{weber2010bacterial}
Weber S~C, Spakowitz A~J and Theriot J~A 2010 {\em Physical review letters\/}
  {\bf 104} 238102

\bibitem{doi1988theory}
Doi M, Edwards S~F and Edwards S~F 1988 {\em The theory of polymer dynamics\/}
  vol~73 (oxford university press)

\bibitem{ashwin2020heterogeneous}
Ashwin S, Maeshima K and Sasai M 2020 {\em Biophysical Reviews\/} {\bf 12}
  461--468

\bibitem{salari2022spatial}
Salari H, Di~Stefano M and Jost D 2022 {\em Genome Research\/} {\bf 32} 28--43

\bibitem{javer2013short}
Javer A, Long Z, Nugent E, Grisi M, Siriwatwetchakul K, Dorfman K~D, Cicuta P
  and Cosentino~Lagomarsino M 2013 {\em Nature communications\/} {\bf 4} 1--8

\bibitem{oliveira2021precise}
Oliveira G~M, Oravecz A, Kobi D, Maroquenne M, Bystricky K, Sexton T and Molina
  N 2021 {\em Nature communications\/} {\bf 12} 1--11

\bibitem{nozaki2017dynamic}
Nozaki T, Imai R, Tanbo M, Nagashima R, Tamura S, Tani T, Joti Y, Tomita M,
  Hibino K, Kanemaki M~T {\em et~al.\/} 2017 {\em Molecular cell\/} {\bf 67}
  282--293

\bibitem{stadler2017non}
Stadler L and Weiss M 2017 {\em New Journal of Physics\/} {\bf 19} 113048

\bibitem{bronstein2009transient}
Bronstein I, Israel Y, Kepten E, Mai S, Shav-Tal Y, Barkai E and Garini Y 2009
  {\em Physical review letters\/} {\bf 103} 018102

\bibitem{bronshtein2015loss}
Bronshtein I, Kepten E, Kanter I, Berezin S, Lindner M, Redwood A~B, Mai S,
  Gonzalo S, Foisner R, Shav-Tal Y {\em et~al.\/} 2015 {\em Nature
  communications\/} {\bf 6} 1--9

\bibitem{shaban2018formation}
Shaban H~A, Barth R and Bystricky K 2018 {\em Nucleic acids research\/} {\bf
  46} e77--e77

\bibitem{shaban2020hi}
Shaban H~A, Barth R, Recoules L and Bystricky K 2020 {\em Genome biology\/}
  {\bf 21} 1--21

\bibitem{barth2020coupling}
Barth R, Bystricky K and Shaban H 2020 {\em Science advances\/} {\bf 6}
  eaaz2196

\bibitem{saintillan2018extensile}
Saintillan D, Shelley M~J and Zidovska A 2018 {\em Proceedings of the National
  Academy of Sciences\/} {\bf 115} 11442--11447

\bibitem{chuang2006long}
Chuang C~H, Carpenter A~E, Fuchsova B, Johnson T, de~Lanerolle P and Belmont
  A~S 2006 {\em Current Biology\/} {\bf 16} 825--831

\bibitem{levi2005chromatin}
Levi V, Ruan Q, Plutz M, Belmont A~S and Gratton E 2005 {\em Biophysical
  journal\/} {\bf 89} 4275--4285

\bibitem{zidovska2013micron}
Zidovska A, Weitz D~A and Mitchison T~J 2013 {\em Proceedings of the National
  Academy of Sciences\/} {\bf 110} 15555--15560

\bibitem{barth2022spatially}
Barth R and Shaban H~A 2022 {\em bioRxiv\/}

\bibitem{park2021mini}
Park S, Lee O~c, Durang X, Jeon J~H {\em et~al.\/} 2021 {\em Journal of the
  Korean Physical Society\/} {\bf 78} 408--426

\bibitem{stracy2021transient}
Stracy M, Schweizer J, Sherratt D~J, Kapanidis A~N, Uphoff S and Lesterlin C
  2021 {\em Molecular cell\/} {\bf 81} 1499--1514

\bibitem{normanno2015probing}
Normanno D, Boudar{\`e}ne L, Dugast-Darzacq C, Chen J, Richter C, Proux F,
  B{\'e}nichou O, Voituriez R, Darzacq X and Dahan M 2015 {\em Nature
  communications\/} {\bf 6} 1--10

\bibitem{stylianidou2014cytoplasmic}
Stylianidou S, Kuwada N~J and Wiggins P~A 2014 {\em Biophysical journal\/} {\bf
  107} 2684--2692

\bibitem{serge2008dynamic}
Serg{\'e} A, Bertaux N, Rigneault H and Marguet D 2008 {\em Nature methods\/}
  {\bf 5} 687--694

\bibitem{metzler2016non}
Metzler R, Jeon J~H and Cherstvy A 2016 {\em Biochimica et Biophysica Acta
  (BBA)-Biomembranes\/} {\bf 1858} 2451--2467

\bibitem{krapf2015mechanisms}
Krapf D 2015 {\em Current topics in membranes\/} {\bf 75} 167--207

\bibitem{masson2014mapping}
Masson J~B, Dionne P, Salvatico C, Renner M, Specht C~G, Triller A and Dahan M
  2014 {\em Biophysical journal\/} {\bf 106} 74--83

\bibitem{torreno2016uncovering}
Torreno-Pina J~A, Manzo C and Garcia-Parajo M~F 2016 {\em Journal of Physics D:
  Applied Physics\/} {\bf 49} 104002

\bibitem{charalambous2017nonergodic}
Charalambous C, Mu{\~n}oz-Gil G, Celi A, Garcia-Parajo M, Lewenstein M, Manzo C
  and Garc{\'\i}a-March M 2017 {\em Physical Review E\/} {\bf 95} 032403

\bibitem{sanz2023broadband}
Sanz-Paz M, van Zanten T~S, Manzo C, Mivelle M and Garcia-Parajo M~F 2023 {\em
  Small\/}  2207977

\bibitem{manley2008high}
Manley S, Gillette J~M, Patterson G~H, Shroff H, Hess H~F, Betzig E and
  Lippincott-Schwartz J 2008 {\em Nature methods\/} {\bf 5} 155--157

\bibitem{cutler2013multi}
Cutler P~J, Malik M~D, Liu S, Byars J~M, Lidke D~S and Lidke K~A 2013 {\em PloS
  one\/} {\bf 8} e64320

\bibitem{weron2017ergodicity}
Weron A, Burnecki K, Akin E~J, Sol{\'e} L, Balcerek M, Tamkun M~M and Krapf D
  2017 {\em Scientific reports\/} {\bf 7} 1--10

\bibitem{yamamoto2017dynamic}
Yamamoto E, Akimoto T, Kalli A~C, Yasuoka K and Sansom M~S 2017 {\em Science
  advances\/} {\bf 3} e1601871

\bibitem{jeon2016protein}
Jeon J~H, Javanainen M, Martinez-Seara H, Metzler R and Vattulainen I 2016 {\em
  Physical Review X\/} {\bf 6} 021006

\bibitem{campagnola2015superdiffusive}
Campagnola G, Nepal K, Schroder B~W, Peersen O~B and Krapf D 2015 {\em
  Scientific reports\/} {\bf 5} 1--10

\bibitem{wang2020non}
Wang D and Schwartz D~K 2020 {\em The Journal of Physical Chemistry C\/} {\bf
  124} 19880--19891

\bibitem{caspi2000enhanced}
Caspi A, Granek R and Elbaum M 2000 {\em Physical Review Letters\/} {\bf 85}
  5655

\bibitem{rennick2021key}
Rennick J~J, Johnston A~P and Parton R~G 2021 {\em Nature nanotechnology\/}
  {\bf 16} 266--276

\bibitem{flores2011roles}
Flores-Rodriguez N, Rogers S~S, Kenwright D~A, Waigh T~A, Woodman P~G and Allan
  V~J 2011 {\em PloS one\/} {\bf 6} e24479

\bibitem{korabel2018non}
Korabel N, Waigh T~A, Fedotov S and Allan V~J 2018 {\em PloS one\/} {\bf 13}
  e0207436

\bibitem{chen2015memoryless}
Chen K, Wang B and Granick S 2015 {\em Nature Materials\/} {\bf 14} 589--593

\bibitem{foret2012general}
Foret L, Dawson J~E, Villase{\~n}or R, Collinet C, Deutsch A, Brusch L, Zerial
  M, Kalaidzidis Y and J{\"u}licher F 2012 {\em Current Biology\/} {\bf 22}
  1381--1390

\bibitem{castro2021fusion}
Castro M, Lythe G, Smit J and Molina-Par{\'\i}s C 2021 {\em Scientific
  reports\/} {\bf 11} 1--13

\bibitem{alexandrov2022dynamics}
Alexandrov D~V, Korabel N, Currell F and Fedotov S 2022 {\em Cancer
  Nanotechnology\/} {\bf 13} 1--13

\bibitem{vicsek1985scaling}
Vicsek T, Meakin P and Family F 1985 {\em Physical Review A\/} {\bf 32} 1122

\bibitem{polev2022large}
Polev K, Kolygina D~V, Kandere-Grzybowska K and Grzybowski B~A 2022 {\em
  Cells\/} {\bf 11} 270

\bibitem{perkins2021intertwined}
Perkins H~T and Allan V 2021 {\em Cells\/} {\bf 10} 2341

\bibitem{georgiades2017flexibility}
Georgiades P, Allan V~J, Wright G~D, Woodman P~G, Udommai P, Chung M~A and
  Waigh T~A 2017 {\em Scientific reports\/} {\bf 7} 1--10

\bibitem{speckner2018anomalous}
Speckner K, Stadler L and Weiss M 2018 {\em Physical Review E\/} {\bf 98}
  012406

\bibitem{sabri2020elucidating}
Sabri A, Xu X, Krapf D and Weiss M 2020 {\em Physical Review Letters\/} {\bf
  125} 058101

\bibitem{etoc2018non}
Etoc F, Balloul E, Vicario C, Normanno D, Li{\ss}e D, Sittner A, Piehler J,
  Dahan M and Coppey M 2018 {\em Nature materials\/} {\bf 17} 740--746

\bibitem{montenegro2012mutations}
Montenegro G, Rebelo A~P, Connell J, Allison R, Babalini C, D’Aloia M,
  Montieri P, Sch{\"u}le R, Ishiura H, Price J {\em et~al.\/} 2012 {\em The
  Journal of clinical investigation\/} {\bf 122} 538--544

\bibitem{gupta2016protein}
Gupta S, Biehl R, Sill C, Allgaier J, Sharp M, Ohl M and Richter D 2016 {\em
  Macromolecules\/} {\bf 49} 1941--1949

\bibitem{cipelletti2011glassy}
Cipelletti L and Weeks E~R 2011 {\em Dynamical heterogeneities in glasses,
  colloids, and granular media\/} {\bf 150} 110

\bibitem{roberts2014therapeutic}
Roberts C~J 2014 {\em Trends in biotechnology\/} {\bf 32} 372--380

\bibitem{xia2020origin}
Xia C, He X, Wang J and Wang W 2020 {\em Physical Review E\/} {\bf 102} 062424

\bibitem{hassani2022multiscale}
Hassani A~N, Haris L, Appel M, Seydel T, Stadler A~M and Kneller G~R 2022 {\em
  The Journal of Chemical Physics\/} {\bf 156} 025102

\bibitem{kampf2012power}
K{\"a}mpf K, Klameth F and Vogel M 2012 {\em The Journal of chemical physics\/}
  {\bf 137} 205105

\bibitem{dieball2022scattering}
Dieball C, Krapf D, Weiss M and Godec A 2022 {\em New Journal of Physics\/}
  {\bf 24} 023004

\bibitem{taloni2010generalized}
Taloni A, Chechkin A and Klafter J 2010 {\em Physical review letters\/} {\bf
  104} 160602

\bibitem{calligari2012toward}
Calligari P and Abergel D 2012 {\em The Journal of Physical Chemistry B\/} {\bf
  116} 12955--12965

\bibitem{booth1997instability}
Booth D~R, Sunde M, Bellotti V, Robinson C~V, Hutchinson W~L, Fraser P~E,
  Hawkins P~N, Dobson C~M, Radford S~E, Blake C~C {\em et~al.\/} 1997 {\em
  Nature\/} {\bf 385} 787--793

\bibitem{carrick2005internal}
Carrick L, Tassieri M, Waigh T, Aggeli A, Boden N, Bell C, Fisher J, Ingham E
  and Evans R 2005 {\em Langmuir\/} {\bf 21} 3733--3737

\bibitem{hu2020recent}
Hu X, Liao M, Gong H, Zhang L, Cox H, Waigh T~A and Lu J~R 2020 {\em Current
  Opinion in Colloid \& Interface Science\/} {\bf 45} 1--13

\bibitem{cox2018single}
Cox H, Xu H, Waigh T~A and Lu J~R 2018 {\em Langmuir\/} {\bf 34} 14678--14689

\bibitem{cox2019active}
Cox H, Cao M, Xu H, Waigh T~A and Lu J~R 2019 {\em Biomacromolecules\/} {\bf
  20} 1719--1730

\bibitem{mcguckin2011mucin}
McGuckin M~A, Lind{\'e}n S~K, Sutton P and Florin T~H 2011 {\em Nature Reviews
  Microbiology\/} {\bf 9} 265--278

\bibitem{zhang2015particle}
Zhang X, Hansing J, Netz R~R and DeRouchey J~E 2015 {\em Biophysical journal\/}
  {\bf 108} 530--539

\bibitem{wagner2017rheological}
Wagner C~E, Turner B~S, Rubinstein M, McKinley G~H and Ribbeck K 2017 {\em
  Biomacromolecules\/} {\bf 18} 3654--3664

\bibitem{christodoulou2020live}
Christodoulou C, Spencer J~A, Yeh S~C~A, Turcotte R, Kokkaliaris K~D, Panero R,
  Ramos A, Guo G, Seyedhassantehrani N, Esipova T~V {\em et~al.\/} 2020 {\em
  Nature\/} {\bf 578} 278--283

\bibitem{luzhansky2018anomalously}
Luzhansky I~D, Schwartz A~D, Cohen J~D, MacMunn J~P, Barney L~E, Jansen L~E and
  Peyton S~R 2018 {\em APL bioengineering\/} {\bf 2} 026112

\bibitem{graf2008heterogeneity}
Graf T and Stadtfeld M 2008 {\em Cell stem cell\/} {\bf 3} 480--483

\bibitem{passucci2019identifying}
Passucci G, Brasch M~E, Henderson J~H, Zaburdaev V and Manning M~L 2019 {\em
  PLoS computational biology\/} {\bf 15} e1006732

\bibitem{giese1996dichotomy}
Giese A, Loo M~A, Tran N, Haskett D, Coons S~W and Berens M~E 1996 {\em
  International journal of cancer\/} {\bf 67} 275--282

\bibitem{fedotov2007migration}
Fedotov S and Iomin A 2007 {\em Physical Review Letters\/} {\bf 98} 118101

\bibitem{wu2014three}
Wu P~H, Giri A, Sun S~X and Wirtz D 2014 {\em Proceedings of the National
  Academy of Sciences\/} {\bf 111} 3949--3954

\bibitem{huda2018levy}
Huda S, Weigelin B, Wolf K, Tretiakov K~V, Polev K, Wilk G, Iwasa M, Emami F~S,
  Narojczyk J~W, Banaszak M {\em et~al.\/} 2018 {\em Nature communications\/}
  {\bf 9} 1--11

\bibitem{dieterich2008anomalous}
Dieterich P, Klages R, Preuss R and Schwab A 2008 {\em Proceedings of the
  National Academy of Sciences\/} {\bf 105} 459--463

\bibitem{krummel2016t}
Krummel M~F, Bartumeus F and G{\'e}rard A 2016 {\em Nature Reviews
  Immunology\/} {\bf 16} 193--201

\bibitem{miller2003autonomous}
Miller M~J, Wei S~H, Cahalan M~D and Parker I 2003 {\em Proceedings of the
  National Academy of Sciences\/} {\bf 100} 2604--2609

\bibitem{harris2012generalized}
Harris T~H, Banigan E~J, Christian D~A, Konradt C, Tait~Wojno E~D, Norose K,
  Wilson E~H, John B, Weninger W, Luster A~D {\em et~al.\/} 2012 {\em Nature\/}
  {\bf 486} 545--548

\bibitem{fricke2016persistence}
Fricke G~M, Letendre K~A, Moses M~E and Cannon J~L 2016 {\em PLoS computational
  biology\/} {\bf 12} e1004818

\bibitem{jerison2020heterogeneous}
Jerison E~R and Quake S~R 2020 {\em Elife\/} {\bf 9} e53933

\bibitem{maiuri2015actin}
Maiuri P, Rupprecht J~F, Wieser S, Ruprecht V, B{\'e}nichou O, Carpi N, Coppey
  M, De~Beco S, Gov N, Heisenberg C~P {\em et~al.\/} 2015 {\em Cell\/} {\bf
  161} 374--386

\bibitem{cavanagh2022t}
Cavanagh H, Kempe D, Mazalo J~K, Biro M and Endres R~G 2022 {\em Journal of the
  Royal Society Interface\/} {\bf 19} 20220081

\bibitem{lovely1975statistical}
Lovely P~S and Dahlquist F 1975 {\em Journal of theoretical biology\/} {\bf 50}
  477--496

\bibitem{berg2018random}
Berg H~C 2018 Random walks in biology {\em Random Walks in Biology\/}
  (Princeton University Press)

\bibitem{wu2000particle}
Wu X~L and Libchaber A 2000 {\em Physical review letters\/} {\bf 84} 3017

\bibitem{ariel2015swarming}
Ariel G, Rabani A, Benisty S, Partridge J~D, Harshey R~M and Be'Er A 2015 {\em
  Nature communications\/} {\bf 6} 1--6

\bibitem{zonia2009swimming}
Zonia L and Bray D 2009 {\em Journal of the Royal Society Interface\/} {\bf 6}
  1035--1046

\bibitem{koorehdavoudi2017multi}
Koorehdavoudi H, Bogdan P, Wei G, Marculescu R, Zhuang J, Carlsen R~W and Sitti
  M 2017 {\em Proceedings of the Royal Society A: Mathematical, Physical and
  Engineering Sciences\/} {\bf 473} 20170154

\bibitem{figueroa20203d}
Figueroa-Morales N, Soto R, Junot G, Darnige T, Douarche C, Martinez V~A,
  Lindner A and Cl{\'e}ment E 2020 {\em Physical Review X\/} {\bf 10} 021004

\bibitem{itto2021superstatistical}
Itto Y and Beck C 2021 {\em Journal of the Royal Society Interface\/} {\bf 18}
  20200927

\bibitem{huo2021swimming}
Huo H, He R, Zhang R and Yuan J 2021 {\em Applied and environmental
  microbiology\/} {\bf 87} e02429--20

\bibitem{feng2019single}
Feng J, Zhang Z, Wen X, Xue J and He Y 2019 {\em Iscience\/} {\bf 22} 123--132

\bibitem{zaid2011levy}
Zaid I~M, Dunkel J and Yeomans J~M 2011 {\em Journal of The Royal Society
  Interface\/} {\bf 8} 1314--1331

\bibitem{rogers2008microrheology}
Rogers S, Van Der~Walle C and Waigh T 2008 {\em Langmuir\/} {\bf 24}
  13549--13555

\bibitem{hart2019microrheology}
Hart J~W, Waigh T~A, Lu J~R and Roberts I~S 2019 {\em Langmuir\/} {\bf 35}
  3553--3561

\bibitem{worlitzer2022biophysical}
Worlitzer V~M, Jose A, Grinberg I, B{\"a}r M, Heidenreich S, Eldar A, Ariel G
  and Be’er A 2022 {\em Science advances\/} {\bf 8} eabn8152

\bibitem{xu2022autonomous}
Xu H, Huang Y, Zhang R and Wu Y 2022 {\em Nature Physics\/}  1--6

\bibitem{kosztolowicz2020modelling}
Koszto{\l}owicz T, Metzler R, Wasik S and Arabski M 2020 {\em Plos one\/} {\bf
  15} e0243003

\bibitem{kosztolowicz2020diffusion}
Koszto{\l}owicz T and Metzler R 2020 {\em Physical Review E\/} {\bf 102} 032408

\bibitem{leptos2009dynamics}
Leptos K~C, Guasto J~S, Gollub J~P, Pesci A~I and Goldstein R~E 2009 {\em
  Physical Review Letters\/} {\bf 103} 198103

\bibitem{kanazawa2020loopy}
Kanazawa K, Sano T~G, Cairoli A and Baule A 2020 {\em Nature\/} {\bf 579}
  364--367

\bibitem{granek2022anomalous}
Granek O, Kafri Y and Tailleur J 2022 {\em Physical Review Letters\/} {\bf 129}
  038001

\bibitem{li2008persistent}
Li L, N{\o}rrelykke S~F and Cox E~C 2008 {\em PLoS one\/} {\bf 3} e2093

\bibitem{cherstvy2018non}
Cherstvy A~G, Nagel O, Beta C and Metzler R 2018 {\em Physical Chemistry
  Chemical Physics\/} {\bf 20} 23034--23054

\bibitem{alves2016transient}
Alves L~G, Scariot D~B, Guimaraes R~R, Nakamura C~V, Mendes R~S and Ribeiro H~V
  2016 {\em PLoS One\/} {\bf 11} e0152092

\bibitem{peliti2021stochastic}
Peliti L and Pigolotti S 2021 {\em Stochastic Thermodynamics: An
  Introduction\/} (Princeton University Press)

\bibitem{de1992soft}
De~Gennes P~G 1992 {\em Science\/} {\bf 256} 495--497

\bibitem{olson2010linking}
Olson E~N and Nordheim A 2010 {\em Nature reviews Molecular cell biology\/}
  {\bf 11} 353--365

\bibitem{lauga2020fluid}
Lauga E 2020 {\em The fluid dynamics of cell motility\/} vol~62 (Cambridge
  University Press)

\bibitem{hirokawa2010molecular}
Hirokawa N, Niwa S and Tanaka Y 2010 {\em Neuron\/} {\bf 68} 610--638

\bibitem{woringer2020anomalous}
Woringer M, Izeddin I, Favard C and Berry H 2020 {\em Frontiers in Physics\/}
  {\bf 8} 134

\bibitem{hellmann2012enhancing}
Hellmann M, Heermann D~W and Weiss M 2012 {\em EPL (Europhysics Letters)\/}
  {\bf 97} 58004

\bibitem{sumner2021random}
Sumner M~C, Torrisi S~B, Brickner D~G and Brickner J~H 2021 {\em Elife\/} {\bf
  10}

\bibitem{lauga2016bacterial}
Lauga E 2016 {\em Annual Review of Fluid Mechanics\/} {\bf 48} 105--130

\bibitem{podlubny1999fractional}
Podlubny I 1999 {\em Mathematics in science and engineering\/} {\bf 198}
  41--119

\bibitem{gardiner2009stochastic}
Gardiner C 2009 {\em Stochastic methods\/} vol~4 (Springer Berlin)

\end{thebibliography}

\end{document}